\documentclass[journal]{IEEEtran}
\usepackage{cite}

\ifCLASSINFOpdf
  \usepackage[pdftex]{graphicx}
\else
  \usepackage[dvips]{graphicx}
\fi
\usepackage[cmex10]{amsmath}
\usepackage{algorithmic}

\usepackage{array}

\ifCLASSOPTIONcompsoc
  \usepackage[caption=false,font=normalsize,labelfont=sf,textfont=sf]{subfig}
\else
  \usepackage[caption=false,font=footnotesize]{subfig}
\fi
\usepackage{dblfloatfix}

\usepackage{booktabs}

\usepackage{url}

\usepackage{float}
\usepackage{multirow}
\usepackage{footnote}
\usepackage{amsmath, amsthm, amssymb}
\usepackage{xcolor}

\usepackage[para,online,flushleft]{threeparttable}
\begin{document}
%
\title{Machine Learning for the Detection and Identification of Internet of Things (IoT) Devices: A Survey}




\author{Yongxin~Liu,
        Jian~Wang,
        Jianqiang~Li,
        Shuteng Niu,
        and Houbing~Song,~\IEEEmembership{Senior Member,~IEEE}
\thanks{Yongxin Liu and Jianqiang Li are with the College of Computer Science and Software Engineering, Shenzhen University, China}
\thanks{Yongxin Liu, Jian Wang, Houbing Song and Shuteng Niu are with the Embry-Riddle Aeronautical University, Daytona Beach, FL 32114 USA}
\thanks{Corresponding authors: Jianqiang Li, Houbing Song}
\thanks{Manuscript received October 18, 2020; revised XXX.}}

\markboth{IEEE Internet of Things Journal,~Vol.~7, No.~5, May~2020}%
{Shell \MakeLowercase{\textit{et al.}}: Bare Demo of IEEEtran.cls for Journals}
%



\IEEEtitleabstractindextext{%
\begin{abstract}
The Internet of Things (IoT) is becoming an indispensable part of everyday life, enabling a variety of emerging services and applications. However, the presence of rogue IoT devices has exposed the IoT to untold risks with severe consequences. The first step in securing the IoT is detecting rogue IoT devices and identifying legitimate ones. \textcolor{black}{Conventional approaches use cryptographic mechanisms to authenticate and verify legitimate devices' identities. However, cryptographic protocols are not available in many systems. Meanwhile, these methods are less effective when legitimate devices can be exploited or encryption keys are disclosed.} Therefore, non-cryptographic IoT device identification and rogue device detection become efficient solutions to secure existing systems and will provide additional protection to systems with cryptographic protocols. Non-cryptographic approaches require more effort and are not yet adequately investigated. In this paper, we provide a comprehensive survey on \textcolor{black}{machine learning technologies for the identification of IoT devices along with the detection of compromised or falsified ones from the viewpoint of passive surveillance agents or network operators.} We classify the IoT device identification and detection into four categories: device-specific pattern recognition, Deep Learning enabled device identification, unsupervised device identification, and abnormal device detection. Meanwhile, we discuss various ML-related enabling technologies for this purpose. \textcolor{black}{These enabling technologies include learning algorithms, feature engineering on network traffic traces and wireless signals, continual learning, and abnormality detection. }



\end{abstract}

}

\maketitle

\IEEEdisplaynontitleabstractindextext

%
\IEEEpeerreviewmaketitle

\section{Introduction}
%
%
%
%
\IEEEPARstart{A}{s} a rapidly evolving field, the Internet of Things (IoT) involves the interconnection and interaction of smart objects, i.e., IoT devices with embedded sensors, onboard data processing capabilities, and means of communication, to provide automated services that would otherwise not be possible \cite{IIOT17}. Trillions of network-connected IoT devices are expected to emerge in the global network around 2020 \cite{IoT-BDA}. The IoT is becoming pervasive parts of everyday life, enabling a variety of emerging services and applications in cities and communities \cite{SC17}, including in health \cite{IoT-health}, transportation \cite{TCPS17}, energy/utilities, and other areas. Furthermore, big data analytics enables the move from the IoT to real-time control \cite{BDA19,jiang2020applying,zhang2020tree,liu2019domain}.

However, the IoT is subject to threats stemming from increased connectivity \cite{SP17, IoT-security}. For example, rogue IoT devices, \textit{defined as devices claiming a falsified identity or compromised legitimate devices}, have exposed the IoT to untold risks with severe consequences. Rogue IoT devices could conduct various attacks: forging the identity of trusted entities to access sensitive resources, hijacking legitimate devices to participate in distributed denial of service (DDoS) attacks\cite{IoT-security}, and etc. The problem of rogue devices becomes even more hazardous in wirelessly connected IoT, as the network traffic is easier to be intercepted, falsified, and broadcast broadly. \textcolor{black}{Hence, from the perspective of network operators}, the first step in securing the IoT from risks due to rogue devices is identifying known (or unknown) devices and detecting compromised ones. \textcolor{black}{This survey defines the term \textit{Device Detection and Identification to contain two perspectives: a) Identity verification of known devices. b) Detection of falsified or compromised devices}.}

\textcolor{black}{Conventional cryptographic mechanisms use message authentication code, digital signatures, challenge-response sessions, and etc. to authenticate legitimate peers or verify the identities of message senders. These methods make it mathematically impossible for the malicious to forge the legitimate ones' identities.} Even though cryptographic mechanisms are effective as long as critical keys are securely protected, security requirements may not be fully satisfied in pervasively distributed IoT. Reports have shown that it is possible to use reverse engineering to access encryption keys or conduct further exploitations \cite{wurm2016security,shwartz2018reverse,gupta2019iot,liu2020manually,costa2019vulnerabilities}. Moreover, \textcolor{black}{it is impossible to install cryptographic protocols into the huge amount of insecure systems or devices in a short time. Some of them have already become part of critical infrastructures \cite{pollack2018aviation,manesh2017analysis,dalaklis2018vulnerabilities,ray2015deais,yihunie2020assessing,koubaa2019micro}}. Finally, cryptographic approaches become less effective in dealing with hijacked devices. Therefore, as a supplementary to existing cryptography mechanisms, non-cryptographic Device Identification with Rogue Device Detection functions are needed to secure the IoT ecosystem \textcolor{black}{especially from the perspective of network operators and cybersecurity surveillance agents.}



\begin{figure*}[]
\centering
\includegraphics[width=\linewidth]{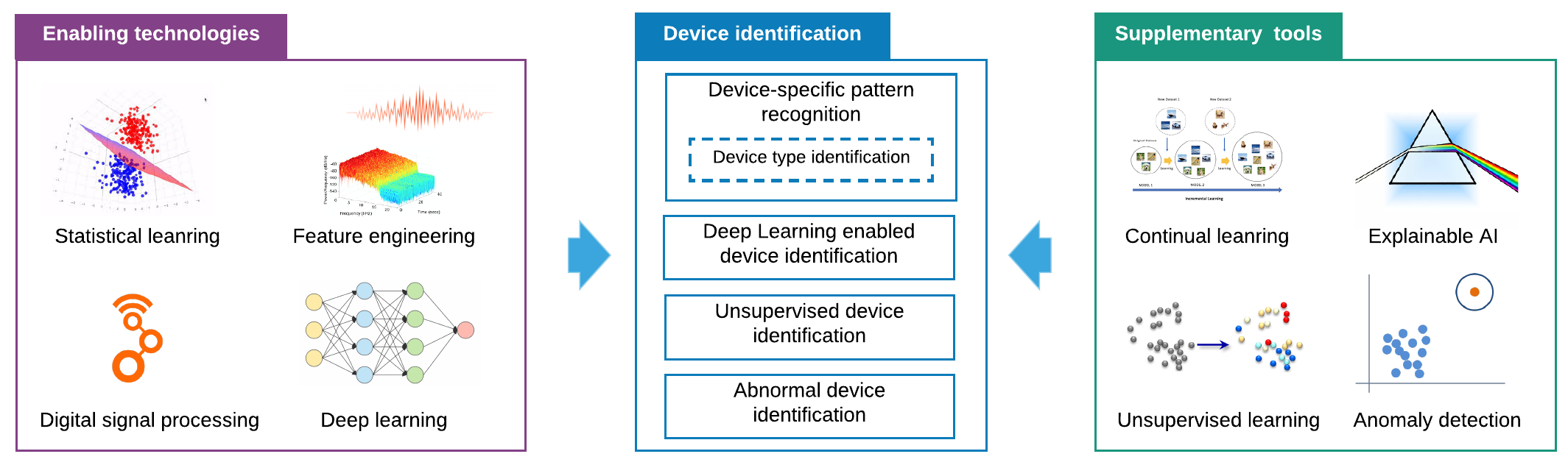}
\caption{Overview of ML for the Detection and Identification of Rogue IoT Devices}
\label{figTopicRelation}
\end{figure*}
Non-cryptographic device identification and rogue device detection have emerged as essential requirements in safety-critical IoT \cite{paul2008physical,wang2016wireless,khattak2019perception}. Compared with cryptographic approaches, non-cryptographic approaches aim to identify known devices and detect rogue devices by exploiting device-specific signal patterns or behavior characteristics \cite{brik2008wireless}. \textcolor{black}{More importantly, non-cryptographic approaches do not require modifications to existing systems that can not be upgraded easily, e.g., ADS-B \cite{riddle1090}, AIS \cite{tetreault2005use} and etc.}

\textcolor{black}{Non-cryptographic device identification and detection} are still challenging. Firstly, the flexible deployment scenarios and diverse specifications of devices make it challenging to provide a general guideline to derive distinctive features from signals or network traffic. Moreover, even though machine learning (ML) and Deep Learning (DL) have the potential to automatically discover distinctive latent features for accurate device identification, state-of-art algorithms require intensive modifications to be utilized in IoT \cite{amanullah2020deep}. Therefore, this domain is not yet thoroughly investigated and motivated us to conduct a comprehensive survey as a summary of existing works and anticipate the future development of this domain from the perspective of machine learning.

\begin{figure}[h]
\centering
\includegraphics[width=0.8\linewidth]{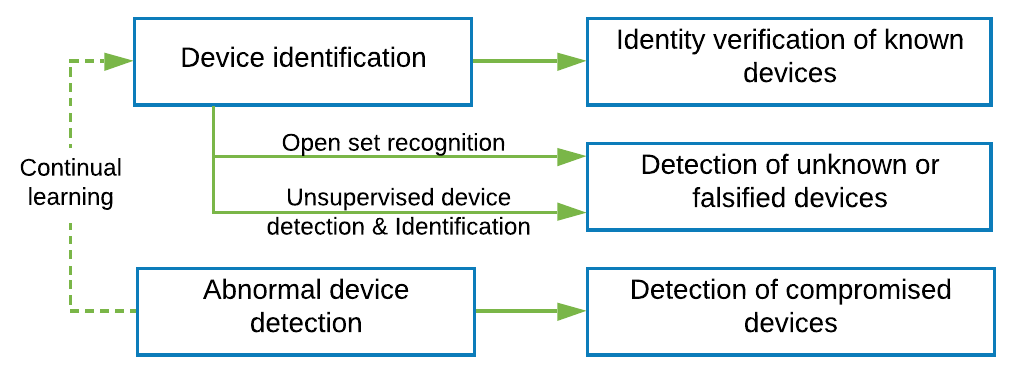}
\caption{\textcolor{black}{Key concepts in this survey}.}
\label{figRelatedTerms}
\end{figure}

The scope of this paper and related surveys are compared in Table~\ref{tabSurveyCompare}. In general, existing surveys focus on presenting broad overviews of threats and countermeasures in IoT. \textcolor{black}{In this paper, we focus on a more specific point by providing a comprehensive survey of machine learning for the detection and identification of devices in IoT using passively collected traffic traces and wireless signals, which are easily accessible to network operators and surveillance agents.} Figure~\ref{figTopicRelation} presents an overview of ML for the detection and identification of IoT devices \textcolor{black}{with relations between key concepts in Figure~\ref{figRelatedTerms}.} We classify the IoT device identification and detection into four categories: device-specific pattern recognition, Deep Learning enabled device identification, unsupervised device identification, and abnormal device detection. We identify various ML-related enabling technologies and tools for this purpose, including statistical learning, feature engineering, digital signal processing, and deep learning. These tools include continual learning, unsupervised learning, and anomaly detection. 
\begin{table}[h]
\centering
\caption{\textcolor{black}{A comparison with existing surveys}}
\label{tabSurveyCompare}
\begin{threeparttable}
\begin{tabular}{lllllll} 
\toprule
\multicolumn{1}{c}{Surveys} & \multicolumn{1}{c}{Year~} & \multicolumn{1}{c}{\begin{tabular}[c]{@{}c@{}}FD\end{tabular}} & \multicolumn{1}{c}{\begin{tabular}[c]{@{}c@{}}DL\end{tabular}} & \multicolumn{1}{c}{\begin{tabular}[c]{@{}c@{}}DT\end{tabular}} & \begin{tabular}[c]{@{}l@{}}UD\end{tabular} & \multicolumn{1}{c}{\begin{tabular}[c]{@{}c@{}}RD\end{tabular}} \\ 
\hline
\begin{tabular}[c]{@{}l@{}}\cite{al2020survey}\end{tabular} & 2020 & $\bullet$ & $\bullet$ &  &  & $\bullet$ \\
\begin{tabular}[c]{@{}l@{}}\cite{wang2019physical}\end{tabular} & 2019 & $\bullet$ &  & $\bullet$ &  & $\bullet$ \\
\begin{tabular}[c]{@{}l@{}}\cite{baldini2017survey}\end{tabular} & 2017 & $\bullet$ & $\bullet$ & $\bullet$ &  &  \\
\begin{tabular}[c]{@{}l@{}}\cite{danev2012physical}\end{tabular} & 2012 & $\bullet$ &  &  &  & $\bullet$ \\
\begin{tabular}[c]{@{}l@{}}\cite{zeng2010non}\end{tabular} & 2010 & $\bullet$ & & & $\bullet$ & $\bullet$ \\
\begin{tabular}[c]{@{}l@{}}This paper\end{tabular} & 2021 & $\bullet$ & $\bullet$ & $\bullet$ & $\bullet$ & $\bullet$ \\

\bottomrule
\end{tabular}
\begin{tablenotes}
\textcolor{black}{FD: Feature-based specific device identification; DL: Deep Learning enabled specific deivice identification; DT: Device type identification; UD: Unsupervised device identification; RD: Rogue device detection.}
\end{tablenotes}
\end{threeparttable}
\end{table}

\textcolor{black}{The remainder of this paper is structured as follows. Section \ref{sectThreatMode} presents a general threat model and attack chain of rogue devices in IoT. In Section \ref{sectLEDID}, we review device type identification (Section \ref{sectDTI}) and statistical learning on device-specific feature identification (Section \ref{sectDSFID}), including conventional radiometric signature and statistical learning. In Section \ref{sectDL} we review state-of-the-art Deep Learning (DL) based methods for device identification with a focus on emerging issues such as continual learning, abnormality detection, hyperparameter, and architecture search. A novel emerging approach, unsupervised device detection, is reviewed in Section \ref{sectUDI}. In Section \ref{sectLEADD}, we present methodologies to detect compromised wireless devices using anomaly detection algorithms, which is complementary to device-specific identification. Section \ref{sectChallengesAndFuture} pinpoints the challenges and future research directions with discussions on enabling technologies.} Section \ref{sectConclusion} concludes this paper.

\begin{figure*}[]
\centering
\includegraphics[scale = 0.8]{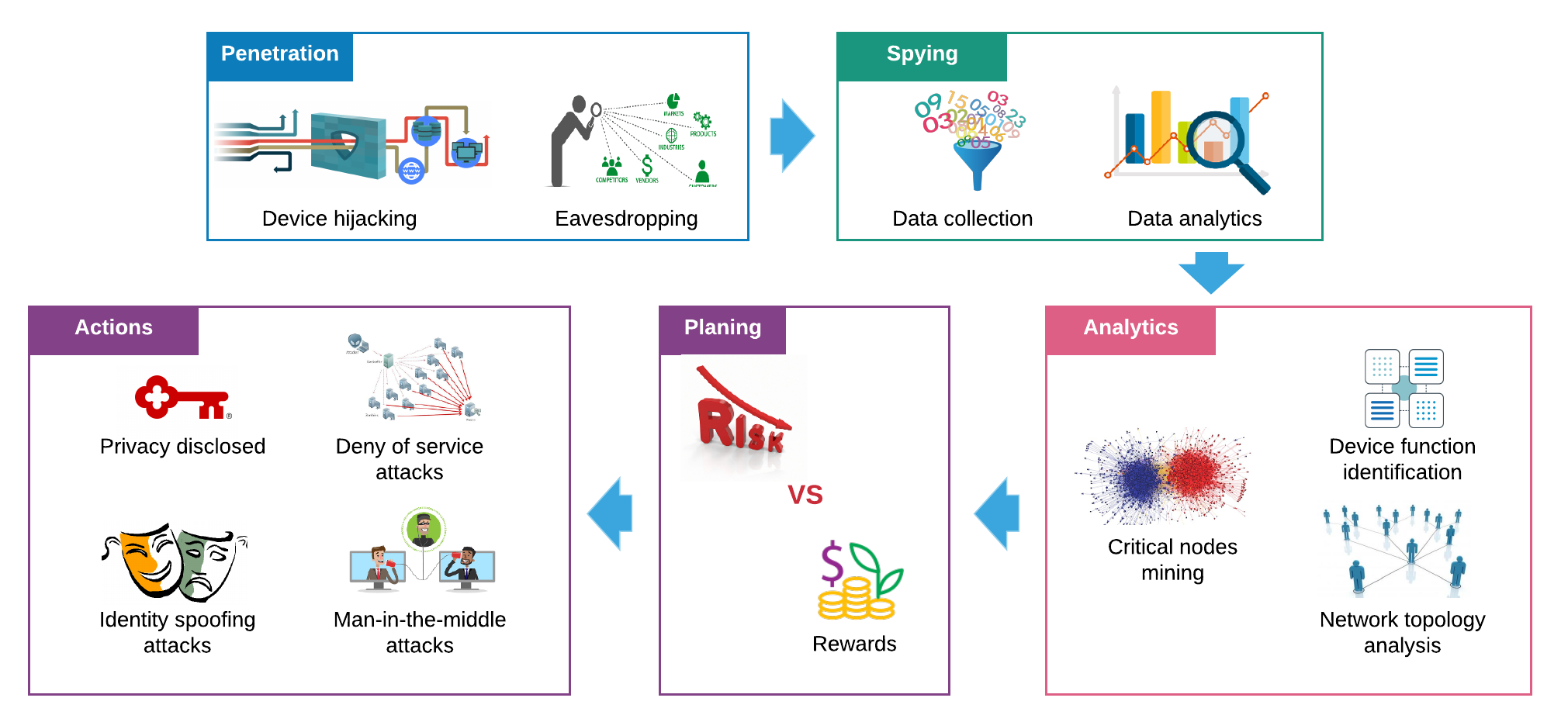}
\caption{Attack chain in the IoT.}
\label{figIoTAttack}
\end{figure*}

\begin{table*}[b]
\centering
\caption{Compare of cryptographic and non-cryptographic countermeasures}
\label{tabCryptographicVsNonCrypto}
\begin{tabular}{@{}clll@{}}
\toprule
Methods & \multicolumn{1}{c}{Principles} & \multicolumn{1}{c}{Advantages} & \multicolumn{1}{c}{Challenges} \\ \midrule
Cryptographic & \begin{tabular}[c]{@{}l@{}}Use shared secrecy to mathematically \\ make the decryption of sensitive \\ information and forge of identity \\computationally expensive.\end{tabular} & \begin{tabular}[c]{@{}l@{}}$\bullet~$Device independent\\$\bullet~$Protects both confidentiality \\ and can verify identity\end{tabular} & \begin{tabular}[c]{@{}l@{}}$\bullet~$Disclosure of secret keys.\\$\bullet~$Re-distribution of secret keys.\\$\bullet~$Needs special adapation to \\ existing systems.\end{tabular} \\ \midrule
Non-cryptographic & \begin{tabular}[c]{@{}l@{}}Extract and verify device-specific \\ features from received messages to \\ assure that messages are from known \\ sources.\end{tabular} & \begin{tabular}[c]{@{}l@{}}$\bullet~$Device-specific.\\$\bullet~$Can identify Hijacked\\ devices with abnormal behaviors. \\$\bullet~$compatible with existing IoT\end{tabular} & \begin{tabular}[c]{@{}l@{}}$\bullet~$Computationally expensive.\\$\bullet~$Identity disclosure.\end{tabular} \\ \bottomrule
\end{tabular}
\end{table*}

\section{Threat mode of rogue devices in IoT}
\label{sectThreatMode}

This section briefly reviews the threat modes of rogue devices along with countermeasures in IoT. \textcolor{black}{We analyze the attack chain and identify the essential requirements of IoT device detection and identification: verifying legitimate devices' identity, detecting unknown or falsified devices, and detecting compromised (hijacked) devices with abnormal behaviors. }

The cyberinfrastructure of IoT allows sharing information and collaborating among devices with different capacities and vulnerabilities. On the one hand, this scheme cultivates a large open system with low entry restrictions. On the other hand, adversaries can conduct rogue activities with great convenience \cite{wang2018security}. Generally, the attack modes of adversaries in IoT are in two folds: passive attack and proactive attacks. In a passive attack, adversaries do not cause damage or performance degradation for a long time. Still, they passively analyze devices' communication and activity patterns, providing road maps for proactive attacks in the future. If we regard passive attackers as spies secretly and peacefully gathering intelligence, the proactive attackers do whatever possible to degrade performances or exploit devices to conduct malicious activities. 
\begin{figure}[]
\centering
\includegraphics[width=\linewidth]{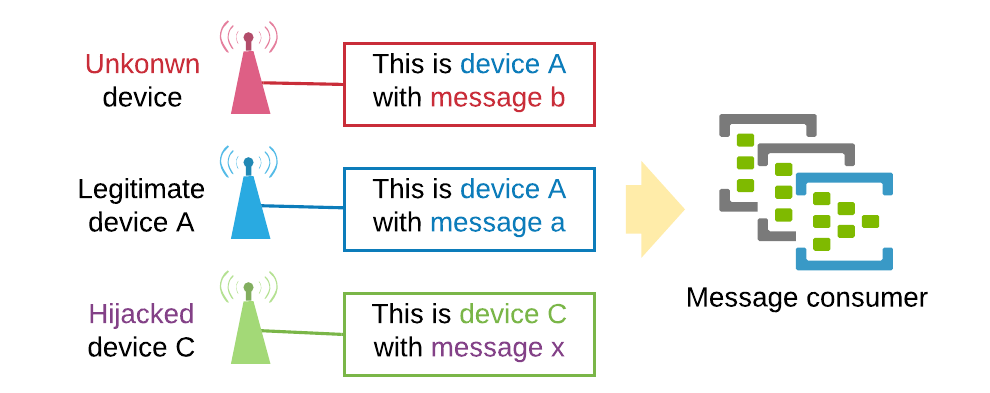}
\caption{Identity spoofing attacks.}
\label{figGeneralAttack}
\end{figure}
In practical attacks, proactive and passive attacks are combined. A typical attack chain to IoT systems is shown in Figure~\ref{figIoTAttack} with a more specific demonstration of identifying spoofing attack depicted in Figure~\ref{figGeneralAttack}. We divided the whole attack chain into five stages, as follows:
\begin{enumerate}
    \item \textit{Penetration: }In this stage, the rogue IoT devices try to eavesdrop on communication channels or attain the control privileges of vulnerable peers for further actions. \textcolor{black}{Research in \cite{loi2017systematically} shows that using ARP (Address Resolution Protocol) spoofing, the malicious can easily observe ongoing traffic generated by connected IoT devices from more than 20 manufacturers. Nowadays, it is still challenging to develop software stacks with assured security \cite{guaman2020systematic}.}
    \item \textit{Spying: }In this stage, the malicious will observe the ongoing activities by exploiting penetrated devices as its agents. \textcolor{black}{As in \cite{loi2017systematically}, more than 50\% of tested popular smart home IoT devices contain at least one vulnerable ports.}
    \item \textit{Data analytics: }The malicious analyses the behaviors and evaluate the vulnerabilities of the IoT from multiple perspectives. \textcolor{black}{An example in \cite{ren2019information} reveals that even if encryption mechanisms are employed, an attacker can still extract sensitive information, such as manufacture, device functionality, and etc.}
    \item \textit{Planning: }In this stage, the adversaries perform strategic planning and wait for the best time to minimize their risk while maximizing the rewards.
    \item \textit{Attack: }In this stage, prevalent attacks are in action. 
\end{enumerate}
Among these stages, passive and proactive attacks are combined in the penetration stage. \textcolor{black}{From the perspective of network operators or cybersecurity surveillance agents}, if we can prevent the adversaries from successfully impersonating legitimate devices in the first stage (penetration) or can identify hijacked devices in the second stage (spying). Network operators and surveillance agents can destroy the whole attack chain. 

Various countermeasures can be applied to secure IoT systems for IoT device identification and detection. Both cryptographic and non-cryptographic methods can be applied. A brief comparison of them is presented in Table~\ref{tabCryptographicVsNonCrypto}. Cryptographic methods are widely used in computer networks and telecommunication systems. However, special modifications are needed to deploy cryptographic protocols to existing systems without cryptographic protocols such as ADS-B, AIS, and etc. Non-cryptographic methods require higher computational capacities to derive device-specific fingerprints, but they are transparently compatible with existing systems. 


\section{Learning-Enabled Device Identification in IoT}
\label{sectLEDID}
\begin{figure}[h]
\centering
\includegraphics[width=0.9\linewidth]{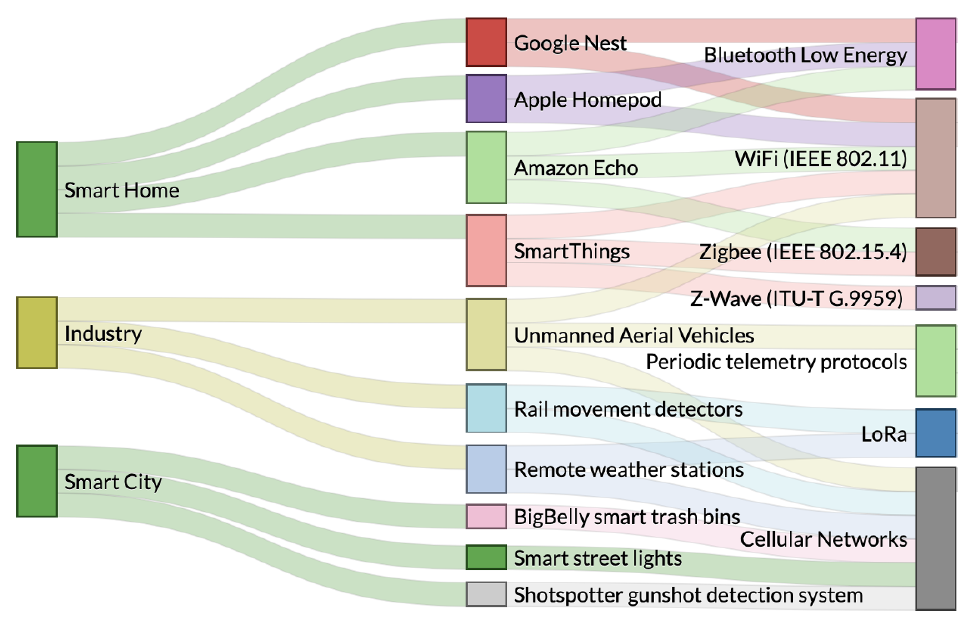}
\caption{\textcolor{black}{Typical IoT devices and protocols.}}
\label{figTypicalIoTdevices}
\end{figure}

\begin{figure}[h]
\centering
\includegraphics[width=\linewidth]{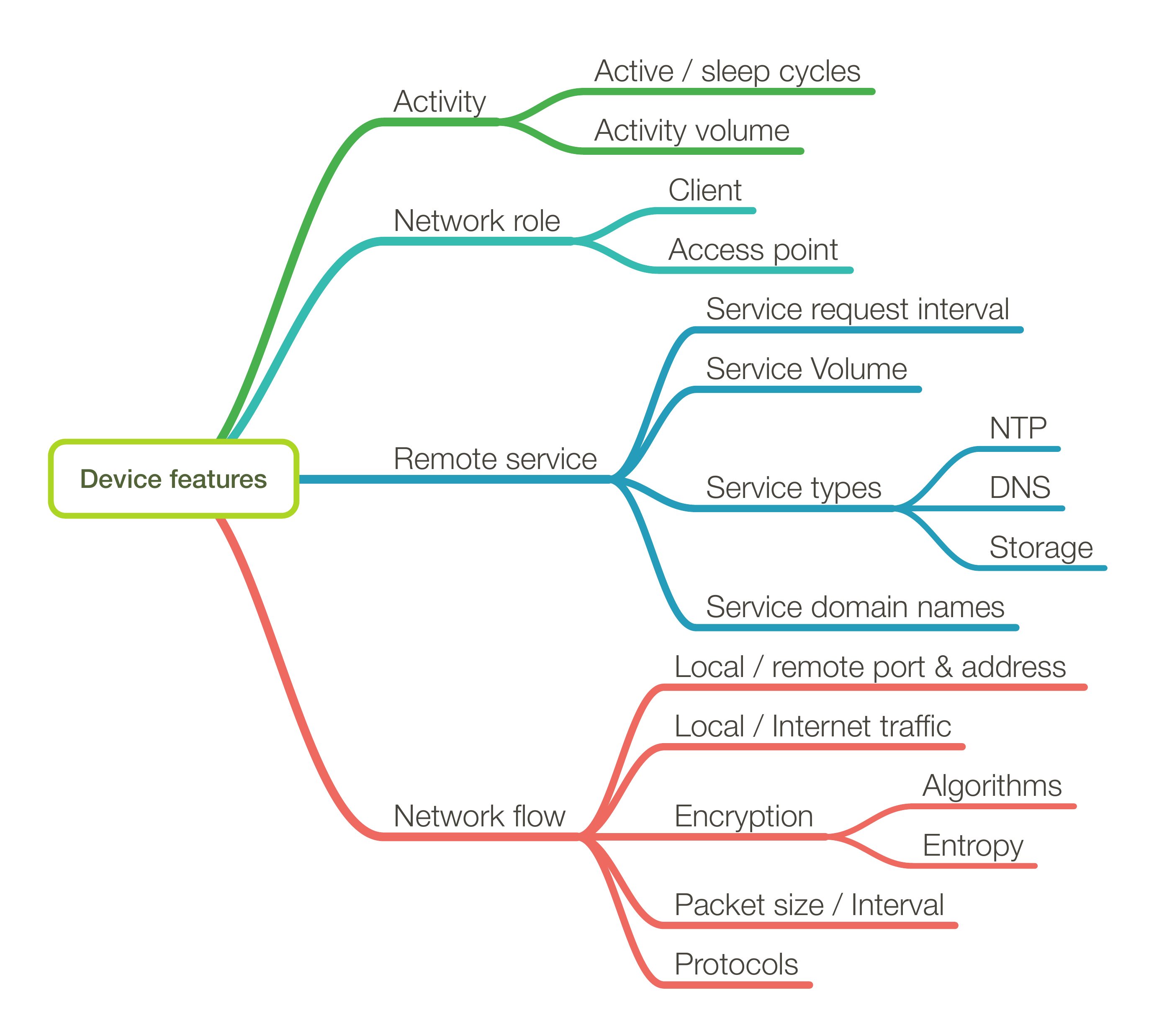}
\caption{\textcolor{black}{Features for device type identification.}}
\label{figDeviceTypeID}
\end{figure}

\textcolor{black}{This section reviews methods to recognize devices' identities and types in IoT. Most of them are based on network traffic and wireless signal pattern recognition. We first review device type identification methods, which are widely used in identifying commercial IoT devices. We then discuss and compare corresponding signal feature-based device recognition approaches. Especially, We discuss Deep Learning in device identification with emerging issues extensively. Finally, we review the unsupervised device identification and its open issues.}
\textcolor{black}{
\subsection{Device type identification}
\label{sectDTI}
Even though device types are not directly related to devices' identities, they still provide essential information for network management and risk control. A brief diagram of typical IoT devices is in Figure~\ref{figTypicalIoTdevices}, and comparisons of their Physical Layer, Data Link Layer as well as aggregated data transmission characteristics are presented in \cite{al2017internet}, \cite{jawhar2018networking} and \cite{metzger2019modeling}, respectively. As in Figure~\ref{figTypicalIoTdevices}, WiFi is pervasively utilized in smart homes while smart cities prefer reliable cellular networks. Device type identifications are frequently performed on network, transportation, and application layers and implemented in Software Defined Network (SDN) controllers or Software Routers \cite{han2010packetshader,hu2014survey,rafique2020complementing}. Device types reveal functionalities and activity profiles. A taxonomy of features for device type identification is presented in Figure~\ref{figDeviceTypeID}. }

\textcolor{black}{As in Figure~\ref{figDeviceTypeID}, remote service is a popular attack surface to disclose the device type or even identity. The reason is that the IoT devices communicate with remote service providers through the REST API \cite{gao2011restful}. Even though sensitive data are encrypted, some unique strings in their Web requests can still be exploited to infer device types. Authors in \cite{sivanathan2020iot} present that using only port numbers, domain names, and cipher suites, a Naive Bayesian classifier can reach high accuracy in classifying 28 commercial IoT devices.}

\begin{figure*}[]
\centering
\includegraphics[scale=0.9]{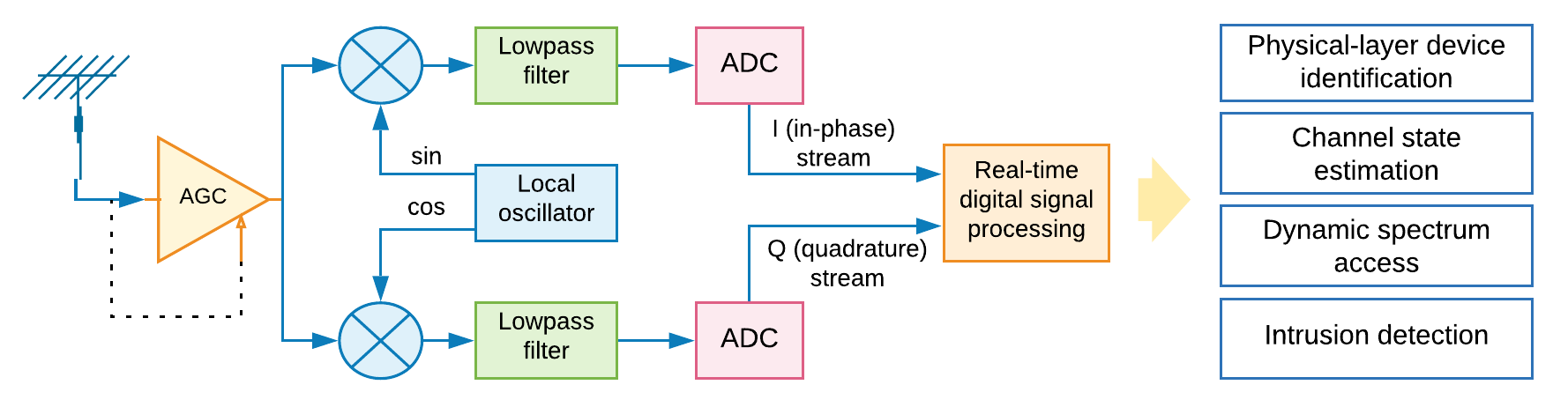}
\caption{General pipeline of Software-Defined wireless signal identification.}
\label{figSDRGe}
\end{figure*}
\textcolor{black}{Even though modeling devices' remote service requests provides promising results in device type identification, these solutions may not work if they interact with anonymous service providers. For alleviation, their activity and data flow patterns can be utilized. Authors in \cite{sivanathan2017characterizing} propose that their Random Forest classifier reaches a high accuracy of 95\% in identifying 20 IoT devices when features of activities, network data flows, and remote service requests are utilized simultaneously. In \cite{marchal2019audi}, devices' types are identified based on the periodicity of activities. The authors first use the Discrete Fourier Transform (DFT) and discrete autocorrelation to find the dominant periods in protocol-specific activities. They then use statistical and stability metrics to model devices' behaviors. Finally, the Bayesian-optimized k-Nearest Neighbor algorithm is employed for classification. In \cite{miettinen2017iot} and \cite{meidan2017profiliot}, the authors extract the protocols and network flow properties within a sliding window to generate fingerprints of devices. They use one-versus-rest classifiers to identify commercial devices. In \cite{meidan2017detection}, The authors first provide a Random Forest classifier using TCP/IP stream features. They incorporate confidence thresholds and averaged decisions within a sliding window to identify known or unknown device types. Similar research is presented in \cite{meidan2017detection} and \cite{aksoy2019automated}. In \cite{sivanathan2020managing}, the authors also present that network traffic, device types, and their operation states (boot, active, and idle) can be inferred simultaneously. }

\textcolor{black}{To automate the processes to derive useful features, in \cite{aksoy2019automated}, the authors propose a Genetic Algorithm (GA) enabled feature selector. Furthermore, a Deep Neural Network approach, which does not require complicated feature engineering, is presented in \cite{kotak2020iot}.} 

\textcolor{black}{An extra benefit of modeling device activity patterns is increasing the chances of identifying behavioral variations. Such benefit directly contributes to the detection of compromised devices or network attacks, which will be discussed in section \ref{sectLEADD}.}

Deriving devices' benign flow characteristics is nontrivial, therefore, the IETF standard Manufacturer Usage Description (MUD) profile \cite{lear2019rfc} is proposed as an initial static profile to describe IoT device network behavior and support the making of security policies. A collection of MUD profiles from 30 commercial devices in \cite{iotanalytics}. The MUD profiles can be used to either verify device types or detect devices under attack or being compromised \cite{hamza2020verifying}. However, one issue of using the static profiles is that longer observation time is needed to make decisions.

Device identifiers based on network flow and activity patterns may encounter emerging issues. First, IoT devices are becoming smart devices where new extensions can be installed, and firmware upgrades can happen periodically, thereby changing activity patterns or network flow statistics, as suggested in \cite{ashibani2018user, ashibani2020design} and \cite{sivanathan2020iot}. Second, device types do not necessarily correlate to their identities. \textcolor{black}{Therefore, behavior-independent specific device identification is of great significance.}

\subsection{\textcolor{black}{Feature-based statistical learning for specific device identification}}
\label{sectDSFID}
\textcolor{black}{IoT device identification can be formalized as a classification problem. In this section, we first introduce the generic pipeline for signal reception and then focus on feature-based statistical learning approaches for specific device identification from raw signals and their open issues. }

\textcolor{black}{\subsubsection{Generic wireless signal reception pipeline for device identification}
Software-Defined Radios (SDR) are multipurpose front-ends to deal with various modulation and baseband encoding schemes in wireless device identification. Fundamental technologies in SDR are quadrature modulation and demodulation \cite{zvonar2001software}. }

Generally, wireless signals of IoT devices can be represented as: $S(t)=\boldsymbol{I(t)}\cdot cos[2\pi (f_c+f')t]+\boldsymbol{Q(t)}\cdot sin[2\pi (f_c+f')t]$, where $\boldsymbol{I(t)}$ and $\boldsymbol{Q(t)}$ are denoted as in-phase and quadrature components, respectively. The key idea is use $\boldsymbol{I(t)}$ and $\boldsymbol{Q(t)}$ to represent different modulation schemes. 

A brief quadrature demodulation pipeline is given in Figure~\ref{figSDRGe}. \textcolor{black}{We denote the reconstructed version of $\boldsymbol{I(t)}$ and $\boldsymbol{Q(t)}$ as $\boldsymbol{\hat{I}(t)}$ and $\boldsymbol{\hat{Q}(t)}$, respectively. We can derive the signals instantaneous amplitude, phase, and frequency by $\hat{m}(t) = \sqrt{\hat{I}^2(t)+\hat{Q}^2(t)}$, $\hat{\phi}(t) = tan^{-1}(\hat{Q}(t)/\hat{I}(t))$ and $\hat{f}(t) = \partial \hat{\phi}(t) / \partial t$. }
\begin{figure}[h]
\centering
\includegraphics[width=\linewidth]{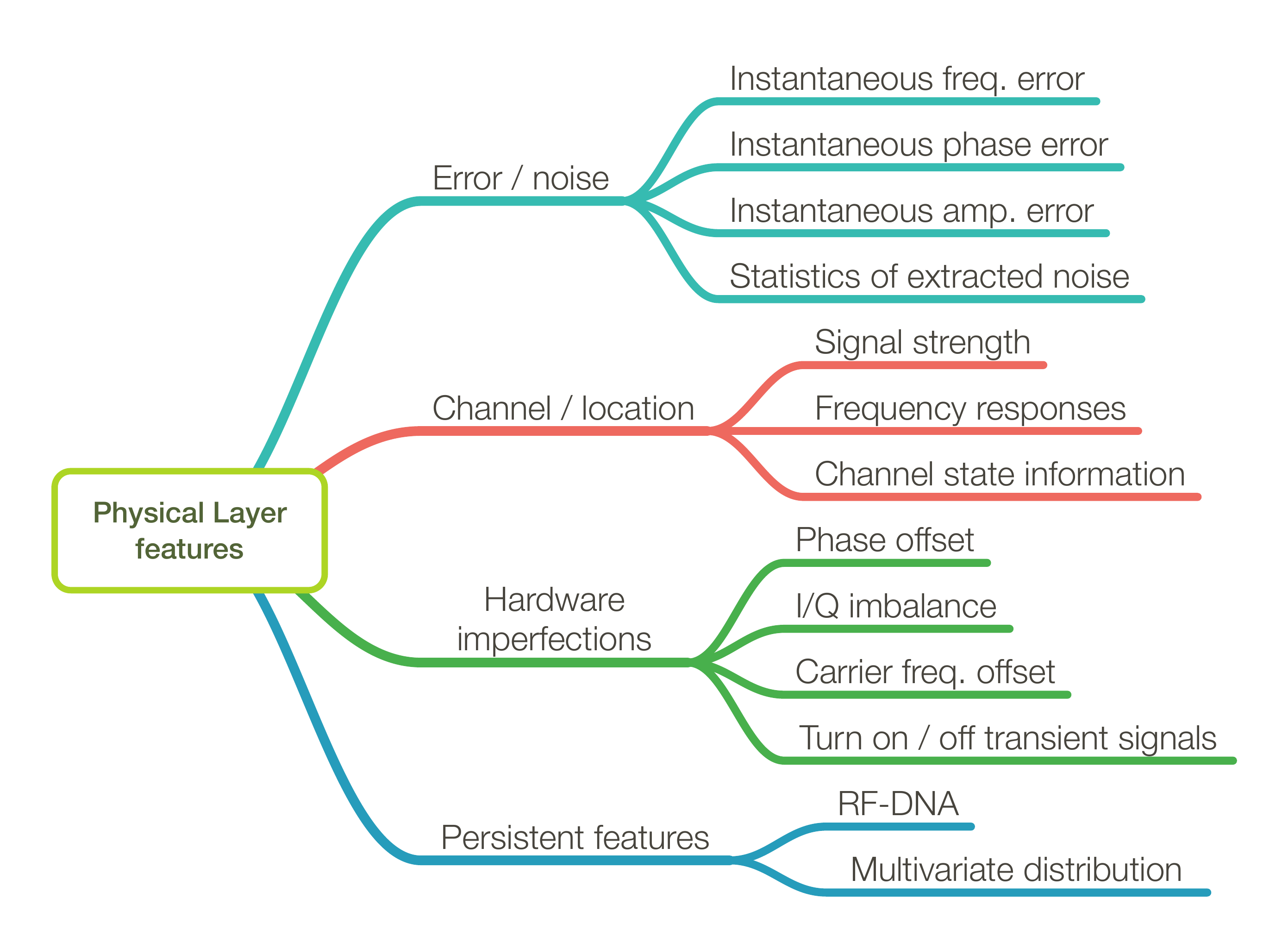}
\caption{\textcolor{black}{Physical Layer device-specific features.}}
\label{figPhysicalLayerFeatures}
\end{figure}
\textcolor{black}{Manufacturing imperfections and channel characteristics can cause $\hat{m}(t)$, $\hat{\phi}(t)$ and $\hat{f}(t)$ to deviate from its original form, providing side channels to identify wireless devices. A brief overview of features for IoT device identity verification \textcolor{black}{using wireless signals in Physical Layer} is given in Figure~\ref{figPhysicalLayerFeatures}. The features for wireless device identification are also named Radiometric Fingerprints.}

\subsubsection{Hardware imperfections}
Heterogeneous imperfections exist in IoT devices' wireless frontends. These imperfections do not necessarily degrade the communication performance but influence signal waveforms, \textcolor{black}{thereby providing a side channel to identify different devices. Such features enclosed in transmitted signals are named Physical Unclonable Features (PUF) \cite{chatterjee2018rf,herder2014physical}) since regular users can not clone or forge the characteristics of these manufacturing imperfections.}

\paragraph{Error / noise patterns}\textcolor{black}{The errors between expected rational signals and actual received signals can disclose useful device-specific information.} In \cite{polak2015wireless} and \cite{azarmehr2017wireless}, the authors use phase errors of Phase Lock Loop (PLL) in transmitters as a distinctive feature. Their simulations indicate promising results even with low SNR (Signal-to-Noise Ratio). In \cite{zhuang2018fbsleuth}, the authors use the instantaneous differences between received I/Q signals and theoretically expected templates to construct error vectors. They then combine error vectors' statistics and time-frequency domain statistics to synthesize the fingerprints of RF transmitters. 

In \cite{peng2016differential,peng2018design,peng2019deep}, the authors use the differential constellation trace figure (DCTF), carrier frequency offset, phase offset, and I/Q offset to identify different Zigbee devices. They develop a low-overhead classifier, which learns how to adjust feature weights under different SNRs. The behaviors of their classifiers are similar to k-NN algorithms. Authors in \cite{zhang2016identification} use odd harmonics of center frequencies as fingerprints for RFID transmitters. Their classification (k-NN) test on 300 RFID cards shows zero error. 

\begin{figure}[]
\centering
\includegraphics[width=1.02\linewidth]{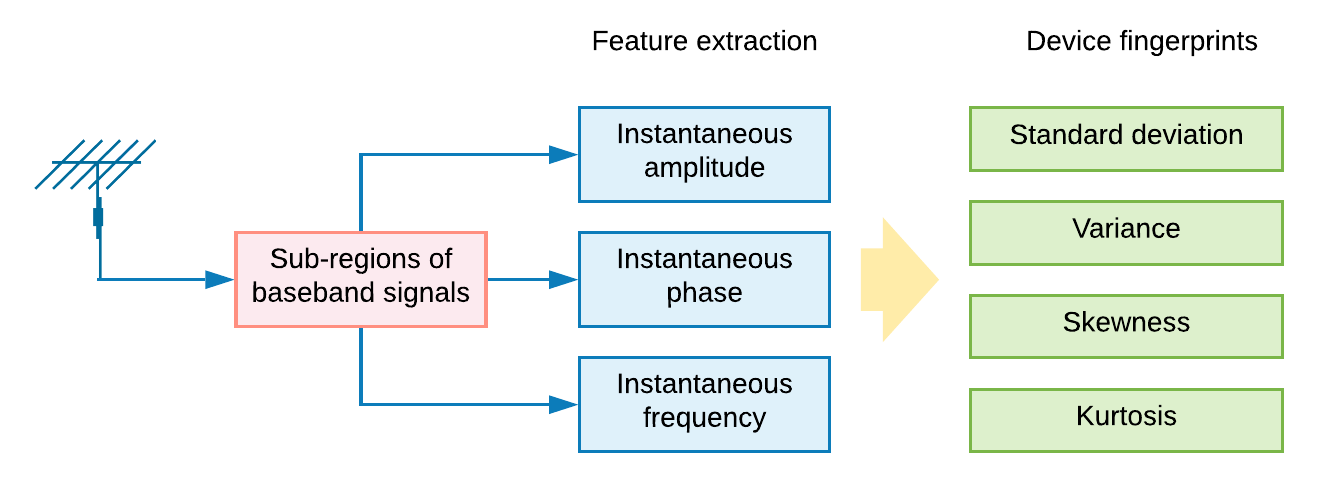}
\caption{A brief dataflow of RF-DNA.}
\label{figRF-DNA}
\end{figure}

\paragraph{Persistent patterns} Persistent pattern recognition assumes that the statistics of consecutive sub-regions in received signals can disclose identity-related information. A typical method is named as RF-DNA (Distinctive Native Attributive \cite{cobb2010physical, wang2017research}. The basic idea is to use the statistical metrics of signals' consecutive subregions to form device fingerprints. A brief dataflow of RF-DNA is given in Figure~\ref{figRF-DNA}. In \cite{bihl2016feature,ramsey2015wireless,dubendorfer2012rf}, the authors capture the preamble of WPAN (Wireless Personal Area Network) signals and extract the variance, skewness, and kurtosis of signals' subregions (bins) as signatures. Research in \cite{harmer2011using} also shows that the idea of RF-DNA can be applied in the Fourier transform of messages' signals. 

\begin{figure}[b]
\centering
\includegraphics[width=0.7\linewidth]{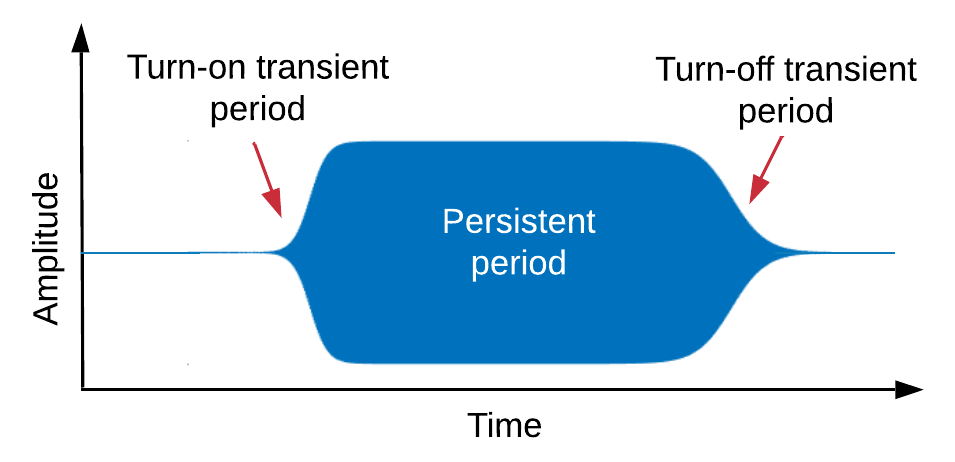}
\caption{Transient periods during wireless communication.}
\label{figTransientSignals}
\end{figure}

From the perspective of the Random Process, a sequence of signal symbols can be regarded as a sample from some multivariate distributions. The parameters of such distribution represent the unique fingerprints of devices' wireless transmitters. With this idea, authors in \cite{zhou2019design} use the Central Limit Theorem and proposed a repetitive stacking symbol-based algorithm. They \textcolor{black}{ model preamble of each packet as a sample from a specific multivariate distribution}. They extract statistics from preambles of ZigBee devices and employ Mahalanobis Distance and nearest neighbor algorithm to identify 50 Zigbee devices. 

\textcolor{black}{Regional statistic vectors from complete messages can unintentionally embed protocol-dependent features and result in unreliable device identification models.} Therefore, if we only extract persistent features from the protocol-agnostic part of signals (e.g., preambles), the resulting device identification model will only focus on signal features rather than communication protocols.

\paragraph{Transient patterns} Compared with persistent statistics of signals' subregions, transient patterns are more difficult to forge in terms of wireless channels \cite{danev2010attacks}. An example of transient periods in wireless communication is given in Figure~\ref{figTransientSignals}. Transient periods are commonly seen at the beginning and end of wireless packet transmission. In \cite{polak2015identification}, the authors employ the nonlinear in-band distortion and spectral regrowth of the received signals (potentially caused by power amplifiers of transmitters) to distinguish the masquerading device. In \cite{kose2019rf}, the authors derive the energy spectrum from transmitters' turn-on transient amplitude envelopes to classify eight different devices. Their experiment shows that frequency-domain features are more reliable than time-domain features. In \cite{zhang2018ensemble} and \cite{tu2019research}, the time-domain statistical metrics and wavelet features of transmitters' turn-on transient signals are transformed into devices' RF fingerprints. Finally, it is notable that the authors in \cite{ali2019assessment} capture the turn-on transient signal of Bluetooth devices and extract 13 time-frequency domain features (via Hibert-Huang spectrum) to construct devices' fingerprints. Their experiments show that well-designed fingerprints provide promising results even without using complicated machine learning models. 

The merit of transient features is that an adversary could not forge such nonlinear features unless they can accurately forge the coupled characteristics of pair-wise wireless channels and RF front-ends between victims and surveillance agents. \textcolor{black}{In other words, the transient features can be influenced by the locations of devices, as different locations can result in variation of RF channel characteristics, e.g., transient responses, machine learning algorithms can produce accurate but unreliable device identification results by exploiting RF channel characteristics rather than learning device-specific features. }

\begin{table*}[]
\centering
\caption{\textcolor{black}{Influential factors for feature-based specific device identification}}
\label{tabDeviceSpecRecog}
\begin{threeparttable}

\begin{tabular}{@{}lcccccll@{}}
\toprule
\multicolumn{1}{c}{Influential factors\tnote{1}} & \begin{tabular}[c]{@{}c@{}}Persistent feature\\ recognition\end{tabular} & \begin{tabular}[c]{@{}c@{}}Transient feature\\ recognition\end{tabular} & \begin{tabular}[c]{@{}c@{}}Channel status\\ recogniton\end{tabular} & \begin{tabular}[c]{@{}c@{}}Cross-domain\\ recognition\end{tabular} & \begin{tabular}[c]{@{}c@{}}Hybrid\\ approaches\end{tabular} & Countermeasures & Reference \\ \midrule
Stationary noise & \begin{tabular}[c]{@{}c@{}}Median \\ (Exc. noise pattern)\end{tabular} & Median & Low & Median & Low & \begin{tabular}[c]{@{}p{3cm}@{}}$\bullet~$Denoise filtering.\\$\bullet~$Data argumentation\\ \cmidrule(r){1-1}  \end{tabular} &\cite{yu2019radio,zhou2019design}  \\
Rx imperfections & Median & Median & Median & Median & Median & \begin{tabular}[c]{@{}p{3cm}@{}}$\bullet~$Adaptive filtering.\\$\bullet~$Calibrations\\ \cmidrule(r){1-1} \end{tabular} & \cite{romano2018unsupervised,restuccia2019deepradioid} \\
Co-channel devices & High & High & Low & High & High & \begin{tabular}[c]{@{}p{3cm}@{}}$\bullet~$MIMO receivers.\\ $\bullet~$Blind signal separation\\ \cmidrule(r){1-1} \end{tabular} & \cite{xu2018research,baliarsingh2016adaptive} \\
\begin{tabular}[c]{@{}l@{}}Channel features\end{tabular}& Median & Median & High & Low & Low &$\bullet~$Adaptive filtering & \cite{romano2018unsupervised} \\
Baseband patterns & \begin{tabular}[c]{@{}c@{}}Median\\ (Exc. noise pattern)\end{tabular} & Low & Median & Low & Low & \begin{tabular}[c]{@{}p{3cm}@{}}\cmidrule(r){1-1}$\bullet~$Message-independent\\ features\end{tabular} & \cite{hanna2019deep} \\ \bottomrule
\end{tabular}
\begin{tablenotes}
\item[1] High: solutions include hardware modifications; Median: solutions are software-based but require high capacity processors; Low: Software-based optimal solutions are available and compatible with regular processors;
\end{tablenotes}
\end{threeparttable}

\end{table*}

\subsubsection{Channel state features: }From the perspective of signal propagation, the nonlinear characteristics of radio channels can cause recognizable distortions to received signals. Those distortions can become unique profiles of transmitters. Therefore, the channel state recognition approach's basic idea is to: a) mathematically or statistically describe the nonlinear characteristics of the propagation channel within receivers and transmitters. b) Estimate whether a wireless device's signals' distortions comply with specific channel characteristics. Typical work is presented in \cite{zheng2019fid}, the authors use a kernel regression method to model the nonlinear pattern of signals' propagation channels. Their basic idea is that the combination of frequency offsets and special channel characteristics may not be forged easily, and therefore, can be used as a profile for wireless devices.

Channel state features are commonly seen in Orthogonal Frequency-Division-Multiplexing OFDM modulated communication systems. In the OFDM and MIMO schemes of wireless communication, the channel state information (CSI) \cite{halperin2011tool, ma2019wifi} can provide rich information on the time-varying characteristics of radio channels. IEEE 802.11 receivers estimate CSI during the reception of each packet's preamble. For each packet, its CSI is expressed as a complex-valued $T_n$ by $R_m$ by $K$ matrix $\boldsymbol{H}$ along with a noise component $\boldsymbol{n}\sim\mathcal{CN}(\boldsymbol{0},\boldsymbol{S})$, where $T_n$ denotes the number of transmitter' antennas, $R_n$ denotes the number of receivers' antennas, $K$ denotes the number of sub-carriers and $n$ denotes the complex-valued Gaussian random variable with mean zero and covariance matrix $\boldsymbol{S}$. Each complex-valued element in $\boldsymbol{H}$ provides instantaneous phase and amplitude response of antenna-wise channels at specific subcarriers. 

Channel state information directly reveals the phase, frequency, and amplitude responses of radio channels and has been utilized to identify fixed-position wireless transmitters. Specifically, CSI is affected by propagation obstacles, signal reflections, and even baseband data patterns \cite{ma2019wifi}. In \cite{liu2014practical}, a CSI based device identification scheme is proposed. The authors use averaged CSI to construct an SVM based profile for each legitimate device to prevent and identify spoofing attacks. They compare CSI and RSS based approaches and demonstrate the superiority of CSI. Another merit of their solution is utilizing the two-cluster k-means algorithm to detect the presence of rogue IoT transmitters when constructing legitimate devices' profiles. Similar research is presented in \cite{zaman2018deep}, legitimate devices' CSI from multiple locations are collected to train a more robust device identification model. Comparably, in \cite{zou2017tagfree}, the authors use the information from CSI to model the radiometric signatures of obstacles within the signals' propagation path. They provide an iterative differentiation approach to derive the weights and factor out the multipath components in received signals. The weights of reflection signals can be used as a location-based signature of transmitters.

Except for wireless channel characteristics, CSI can disclose RF transmitter-specific information for persistent feature-based device identification. Related researches are as follows:
\begin{itemize}
    \item \textit{Carrier Frequency Offsets (CFOs): }In \cite{hua2018accurate}, the authors propose to derive Carrier Frequency Offsets (CFOs) from CSI as devices' fingerprints. Their primitive hypothesis is that the constant CFOs can cause a linearly varying trend in instantaneous phases in received signals. Specifically, the authors first use phase measurements on specifically selected subcarriers to eliminate phase shifts at the receiver of the device identification oracle. They then use the differentiated phases from adjacent packets to eliminate the phase shifts introduced by the relative positions of transmitters. Finally, they derive the carriers' frequency offsets by the slope (relative to the time intervals of adjacent packets) of the purified instantaneous phase. 
    \item \textit{Phase errors: }Authors in \cite{liu2019real} use summation of selected subcarriers' instant phases to extract the rationale arrival phases of subcarriers. They then estimate and subtract the rationale arrival phases and receivers' insertion phase lag to derive the phase error caused by transmitters' internal imperfections. A drawback of their approach is they need to estimate the Time of Flight (ToF) of received packets.
\end{itemize}

\begin{figure}[]
\centering
\includegraphics[width=\linewidth]{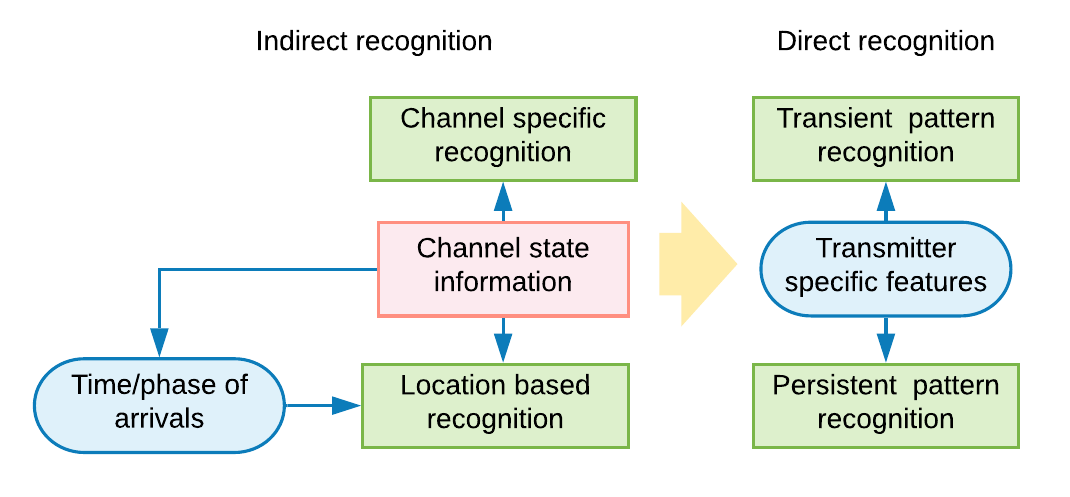}
\caption{A brief overview of channel state recognition and related approaches.}
\label{figCSI}
\end{figure}

A summary of device identification based on channel state features is in Figure~\ref{figCSI}. The drawbacks of channel state features are apparent. For one thing, researches show that channel state features can even be influenced by the motions of obstacles in subcarriers' propagation path \cite{wang2018wi,hong2016wfid,zeng2014your}. On the other hand, the channel characteristics are environment-oriented. Therefore, using channel state features based device identifier in indoor or mobile environments with human activities is still challenging \cite{wang2014eyes,wang2016wifall}.

\textcolor{black}{It should be noted that a great majority of CSI enabled researches depend on limited categories of Network Interface Cards (NICs) for data collection, owing to the limitation of CSI Tools \cite{halperin2011tool}. However, the authors in \cite{li2019location} provide a new way. They use generic SDR transceivers to extract the Long Training Sequences (LTS) in the preambles of IEEE 802.11n pilot carriers to identify more than 50 Network Interface Cards. They show that by exploiting the frequency offsets and comparing LTS frequency responses of adjacent pilot carriers, they can even derive a location-agnostic device identification model.}
\subsubsection{Cross domain features}
Many researchers convert signals to other domains that are more distinguishable. A straightforward way is to remap signals into the time-frequency domain \cite{xu2020rf}. In \cite{chen2017identification}, the authors use the STFT (Short-Time Fourier Transform) with the SVM algorithm to identify four different transceivers. This research is comparable to \cite{reising2015authorized}, where Discrete Gabor Transform (Gaussian windowed STFT) is employed. 

Other domains can also be utilized as long as they can model devices' signal patterns. In \cite{li2018low,wu2019specific}, the authors utilize the wavelet transform as well as classifiers (SVM and Probabilistic Neural Network) to construct a device identifier, compared with \cite{chen2017identification}, they further use the PCA algorithm to reduce the redundancy of the extracted data. In \cite{sun2016specific}, the authors provide a normal frequency-based method along with PCA and SVM to distinguish devices in the GSM band. They compare their methods with Hibert-Huang Transform based method in \cite{yuan2014specific}. Similar work presented in \cite{satija2018specific}, shows that Variation Mode Decomposition theoretically provides even better performance than the conventional EMD method even for relaying scenarios.

It is notable that Bispectrum is also widely utilized. In \cite{wang2017radio}, the energy entropy and color moments of the Bispectrum combined with Support Vector Machine (SVM) are employed to simulate the possibility of device identification. Their results indicate that higher-order statistics can theoretically improve the performance of identification under low SNR. However, other authors \cite{lei2016individual} claim that compared to Bispectrum, the squared integral bispectra (SIB) is more robust to noise while providing the same amount of information as the Bispectrum. In \cite{sun2017rf}, the authors employed singular values of the Axial Integrated Wigner bispectrum (AIWB) feature to identify spoofing signals from genuine signals in navigation satellite systems (GNSS). 


\begin{table}[b]
\centering
\caption{A brief compare of classifiers in deployable wireless transmitter identification systems}
\label{tabStatisticalLearn}
\begin{tabular}{@{}clcc@{}}
\toprule
Approach & \multicolumn{1}{c}{\begin{tabular}[c]{@{}c@{}}Application \\ overhead\end{tabular}} & \begin{tabular}[c]{@{}c@{}}Continual \\ learning\end{tabular} & \begin{tabular}[c]{@{}c@{}}Abnormality \\ detection\end{tabular} \\ \midrule
k-NN & \begin{tabular}[c]{@{}l@{}}Depends on the size of \\ fingerprint library.\end{tabular} & \begin{tabular}[c]{@{}l@{}}Natively \\ supported\end{tabular} & \begin{tabular}[c]{@{}c@{}}Clustering or\\ statistical models\end{tabular} \\ \cmidrule(r){1-4}
SVM & \begin{tabular}[c]{@{}l@{}}Depends on the number\\ of feature dimensions\end{tabular} & \begin{tabular}[c]{@{}c@{}}Knowledge \\replay\end{tabular} & \begin{tabular}[c]{@{}c@{}}One-class \\ SVM\end{tabular} \\ \cmidrule(r){1-4}
\begin{tabular}[c]{@{}c@{}}Random \\ forest\end{tabular} & \begin{tabular}[c]{@{}l@{}}Depends on the number\\ of decision trees.\end{tabular} & \begin{tabular}[c]{@{}c@{}}Knowledge \\replay\end{tabular} & \begin{tabular}[c]{@{}c@{}}Isolation \\forest\end{tabular} \\ \cmidrule(r){1-4}
\begin{tabular}[c]{@{}c@{}}Neural \\ network\end{tabular} & \begin{tabular}[c]{@{}l@{}}Depends on structural \\complexity\end{tabular} &\begin{tabular}[c]{@{}c@{}}Section \\ \ref{sectNeuralContimualLearning}\end{tabular}  &\begin{tabular}[c]{@{}c@{}}Section \\ \ref{sectNeuralAnomaly}\end{tabular} \\

\bottomrule
\end{tabular}
\end{table}

\subsubsection{Hybrid methods}
\label{sectDF}

A large number of device-specific features have been discovered along with different signal transform techniques. Hybrid methods aim to find the optimized combinations of features from different domains to derive robust identification models. In \cite{zhang2018research}, the authors extract the signals' energy distribution from wavelet coefficients, and marginal spectrum \cite{huang2014hilbert} and use k-NN and SVM to identify eight devices. Their tests show that this k-NN requires higher SNR than SVM. In \cite{wang2017specific}, the authors apply Intrinsic Time-Scale Decomposition (ITD) \cite{frei2006intrinsic} to input signals. They extract factual, bispectrum, and energy features to all subchannels of ITD decomposition sub-signals, their test on SVM shows that more features can significantly improve device identifiers' performance.

Although integrating signals' features from multiple domains can provide promising device identification results, the redundant information within the integrated features requires complicated models and considerable processing capacity. Therefore, automatic feature selection is introduced and becomes an indispensable part. Research in \cite{bihl2016feature} demonstrates that properly selected features, particularly from the F-test and MLF methods, enable a significant (80\%) reduction of redundancy. In \cite{shi2011improved}, the authors capture the pilot tones of the OFDM signals and extract a series of features relative to the rational signal. They use an information-theoretic approach to select useful features for device identification. In \cite{vo2016fingerprinting}, four types of features, scramble seed similarity, carrier frequency offset, sampling clock offset, and transient pattern, are suggested for the physical layer fingerprints of WiFi devices. The authors also claim that by combining all these features, their device identification accuracy reaches 95\%. 

A comparison of device-specific feature-based approaches in Table~\ref{tabDeviceSpecRecog}, hybrid approaches have superior performance under various influential factors, since the automatic feature selection methods can remove irrelevant information and provide an optimal combination of features. However, hybrid features could bring side effects, especially in statistical learning algorithms: a) The complicated combination of a large number of features can result in a highly accurate identifier with its internal mechanism not interpretable. b) High dimension features can potentially result in complicated models that are computationally difficult to retrain for operational variations. We can make better use of hybrid features in Deep Neural Networks, which will be discussed in Section \ref{sectDL}.

\subsubsection{Open issues}

\begin{table*}[b]
\centering
\caption{Countermeasures to prevent learning from trivial features}
\begin{tabular}{llll}
\toprule
Reference & Methodology & Description & Challenges \\ \midrule
\cite{agadakos2019deep} & Fragmenting & \begin{tabular}[c]{@{}l@{}}The raw I/Q signals are split into \\ small signal fragments whose duration \\ is shorter than the duration of trivial parts \\ or just use the preambles of packets..\end{tabular} & \begin{tabular}[c]{@{}l@{}}Long range dependent features \\ will be destroyed after fragmenting\end{tabular} \\ \cmidrule(r){1-4}
\cite{yu2019robust} & Masking & \begin{tabular}[c]{@{}l@{}}One can directly mask or remove the \\ trivial parts in raw signals.\end{tabular} & \begin{tabular}[c]{@{}l@{}}The position and length of the \\ masking bits or discontinuity can\\ leak protocol information\end{tabular} \\ \cmidrule(r){1-4}
\cite{merchant2018deep,riyaz2018deep} & Randomization & \begin{tabular}[c]{@{}l@{}}One can let transmitters send random \\ contents\end{tabular} & \begin{tabular}[c]{@{}l@{}}One has to gain the access of \\ large number of transmitters to \\ train a reliable classifier.\end{tabular} \\ \bottomrule
\end{tabular}
\label{tabPreventTrivival}
\end{table*}
In general, the following issues need to be investigated in feature-based statistical learning for specific device identification: 
\begin{enumerate}
    \item These methods require efforts to manually extract features or high-order statistics, the quality of handcraft features dominates device identification performances. E.g., authors in \cite{huang2017specific} show that the combination of permutation entropy \cite{bandt2002permutation} and K-NN even surpasses combination of bispectrum \cite{nikias1987bispectrum} and SVM in \cite{wang2017radio}.
    \item Experiments are conducted in rational environments with a limited number (less than 30) of IoT devices. Therefore, publicly available datasets containing signals from a larger number of IoT devices are needed to provide a reliable benchmark. Currently, publicly available datasets for IoT device identification from wireless signals are still limited. \textcolor{black}{Some small datasets are provided in \cite{morin2019transmitter,allahham2019dronerf} and \cite{GENESYS} while a larger dataset but with only ADS-B signals is in \cite{gt9v-kz32-20}.}
    \item There's no guarantee whether a specific type of feature is time-invariant. Therefore, this type of system should incorporate wireless channel estimation approaches to identify real device-specific patterns.
    \item A brief comparison of the device-specific feature-based wireless device identification with influential factor is given in Table~\ref{tabDeviceSpecRecog}, co-channel devices have the most significant impacts among all solutions. Unfortunately, there's limited research in dealing with it.
    \item \textcolor{black}{A deployable wireless device identification system should have the capacity to report unknown abnormalities and continually evolve and adapt to operational variations. A comparison of frequently employed statistical learning algorithms on continual learning and abnormality detection is in Table~\ref{tabStatisticalLearn}. Among these algorithms, only k-NN provides intuitive and native supports for continual learning and abnormality detection. However, k-NN is insufficient in handling complicated features. Though SVM or Random Forest could handle more complicated features, they lack the continual learning and abnormality detection abilities and explainability. }

\end{enumerate}

\subsection{\textcolor{black}{Deep Learning enabled specific device identification}}
\label{sectDL}

\textcolor{black}{The feature-based statistical learning approaches require manual selection of useful transforms or features. In contrast, deep neural networks (DNN) can incorporate existing features or directly deal with raw inputs and derive latent distinctive features. Therefore, Deep Learning enabled device identification mechanisms are increasingly investigated. A brief comparison of device-specific feature-based statistical learning and deep learning based approaches are presented in Table \ref{tabGeneralCompare}. In this section, we discuss typical deep learning enabled wireless device identification solutions and then focus on open issues that impede the application of deep learning in IoT device identification}.

\begin{figure}[]
\centering
\includegraphics[width=\linewidth]{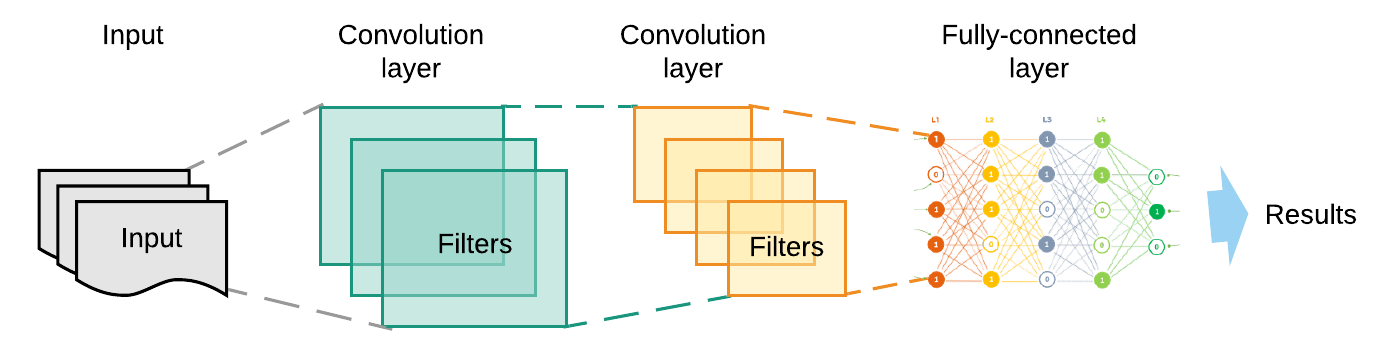}
\caption{Typical architecture of deep neural network classifiers}
\label{figDeepLearningArch}
\end{figure}
\subsubsection{Case studies and comparisons}
A typical Deep Neural Network enabled classifier is depicted in Figure~\ref{figDeepLearningArch}. Generally, It employs convolutional layers to extract latent features and uses fully connected dense layers to produce final results. Deep Neural Networks with convolutional layers are also referred as Covolutional Neural Networks (CNN).

Deep neural networks can be seamlessly integrated with existing feature engineering methods. In \cite{merchant2018deep}, the authors use the differential error between re-constructed rational signals and received signals to train Deep Neural Networks to distinguish Zigbee transceivers. In \cite{baldini2018comparison}, the authors compare the effects of short-time Fourier features and wavelet features for device identification, and their results show that wavelet features can outperform Fourier features. In \cite{yu2019robust}, the authors extract the 1-D Regions of Interest (ROIs) from 54 Zigbee devices' preambles under different SNRs and then resample signals within ROIs into various substreams with different sample rates. Finally, the substreams are fed into a convolutional neural network for identification. Similar ideas are proposed in \cite{agadakos2019deep,roy2019rfal} and \cite{gopalakrishnan2019robust}.

Compared with the conventional fully-connected neural network, convolutional layers apply filters (a.k.a. kernels) with much fewer parameters to obtain distinctive information. In \cite{yu2019radio}, the authors propose a combined solution to denoise signals and identify devices simultaneously using an autoencoder and a CNN network. The authors use their encoder to automatically extract relevant features from the received signals and use the derived features to train another deep neural network for device identification. Similar methods are presented in \cite{huang2017communication}. In \cite{riyaz2018deep}, the authors provide an optimized Deep Convolutional Neural Network approach to classify wireless devices in 2.4 GHz channels and compare the performance with SVM and Logistic Regression. Their results show that, even by using raw I/Q digital baseband signals, CNN can achieve high accuracy and surpass the best performance of SVM and Logistic Regression. In \cite{morin2019transmitter}, neural networks were trained on raw IQ samples using the open dataset\footnote{\url{https://wiki.cortexlab.fr/doku.php?id=tx-id}} from CorteXlab. Their results also show that CNN can achieve promising results even on raw I/Q signals, but the movement of devices and the varying amplitudes will degrade CNN's performance.

An extensively discussed topic for Deep Learning based device identification is preventing the network from learning only trivial features, such as protocol identifiers, unique identifiers, etc. Generally, three types of countermeasures are applied, and their comparisons are provided as in Table~\ref{tabPreventTrivival}.

Compared with feature-based device identification approaches, Deep Learning methods usually require a much larger dataset to initialize the network. To know how large the training data is needed. In \cite{oyedareestimating}, CNN models are used to classify different devices' signals with controlled difficulty levels. The classification accuracy of a fixed CNN network with different dataset sizes is predicted using a power-law model and the Levenberg-Marquardt algorithm. Their results show that the dataset size should be at least 10,000 to 30,000 times the number of devices to be identified. However, this conclusion is only a rough estimation.

New architectures in Deep Learning are emerging and can significantly influence the performance of device identification systems. In \cite{agadakos2019deep}, the authors use Convolutional Deep Complex-valued Neural Network (CDCN) and Recurrent Deep Complex-valued Neural Network \cite{DBLP:journals/corr/abs-1905-12321} to address the device identification problem. Their networks utilize fragments of raw I/Q symbols as input, and their test is conducted on both WiFi and ADS-B datasets. Their experiments show that the Complex-valued neural networks surpass regular real-valued deep neural networks. \textcolor{black}{In \cite{liu2020zero, liu2020deep}, a zero-bias dense layer is proposed. The authors show that their solution enables deep neural networks' final decision stage to be interpretable. Their solution maintains equivalent identification accuracy and outperforms regular DNN and one-class SVM in detecting unknown devices.}
\begin{table*}[b]
\centering
\caption{Methods for unknown device recognition}
\label{tabUnknownRecognition}
\begin{threeparttable}

\begin{tabular}{@{}llclll@{}}
\toprule
Methods & Description & Complexity & \multicolumn{1}{c}{Memory} & Pros \& Cons & Reference \\ \midrule
GAN & \begin{tabular}[c]{@{}l@{}}Use the discriminator from GAN model as\\  an outlier detector.\end{tabular} & High\tnote{1} & \begin{tabular}[c]{@{}l@{}}Depends on final\\ network\end{tabular} & \begin{tabular}[c]{@{}l@{}}$\bullet$ Can catch deep latent features.\\ $\bullet$ Hard to design and train.\end{tabular} & \cite{roy2019rfal,li2018generative}\\ \cmidrule(r){1-6} 
Autoencoder & \begin{tabular}[c]{@{}l@{}}Train a deep Autoencoder on known signals \\ and use its reconstruction error to judge \\ outliers.\end{tabular} & High\tnote{1} & \begin{tabular}[c]{@{}l@{}}Depends on final\\ network\end{tabular} & \begin{tabular}[c]{@{}l@{}}$\bullet$ Can catch deep latent features.\\$\bullet$ Easier than GAN to design\\  and train\end{tabular} & \cite{marchi2015novel,khan2017detecting}\\ \cmidrule(r){1-6}
\begin{tabular}[c]{@{}l@{}}Statistic metrics\end{tabular} & \begin{tabular}[c]{@{}l@{}}Measure the confidence of whether a signal \\ or its fingerprint is generated by a given \\ category. \end{tabular} & Low & Low & \begin{tabular}[c]{@{}l@{}}$\bullet$ Provide explainable results.\\ $\bullet$ Accuracy depends on the \\ fingerprinting methods.\end{tabular} & \cite{shi2019deep,gritsenko2019finding,kim2018identifying,liu2020deep} \\ \cmidrule(r){1-6}
Clustering & \begin{tabular}[c]{@{}l@{}}Perform clustering analysis on known signals' \\ fingerprints to judge whether it is in identical \\ cluster as known ones.\end{tabular} & Median\tnote{2} & \begin{tabular}[c]{@{}l@{}}Depends on the \\ number of existing\\ fingerprints.\end{tabular} & \begin{tabular}[c]{@{}l@{}}$\bullet$ Provide explainable results\\$\bullet$ Accuracy depends on the \\ fingerprinting methods.\end{tabular} & \cite{shi2019deep,wong2018clustering} \\ \bottomrule
\end{tabular}
\begin{tablenotes}
\item[1] Needs to specify both network architecture and hyperparameters.
\item[2] Needs to specify the clustering algorithms to use.
\end{tablenotes}

\end{threeparttable}

\end{table*}

\subsubsection{Open issues in Deep Learning for IoT device identification}
Deep Learning is becoming a promising technology in this domain. However, as in other domains, Deep Learning encounters several challenges. Although researches in IoT device identification rarely cover the issues, we briefly discuss their current states and solutions. 

\paragraph{Hyperparameter searching} \label{sectDLHPP}
One critical problem for using deep neural networks is hyperparameter tuning. Hyperparameters such as learning rate, mini-batch size, dropout rate, etc. are used to initialize the training process. Hyperparameters can significantly impact the performance of deep neural networks. For instance, in \cite{jafari2018iot}, the authors compare the performance of Deep Neural Networks, Convolutional Neural Network, and the LSTM (Long Short Term Memory) in device identification using the raw I/Q signals directly. Their results show that CNN has the best performance, followed by DNN and LSTM. They point out that the hyper-parameters of Deep Learning, especially for network architectural parameters, significantly influence the upper bound of performance. 

Obtaining optimized hyperparameters is computationally expensive. Several strategies are proposed for efficient hyperparameter searching, such as grid search, random search, prediction-based approaches, and evolutionary algorithms. Their characteristics are as follows:
\begin{itemize}
    \item \textbf{Grid search: }Grid search divides the whole parameter space into identical intervals and performs brute-force trials to find optimal parameter combinations. However, this strategy is inefficient since useless combinations of parameters can not be pruned rapidly. 
    \item \textbf{Random search: }In random search, sample points are distributed uniformly in search space. This strategy increases the variation and outperforms the grid search when only a small number of parameters can impact the network performance. 
    \item \textbf{Prediction-based:} In prediction-based approaches, the algorithms first perform random trials at the beginning to model the relation between the network performances with hyperparameters. Then the algorithms perform new trials based on parameters that are more probable to yield better results. Such trial-model-predict paradigm is conducted repeatedly  \cite{bergstra2011algorithms}. A typical prediction strategy is the Bayesian optimization process \cite{pelikan1999boa}, in which the algorithms model the target outcome space as Gaussian processes. 
    \item \textbf{Evolution based: }In evolutionary algorithm based approaches, the heuristic searches are performed as in other nonlinear optimization problems. In \cite{young2015optimizing}, the authors utilize the Genetic Algorithm to find the optimal hyperparameters of a neural network. Compared with prediction-based approaches, evolutionary algorithms provide the best-guess with bio-inspired strategies. However, there is no guarantee for the performances of evolutionary algorithms.
\end{itemize}

\paragraph{Neural network Architecture search} Network Architecture Search (NAS) is another challenging task in designing neural networks. Network architecture defines the flow of tensors and could significantly affect the complexity and performance of neural networks \cite{elsken2018neural,liu2017survey}. At the current stage, most network architectures are specified manually or with trial-and-error. 

\textcolor{black}{Architecture searching algorithms are provided by several Automatic Machine Learning (AutoML) platforms. A brief comparison of their functionality and performance on different datasets is in \cite{truong2019towards}.} A collection of recent literature and open-source tools are given in \cite{AutoMLNeuralArchSearch} and \cite{AwesomeNASDxy} respectively. These efforts can be classified into three categories: (i) network pruning \cite{zanchettin2006methodology}, (ii) progressively growing \cite{islam2009new}, and (iii) heuristic network architecture search \cite{lam2001tuning}. Their features are as follows:
\begin{itemize}
    \item \textbf{Network pruning: }Network pruning algorithms use group sparsity regularizers \cite{alvarez2016learning} to remove unimportant connections from a regularly trained network. Then the pruned network will be retrained to fine-tune the weights of the remaining connections \cite{guo2016dynamic,han2015learning}. A key benefit of network pruning is that it can greatly compress neural networks and make them suitable to deploy in low capacity IoT devices.
    
    \item \textbf{Progressively growing: }This strategy grows a neural network architecture during training. It is effective in simple networks with only one hidden layer \cite{islam2003constructive,  narasimha2008integrated}. More recent advances employ growing strategies to progressively add nodes and layers to increase the network's approximation ability \cite{ma2003new,cortes2017adanet}. 
    
    \item \textbf{Heuristic network search: }In heuristic network search, the architecture of the Deep Neural Network (can either be block-wise \cite{liu2018progressive} or element-wise \cite{stanley2002efficient}) can first be represented in a high dimension space with billions of parameters. Next, heuristic searching algorithms are applied to transverse this search space to find optimal solutions. Examples are given in \cite{stanley2002evolving,lam2001tuning} and \cite{liang2019evolutionary}. The authors make use of the Genetic Algorithm to find the possible structure of neural networks. Notably, the Genetic Algorithm fits perfectly in NAS problems since it allows using length-varying variables (genes) to encode the candidate solutions. An empirical example is the NeuroEvolution of Augmenting Topologies (NEAT) algorithm \cite{stanley2002evolving}. 
    
    \item \textbf{Reinforcement Learning: }Reinforcement learning (RL) has become a popular strategy in NAS \cite{baker2016designing,tan2019mnasnet,hsu2018monas}. Its basic idea is to let a deep learning-enabled agent explore network architectures' representative space and use validation accuracy or other metrics as rewards to adjust the agents' solutions. Ideally, as an RL process moves on, an agent can find an optimal searching strategy and discover a novel architecture. Intuitively, evolution algorithms use a fixed strategy to discover the optimal architecture while RL agents learn their own strategies and have better capabilities in avoiding bad solutions. 
    
    \item \textbf{Differentiable space search: }Aforementioned, NAS strategies use discrete space to encode the architecture of neural networks, which is not differentiable and lacks efficiency. Therefore, differentiable spaces to represent the Neural Networks' architectures are proposed, in which efficient off-the-shelf optimization algorithms can be used. Typical solutions are given in \cite{liu2018darts,chen2019progressive}. The algorithm, DART (Differentiable Architecture search), is presented. The authors use the Softmax function to represent discrete selections in a numerically continuous domain. They then use a gradient descent algorithm to explore the search space. Similar work with an enhanced stochastic adaptive searching strategy is presented in \cite{dong2019searching}. Block-wise representations of the neural network and differentiable searching space together are bringing NAS to practice.
\end{itemize}

Network architecture search has become an emerging topic for deep neural network research with publicly available benchmarking tools in \cite{ying2019bench} and \cite{dong2020bench}, respectively.

\paragraph{Openset recognition}
\label{sectNeuralAnomaly}
A critical problem for learning based device identification is that classifiers only recognize pretrained devices' signals but can not deal with novel ones that are not in the training dataset. In \cite{wong2018clustering}, the authors address it as a semi-supervised learning problem. They first train a CNN model with the last layer as a Softmax output on a collection of known devices. They then remove the Softmax function and turn the neural network into a nonlinear feature extractor. Finally, they use the DBSCAN algorithm to perform cluster analysis on the remapped features of raw I/Q signals. Their results show that such a semi-supervised learning method has the potential of detecting a limited number of untrained devices. Comparably, in \cite{youssef2018machine}, the authors use an incremental learning approach to train neural networks to classify newly registered devices. 

From the perspective of Artificial Intelligence, this issue is categorized to the Open Set Recognition \cite{scheirer2012toward,bendale2016towards} and the Abnormality Detection problem. The taxonomy of existing approaches is given in table \ref{tabUnknownRecognition}. In \cite{roy2019rfal}, the authors use the Generative Adversarial Network (GAN) to generate highly realistic fake signals. Then they exploit the discriminator network to distinguish whether an input is from an abnormal source. In \cite{shi2019deep}, the authors provide two methods to deal with unknown devices: i) Reuse trained convolutional layers to transform signals to feature vectors, and then use Mahalanobis distance to judge the outliers. ii) Reuse pretrained convolutional layers to transform signals to feature vectors, and then perform k-means (k = 2) clustering to group outliers. 

\begin{figure}[]
\centering
\includegraphics[width=\linewidth]{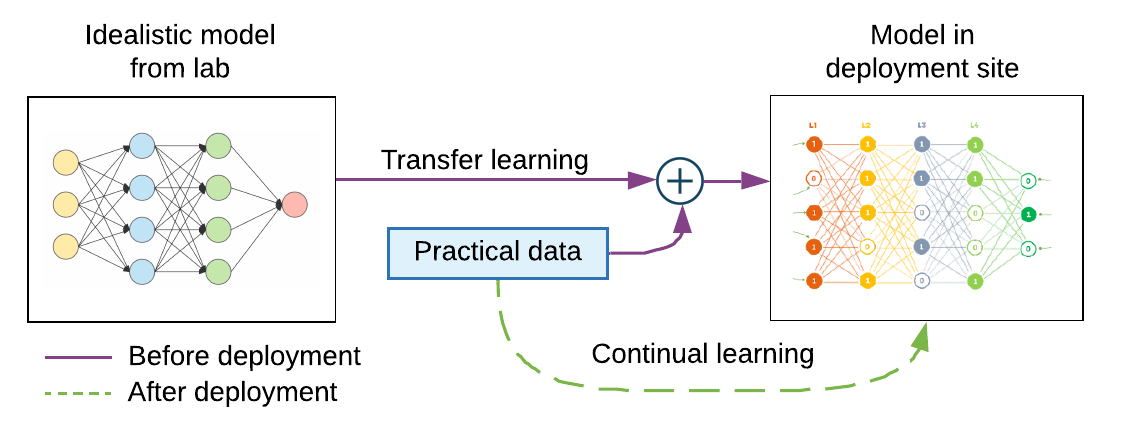}
\caption{\textcolor{black}{Transfer learning and continual learning.}}
\label{figTransferVsContinuous}
\end{figure}

\begin{table*}[h]
\centering
\caption{\textcolor{black}{Brief compare of IoT device identification and detection methods}}
\label{tabGeneralCompare}

\begin{threeparttable}
\begin{tabular}{@{}llccccl@{}}
\toprule
\multicolumn{1}{c}{\begin{tabular}[c]{@{}c@{}}Device identification\\ approaches\end{tabular}} & \begin{tabular}[c]{@{}l@{}}Technology\\ branch\end{tabular} & \begin{tabular}[c]{@{}c@{}}Feature\\ requirement\end{tabular} & \begin{tabular}[c]{@{}c@{}}Model \\ explanability\end{tabular} & \begin{tabular}[c]{@{}c@{}}Continuous \\ learning \end{tabular} & \begin{tabular}[c]{@{}c@{}}Anomaly detection\\ \end{tabular} & \multicolumn{1}{c}{Challenges} \\ \midrule 
\textcolor{black}{\begin{tabular}[c]{@{}l@{}}Feature based \\ device identification\end{tabular}} & \begin{tabular}[c]{@{}l@{}}Supervised\\ learning\end{tabular} & High\tnote{1} & \begin{tabular}[c]{@{}c@{}}Strong (k-NN) / \\ median (SVM)\end{tabular} & \begin{tabular}[c]{@{}c@{}}Easy (k-NN) /\\ median (PCA-SVM)\end{tabular} & \begin{tabular}[c]{@{}c@{}}Low (k-NN)\\ Median (k-Means)\end{tabular} & \begin{tabular}[c]{@{}l@{}}Can not discover\\ latent feature.\end{tabular} \\ \cmidrule(r){1-7} 
\begin{tabular}[c]{@{}l@{}}Deep learning enabled\\ device identification\end{tabular} & \begin{tabular}[c]{@{}l@{}}Supervised\\ learning\end{tabular} & Low & Weak\tnote{2} & Hard (EWC)\tnote{3} & \begin{tabular}[c]{@{}c@{}}High (Autoencoder) /\\ Median (clustering)\end{tabular} & \begin{tabular}[c]{@{}l@{}}Learning from \\ trivial features\end{tabular} \\ \cmidrule(r){1-7} 
\textcolor{black}{\begin{tabular}[c]{@{}l@{}} Unsupervised device\\ detection and identification\end{tabular}} & \begin{tabular}[c]{@{}l@{}}Unsupervised\\ learning\end{tabular} & High\tnote{1} & Strong & N/A & Low & \begin{tabular}[c]{@{}l@{}}Can not be applied to \\ complex environment\end{tabular} \\ \bottomrule
\end{tabular}
\end{threeparttable}{}
\begin{tablenotes}
\item[1] Requires an extra feature engineering process.
\item[2] Please refer to Explainable AI (XAI) in \cite{gunning2017explainable}
\item[3] Please refer to section \ref{sectNeuralContimualLearning}
\end{tablenotes}
\end{table*}

\paragraph{Continual learning}
\label{sectNeuralContimualLearning}
In practical scenarios, deep neural networks would have to evolve to adapt to operational variations continuously. Intuitively, a deep learning enabled IoT device identifier has to learn new devices' characteristics during its life cycle. Therefore, such functionalities are defined as lifelong learning. Generally, there are two ways to achieve this goal: Transfer Learning (TL) and Continual Learning (CL). In Transfer Learning, neural networks are pre-trained in the lab and then fine-tuned for deployment using practical data \cite{tan2018survey}. In continual learning, neural networks are trained incrementally as new data come in progressively \cite{parisi2019continual}. Continual learning does not allow neural networks to forget what they have learned in the early stages compared with transfer learning. The phenomenon in which a neural network forgets what it has previously learned after training on new data is named Catastrophic Forgetting. Therefore, transfer learning is useful when deploying new signal identification systems, and continual learning is useful in regular software updates and maintenance, as depicted in Figure~\ref{figTransferVsContinuous}. The strategies to implement continual learning for deep neural networks are as follows:
\begin{itemize}
    \item \textbf{Knowledge replay: }An intuitive solution for continual learning is to replay data from old tasks while training neural networks for new tasks. However, such a solution requires longer training time and larger memory consumption. Besides, one can not judge how many old samples are enough to catch sufficient variations. Therefore, some studies employ data generator networks to replay data from old tasks. For instance, in \cite{shin2017continual}, Generative Adversarial Network (GAN) based scholar networks are proposed to generate old samples and mixed with the current task. In this way, the deep neural network could be trained on various data without using huge memories to retain old training data.
    

    \item \textbf{Regularization: }Initially, regularization is employed to prevent models from overfitting by penalizing the magnitude of parameters \cite{girosi1995regularization}. In continual learning, regularization is employed to prevent model parameters from changing dramatically. In this way, the knowledge (represented by weights) learned from the old tasks will be less likely to vanish when an old network is trained on new tasks. There are two types of regularization strategies: global regularization and local regularization. Global regularization penalizes the whole network's parameters from rapid change but impedes the network from learning new tasks. In local regularization strategies, such as Elastic Weight Consolidation (EWC) \cite{kirkpatrick2017overcoming}, the algorithms identify important connections and protect them from changing dramatically, in which non-critical connections are used to learn new tasks. 
    \item \textbf{Dynamic network expansion: }Network expansion strategies lock the weights of existing connections and supplement additional structures for new tasks. For instance, the Dynamic Expanding Network (DEN) \cite{yoon2017lifelong} algorithm first trains an existing network on a new dataset with regularization. The algorithm compares the weights of each neuron to identify task-relevant units. Finally, critical neurons are duplicated and to allow network capacity expansion adaptively.
\end{itemize}

Continual learning algorithms, as well as abnormality detection, together provide great potential for deploying the neural networks in complex, uncertain scenarios. 
\paragraph{Summary}

A brief comparison of Deep Learning and other statistical learning methods is given in Table~\ref{tabStatisticalLearn}. Compared with statistical learning, Deep Learning is not yet an idealistic solution. However, its unified development pipeline, and the capability of dealing with high dimension features are making it easy to use. Furthermore, for practical issues such as continual learning and abnormality detection, deep learning provides better performances than the majority of statistical learning algorithms. In one word, although deep learning is not theoretically novel, it gains its place by providing the most balanced merits.

\subsection{\textcolor{black}{Unsupervised device detection and identification}}
\label{sectUDI}
\begin{figure*}[h]
\centering
\includegraphics[width=0.7\linewidth]{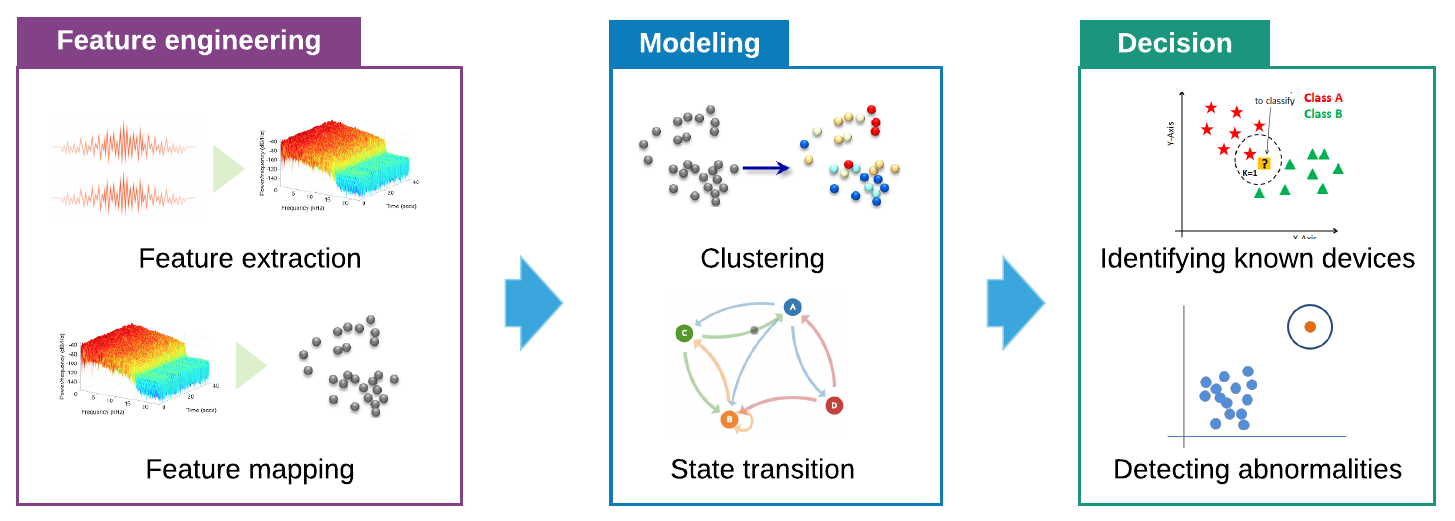}
\caption{\textcolor{black}{Unsupervised device detection and identification}}
\label{figUnsuperviseDeviceID}
\end{figure*}
\textcolor{black}{Feature-based statistical learning and deep learning are supervised learning schemes, where classifiers must learn the features of legitimate devices in advance. Unsupervised device detection and identification are required in scenarios where the identities of devices are not directly available \cite{axell2012spectrum}}. Generally, the methods in this topic can be divided into two folds, device behavior modeling and signal propagation pattern modeling. \textit{the essence of unsupervised device detection is to map devices' signals or activity profiles into latent representative spaces, where different devices are represented by separated clusters or probabilistic distributions}. \textcolor{black}{If behavior or signal propagation patterns are strictly correlated with specific devices, extracted behavior or signal features can be used to verify the identity of devices.} Comparisons of supervised and unsupervised learning based device identification are (also in Table~\ref{tabGeneralCompare})): 
\begin{itemize}
    \item \textcolor{black}{The training data does not directly indicate device specific information (device identifier, device type, and etc.).}
    \item The number of devices may not be known in advance.
\end{itemize}
As depicted in Figure~\ref{figUnsuperviseDeviceID}, the work flow of unsupervised learning enabled device detection and identification is made up of three steps: a) Feature engineering on IoT devices' signals or behavior profiles, including feature selection and mapping. b) Modeling the latent spaces, this step finds out cluster centers, probabilistic distributions, related decision boundaries, or state transition models. c) Matching of input signal or behavior profiles to the most likely clusters or report abnormalities. 

\subsubsection{Device behavior modeling}
\textcolor{black}{Device behavior modeling extracts distinctive features from the input data and finds out the number of different devices using unsupervised learning algorithms. However, the physical layer does not provide much information for device behavior modeling. Therefore, the methods are more frequently employed in the upper layers with related techniques employed are unsupervised feature engineering, clustering, and Software-Defined Networking \cite{rafique2020complementing}. }

\textcolor{black}{In \cite{sivanathan2020detecting} and \cite{sivanathan2019inferring}, the data traffic attributes are obtained from flow-level network telemetry to recognize different IoT devices. The authors utilize Principle Component Analysis along with an adaptive one-class clustering algorithm to find the optimal representative components and cluster centers for each device. They provide a conflict resolution mechanism to associate different types of devices to corresponding cluster centers in the representative spaces. A similar approach using Deep Learning is presented in \cite{ortiz2019devicemien}. The authors use TCP data traffics for each device to train an LSTM-enabled autoencoder to map inputs into a representative feature space. They then use a clustering algorithm to divide the training samples into their natural clusters. Finally, they use probabilistic modeling to associate new data with known clusters for device identification. Unfortunately, their experiments show that unsupervised behavior identification may not work once there are devices from an identical model.}

\subsubsection{Signal propagation pattern modeling}
\begin{table*}[h]
\centering
\caption{Comparison of device localization methods in IoT}
\label{tabLocalization}

\begin{threeparttable}
\begin{tabular}{@{}clclll@{}}
\toprule
Methods & \multicolumn{1}{c}{Requirements} & Unit cost\tnote{1}& \multicolumn{1}{c}{Precision} & \multicolumn{1}{c}{Weakness} & References \\ \midrule
\begin{tabular}[c]{@{}c@{}}Signal propagation \\ modeling\end{tabular} & \begin{tabular}[c]{@{}l@{}}Multiple collaborative transmitters\\ to construct signal strength map.\end{tabular} & Low & \begin{tabular}[c]{@{}l@{}}Depends on environmental\\ features and refresh rate of\\ respondent data.\end{tabular} & \begin{tabular}[c]{@{}l@{}}$\bullet$ Depends highly on signal \\propagation models of certain area.\\$\bullet$ Results do not directly indicate \\certain device types or identities.\end{tabular} & \cite{liu2019synchronization} \\ \cmidrule(r){1-6} 
Coherent TDoA & \begin{tabular}[c]{@{}l@{}}At least 4 coherent receivers and 5\\ receivers are recommended to \\ linearize computational process.\end{tabular} & Median & \begin{tabular}[c]{@{}l@{}}Depends on the estimation \\ of signals' Time of Arrival \\ (ToA).\end{tabular} & \begin{tabular}[c]{@{}l@{}}$\bullet$ Receivers needs to be strictly\\ synchronized.\end{tabular} & \cite{monteiro2015detectinglb} \\ \cmidrule(r){1-6} 
\begin{tabular}[c]{@{}c@{}}Sync-free TDoA\end{tabular} & \begin{tabular}[c]{@{}l@{}}At least 4 receivers and receivers \\ are able to communicate mutually.\end{tabular} & Median & Same as coherent TDoA & \begin{tabular}[c]{@{}l@{}}$\bullet$ Needs specific hardware\\ with known response latency.\end{tabular} & \cite{xu2012high,schafer2014bringing} \\ \cmidrule(r){1-6} 
\end{tabular}
\begin{tablenotes}
\item[1] Low: Does not require extra RF receivers; Median: Requiring commercially available RF receivers with unit cost less than \$1000; High: Requiring special hardware and specific processing stacks.
\item[2] Requiring multiple distributed receivers.
\end{tablenotes}
\end{threeparttable}
\end{table*}
In the Physical Layer, signal propagation patterns provide information for device identification. \textcolor{black}{On the one hand, if devices positions are unique and known in advance, we may directly use wireless localization algorithms to specify whether a received data packet is from its claimed identity. Corresponding surveys on wireless device localization are given in \cite{liu2007survey, guvenc2009survey, han2013localization}, and we provide a brief comparison of the widely employed methods in Table~\ref{tabLocalization}.}

On the other hand, signal propagation modeling derives the path loss or attenuation patterns of received signals to detect different devices using unsupervised learning algorithms\cite{zeng2010non}. In \cite{li2017fingerprints}, the authors use the signals' propagation path effect, and they discover that the received signal strength from transmitters in the same location would have very similar varying trends. They convert signal strength metrics into time series and incorporate the Dynamic Time Warping algorithm to align and find differences between received signals. Finally, they apply a clustering algorithm to identify signals from active transmitters. In \cite{zheleva2015txminer}, the authors assume that the received signals' Power Spectrum Density coefficients of each device, in a specific time window, form a mixture model dominated by a weighted sum of Gaussian distributions and propagation path related Relay distributions. In this way, they use the Expectation-Maximum algorithm to estimate the composition (different transmitters) of received signals. 

\textcolor{black}{Signal propagation pattern modeling only provides an indirect evaluation on whether specific signals come from devices in close locations or with similar propagation paths. Although related methods are not widely utilized in commercial IoT devices owing to their complicated deployment environments, the methods provide a useful solution in preventing identity spoofing attacks in ADS-B systems \cite{nijsure2015adaptive,monteiro2015detecting}.}

\subsubsection{Open issues} 
\textcolor{black}{Unsupervised device identification provides a novel solution when the identities of devices are not directly available. In essence, the unsupervised device identification and detection are similar to automatic knowledge discovery with the following issues to be addressed:
\begin{enumerate}
    \item \textbf{Feature engineering: } Unsupervised device identification relies on feature engineering since representative vectors of devices are supposed to form distinctive clusters. Feature selection is still conducted manually, and there is no guarantee on whether the outputs of the mapped feature can form distinctive clusters. 
    \item \textbf{Clustering: }Clustering in the latent space can be challenging if the number of devices is unknown. Although one may use adaptive algorithm such as DBSCAN \cite{schubert2017dbscan}, Optics \cite{ankerst1999optics} or X-Means \cite{pelleg2000x}, the proper configurations of these algorithms to adapt to the latent space are still difficult, similar obstacles are seen in setting hyperparameters in Deep Neural Networks (section \ref{sectDLHPP}). 
    \item \textbf{Decision boundaries} Even if we know the number of devices, we can still get clusters with uncertain shapes or density, in which decision boundaries between different devices are difficult to define, as indicated in \cite{sivanathan2020detecting}.
    \item \textbf{Direct identity verification: }\textcolor{black}{Researches on unsupervised device identification using behavior-independent and location-agnostic device specific features are still limited. Although unsupervised behavioral modeling has shown promising results in identifying different types of devices, whether these methods are still effective in distinguishing devices from the same model needs further investigation.}
\end{enumerate}
Therefore, we believe learning-based unsupervised device detection is promising with great novelty, but the topic needs substantial investigation. }

\section{Learning-Enabled Abnormal Device detection}
\label{sectLEADD}

\textcolor{black}{Previous sections discussed methods to identify specific IoT devices. Except for device identity verification, detection of compromised devices with abnormal behaviors is needed to alert ongoing attacks and discover system vulnerabilities.}

\textcolor{black}{In general, abnormal device detection algorithms are deployed in network and application layers. The detection algorithms first collect a certain amount of normal operation data from devices to create reference models (or signatures). Then IoT devices' operational data are collected and compared with reference models to judge whether significant deviations appear. Compared with device-specific identification schemes, the key idea is abnormality detection with both unsupervised learning approaches \cite{hwang2020unsupervised} and supervised learning with confidence thresholds \cite{hwang2019detecting}.} 

\subsection{Statistical Modeling}
\label{sectLEADDSM}

Statistical modeling aims to judge whether devices are in abnormal situations. In \cite{alipour2015wireless}, Markov models are utilized to judge whether IEEE 802.11 devices are compromised by calculating the probabilities of its sequential transitions of the protocol state machines. In \cite{sehatbakhsh2018syndrome}, the authors model the Electronic Magnetic (EM) harmonics peaks of medical IoT devices as probabilistic distributions to assess whether a specific device is under attack. They assume that when devices are operated under an abnormal scenario (with rogue shellcodes executing), its EM radiometric signals can deviate from known scenarios. However, statistical modeling requires manual selection of potentially informative features and define their importance.

To reduce the cost of modeling IoT devices' normal behavior, Manufacturer Usage Description (MUD) profile \cite{lear2019rfc} is proposed. A collection of MUD profiles for 30 commercial devices is provided in \cite{iotanalytics}. The MUD profiles enable operators to know devices' network flow patterns and dynamically monitor their behavioral changes. Several open-source tools are provided to dynamically generate, validate, and compare IoT devices' MUD profiles in \cite{hamza2020verifying}. Besides, the authors presented that by comparing the deviation of devices' run-time MUD profiles with static ones, we can identify their behavioral deviations or even identify device types. In \cite{hamza2019detecting}, MUD profiles of devices are translated into flowtable rules and contribute to select appropriate features. The authors then use PCA to map each device's data traffic from side windows into its own representative one-class space, where X-Means \cite{pelleg2000x} and Markov chains are used to partition the space and model the state transition in cluster centers. Finally, an exception is triggered by a specific detector on either the mapped traffic pattern is out of boundaries or the state transitions do not comply with the reference model. Their experiments show the accurate detection of several types of volumetric attacks. 

\subsection{Reconstruction Approaches}
Reconstruction approaches aim to learn and reconstruct domain-specific patterns from devices' normal operation records. In other words, we need to develop a model to "memorize" the normal schemes of IoT devices by producing low reconstruction errors. Simultaneously, the model is supposed to produce high reconstruction errors for unknown scenarios or encounters behavioral deviations. 
This goal is generally achieved using deep autoencoders. Since an encoder removes a great amount of information, a decoder needs to reconstruct lost information according to domain-specific memories. Consequently, once abnormal inputs are given to a well-trained autoencoder, its decoder would not be able to reconstruct such unknown inputs and yields a high abnormal score (reconstruction error). In \cite{thing2017ieee, sridharan2018wadac,luo2018distributed}, the authors utilize autoencoder to detect abnormal activities by modeling the data traffic and content of IoT devices once abnormal activities are detected. In \cite{meidan2018n}, the authors show that compared with other anomaly detection methods (one-class SVM \cite{chen2001one}, Isolation Forest \cite{ding2013anomaly} and Local Outlier Factor \cite{kriegel2009loop}), deep autoencoder yields the best result in terms of reliability and accuracy.

\subsection{Prediction Approaches}
\textcolor{black}{Prediction approaches utilize temporal information in devices' operation records. Corresponding methods model each IoT device's operational data as multi-dimension time series. Then, device-specific prediction models are trained using time series from normal schemes. When devices are hijacked for rogue activities, they are not supposed to behave as predicted, causing the corresponding time series predictors to output high prediction errors. }

In \cite{parwez2017big}, the authors employ a CNN based predictor to analyze the abnormal behaviors in devices' network traffics. They show that predictors trained without abnormal data are sensitive (yield high prediction error) to anomalies. Similar work is shown in \cite{wei2012intrusion}, and the authors use an autoregression model to capture the normal varying trend of devices' traffic volumes. However, modeling a single variable can not be sufficient in dealing with complicated scenarios. Recent studies combine deep Autoencoder with Long Short Term Memory (LSTM) to derive abstracted representations of complex scenarios and make predictions. In \cite{tandiya2018deep} and \cite{rajendran2018saife}, Deep Predictive Coding Neural Network \cite{lotter2016deep} is used to predict consecutive frames of time-frequency video streams of wireless devices. They can even specify the type of attacks using the spatial distribution of error pixels in the reconstructed frames. 

\subsection{Open issue}
\textcolor{black}{Methods in this topic overlap with the methods of open set recognition in Deep Learning. We briefly list several open issues in this topic:}

\begin{itemize}
    \item \textbf{Selection of behavioral features: }Many researches use manual feature selection along with dimension reduction. A concern is that we can not guarantee the selected features are sensitive to unknown intrusions in the future. 
    \item \textbf{Processing of abnormality metrics: }Generally, intrusion detection approaches provide metrics corresponding to the degree of deviation. However, the output error metrics require a posterior process, e.g., selecting appropriate decision thresholds or aggregation window length, which balances between the true positive, false negative, and response latency. One solution is to regard the corresponding parameters as hyperparameters and use cross-validation to tune them. The processing of error metrics remains a case-specific open issue.
\end{itemize}

\section{Challenges and Future Research Directions}
\label{sectChallengesAndFuture}
Our literature review has shown that device detection and identification provide another layer of security features to IoT. However, the existing solutions are still far from perfect. This section summarizes the existing challenges of IoT device identification and detection as well as future research directions.
\textcolor{black}{
\subsection{Challenges in machine learning models}}

\subsubsection{Unknown device recognition}Existing works focus on the accuracy they can obtain using a fixed dataset with all devices labeled, in which Black-Box models (e.g., Deep Learning and SVM) are commonly employed. In practical scenarios, these models can produce wrong answers when encountering novel devices. Additional mechanisms are needed to identify unknown signals. Although we can use the one-versus-rest technique to train a group of classifiers and avoid producing results on unknown devices. However, once we have new devices to register, all classifiers in the group are supposed to be retrained from scratch. Therefore, we need to provide a solution to verify the known devices. Meanwhile, we need to identify:

\begin{itemize}
    \item \textcolor{black}{Devices that are exactly not in the scope of the identification system. }
    \item \textcolor{black}{Unknown devices that are from identical manufacturers. Devices of the same model from an identical manufacturer can share similar behavior patterns, e.g., network flow characteristics. Such similarities can impede identity verification in the network, transportation, or application layers.}
\end{itemize}
\textcolor{black}{The latter is more challenging and requires extracting behavior-independent characteristics. We believe that without the capability of unknown device recognition, these types of systems are still far from practice. }

\subsubsection{Continual learning on new devices} Continual or incremental learning \cite{parisi2019continual} in this domain emphasizes that an identification or detection model should be able to learn newly registered devices without retraining on a large dataset containing new and old devices. Because retaining the old dataset or deriving generators for knowledge replay is computationally expensive. This topic  faces several challenges:
\begin{itemize}
    \item Knowing the capacity or the maximum number of devices a model can memorize, especially for the Black-Box models, e.g., the Deep Neural Networks. 
    \item  Expanding models dynamically as new devices are being added. Continual learning is natively supported in Nearest Neighbor algorithms but is challenging to implement in Deep Neural Networks. 
\end{itemize}

\subsubsection{Deployment of device identification models}The deployment sites and model providers' lab can differ dramatically, in which identification accuracy can be impaired. This issue is more severe in device identification models using wireless signals due to the difference of wireless channel characteristics. For alleviation, extra works are needed:
\begin{itemize}
    \item \textcolor{black}{Deriving features that are independent of wireless channels or deployment sites. Researches in \cite{sankhe2019oracle,sankhe2019no} suggest that neural networks can only learn about channel-specific features rather than device-specific features.}
    \item \textcolor{black}{Occasional site fine-tunes are needed with the help of continual or transfer learning to adapt to variations.}
    \item \textcolor{black}{Model providers need to use data augmentation methods to simulate operational variations during lab training, as suggested in \cite{soltanimore}.}
    \item \textcolor{black}{Model providers can use multi-domain training to derive multi-purpose feature extractors, which will be utilized as building blocks for domain-specific device identification models. Diverse training from different domains could provide more robust feature extractors.}
\end{itemize}

\subsubsection{Reliable benchmark datasets}The IoT device identification is a pattern recognition problem on signals or behaviors. A common benchmark dataset is critical for comparing various methods in device identification and rogue device detection reliably. However, \textcolor{black}{by the end of this survey, we only find a limited number of datasets providing devices' raw signals or network traffic traces in diverse scenarios.} Some datasets are provided in \cite{morin2019transmitter}, \cite{allahham2019dronerf} and \cite{GENESYS}, respectively. For physical layer device identification, a larger dataset containing raw signals from more than 100 airborne transponders are provided in \cite{gt9v-kz32-20}, but it only contains ADS-B protocol. \textcolor{black}{Another dataset containing more than 30 IoT devices' traffic traces under volumetric attack and benign scenarios are in \cite{iotanalytics}.} \textcolor{black}{Such dataset are important because they provide fair comparisons between algorithms. Additionally, models trained on large datasets can be efficiently transferred to more specific applications \cite{pan2009survey,niu2020transfer}.}

\textcolor{black}{\subsection{Challenges in feature engineering}}

\textcolor{black}{\subsubsection{The robustness of features}
Although many existing works claim the effectiveness of their discovered features, only very few evaluate the features' robustness under various scenarios in terms of device mobility pattern, temperature, obstacles, etc. Feature robustness has a limited influence on device type identification in the network or higher layers. However, in the Physical Layer identification of wireless devices, the robustness of features would severely impair the final model. Currently, a popular way to enforce robust feature discovery is through data augmentation to simulate various scenarios. Besides, in neural networks, regularization and dropout methods can encourage models to make full use of input data and discover robust latent features, but their effectiveness needs further study.}

\subsubsection{Making use of time-varying features}

\textcolor{black}{Some device detection and identification models make use of protocol-agnostic and behavior-independent features from physical layer wireless signals.} However, in mobile environments, devices' movements can result in time-varying channel conditions, in which device identification methods based on static channel characteristics can be impaired. On the other hand, varying patterns of channels, signal strength, etc. also encode valuable features, e.g., locations, distances, to describe IoT devices \cite{nguyen2017matthan,wang2020counter}. \textcolor{black}{Therefore, both discovering time-invariant features and making use of time-varying features are still an open issue in device identification and detection.}

\subsubsection{Challenges from deep generative attackers: }

The utilization of GAN brings challenges to device identification, especially in the Physical Layer. Using GAN models, an attacker can train highly realistic signal or data packet generators to mimic its victims' signal characteristics. Research in \cite{shi2019generative} shows that GAN can increase the success rate of spoofing attacks from less than 10\% to approximately 80\%. Fortunately, a simple remedy is to use MIMO receivers and wireless localization methods to estimate whether a transmitter is from a reasonable location. \textcolor{black}{Additionally, controlled imperfections can be dynamically imprinted into the devices' signals or data flows, with a Pseudorandom Noise Code driven time-varying manner \cite{sankhe2019no} which is cryptographically impossible to predict.}

\textcolor{black}{
\subsection{Future research trends}
}

\subsubsection{\textcolor{black}{Deep identification models with explainable behaviors and assured performances}}
\textcolor{black}{The conveniences of Deep Neural Network make it a versatile tool to implement IoT device identification and rogue device detection systems, but more efforts have to be made, especially for model explainability and performance assurability. On the one hand, we have limited knowledge of the decision process, especially on how a deep neural network makes its final decisions and corresponding decision boundaries. Without knowing the decision process and decision boundaries, there is no way to assure its performance. On the other hand, researches on the explainability of Deep Neural Networks focus on explaining models' behaviors but do not provide guidelines on deriving assurable performance. Without explainability, we can not assure the performances of models.}

\subsubsection{\textcolor{black}{Unsupervised and continual learning enabled deep identification model}}
With a large number of devices being connected to IoT, device identification and detection models need to continually adapt to operational variation in real-time. \textcolor{black}{A solution can be the seamless integration of the feature abstraction capability of deep neural networks, continual learning and unsupervised learning. The knowledge of using deep neural networks to perform unsupervised learning for IoT device identification and detection is currently limited. Meanwhile, continual learning in deep models for device identification and detection is also rarely investigated.}


\subsubsection{\textcolor{black}{Controlled imprinting of verifiable patterns}}
\textcolor{black}{Compared with passive non-cryptographic device identification and detection methods in this survey, a proactive way is imprinting verifiable patterns into devices' transmitted signals or activity patterns. As suggested in \cite{mohanti2020airid}, controlled imperfections are utilized as verifiable patterns. Embedded these patterns in signals could significantly enhance the performance of device identification. However, a critical concern is how to prevent the adversaries from collecting and learning about the imprinted identity verification information. As suggested in \cite{sankhe2019oracle}, a possible solution is to dynamically change the identity verification patterns according to a pair of synchronized pseudorandom code generators, where the initialization keys are only shared among the device and corresponding device identifiers. Methods are still limited in imprinting verifiable patterns that are difficult to learn.}

\section{Conclusion}
\label{sectConclusion}
\textcolor{black}{
This survey aims to provide a comprehensive on the existing technologies on IoT device detection and identification from passively collected network traffic traces and wireless signal patterns. We discuss existing non-cryptographic IoT device identification mechanisms from the perspective of machine learning and pinpoint several key developing trends such as continual learning, abnormality detection, and deep unsupervised learning with explainability. We found that a multi-perspective IoT wireless device detection and identification framework is needed. Future research for rogue IoT device identification and detection needs to cope with challenges beyond signal processing and borrow ideas from advanced topics in Artificial Intelligence and Knowledge Discovery. }
\section*{Acknowledgment}

This research was partially supported through Embry-Riddle Aeronautical University’s Faculty Innovative Research in Science and Technology (FIRST) Program and the National Science Foundation under Grant No. 1956193.




\bibliographystyle{IEEEtran}

\bibliography{ref.bib}

\begin{thebibliography}{100}
\providecommand{\url}[1]{#1}
\csname url@samestyle\endcsname
\providecommand{\newblock}{\relax}
\providecommand{\bibinfo}[2]{#2}
\providecommand{\BIBentrySTDinterwordspacing}{\spaceskip=0pt\relax}
\providecommand{\BIBentryALTinterwordstretchfactor}{4}
\providecommand{\BIBentryALTinterwordspacing}{\spaceskip=\fontdimen2\font plus
\BIBentryALTinterwordstretchfactor\fontdimen3\font minus
  \fontdimen4\font\relax}
\providecommand{\BIBforeignlanguage}[2]{{%
\expandafter\ifx\csname l@#1\endcsname\relax
\typeout{** WARNING: IEEEtran.bst: No hyphenation pattern has been}%
\typeout{** loaded for the language `#1'. Using the pattern for}%
\typeout{** the default language instead.}%
\else
\language=\csname l@#1\endcsname
\fi
#2}}
\providecommand{\BIBdecl}{\relax}
\BIBdecl

\bibitem{IIOT17}
S.~Jeschke, C.~Brecher, H.~Song, and D.~Rawat, \emph{Industrial Internet of
  Things: Cybermanufacturing Systems}.\hskip 1em plus 0.5em minus 0.4em\relax
  Cham, Switzerland: Springer, 2017.

\bibitem{IoT-BDA}
Y.~{Sun}, H.~{Song}, A.~J. {Jara}, and R.~{Bie}, ``Internet of things and big
  data analytics for smart and connected communities,'' \emph{IEEE Access},
  vol.~4, pp. 766--773, 2016.

\bibitem{SC17}
H.~Song, R.~Srinivasan, T.~Sookoor, and S.~Jeschke, \emph{Smart Cities:
  Foundations, Principles and Applications}.\hskip 1em plus 0.5em minus
  0.4em\relax Hoboken, NJ: Wiley, 2017.

\bibitem{IoT-health}
Y.~{Zhang}, L.~{Sun}, H.~{Song}, and X.~{Cao}, ``Ubiquitous wsn for healthcare:
  Recent advances and future prospects,'' \emph{IEEE Internet of Things
  Journal}, vol.~1, no.~4, pp. 311--318, Aug 2014.

\bibitem{TCPS17}
Y.~Sun and H.~Song, \emph{Secure and Trustworthy Transportation Cyber-Physical
  Systems}.\hskip 1em plus 0.5em minus 0.4em\relax Singapore: Springer, 2017.

\bibitem{BDA19}
G.~Dartmann, H.~Song, and A.~Schmeink, \emph{Big Data Analytics for
  Cyber-Physical Systems: Machine Learning for the Internet of Things}.\hskip
  1em plus 0.5em minus 0.4em\relax Elsevier, 2019.

\bibitem{jiang2020applying}
Y.~Jiang, Y.~Liu, D.~Liu, and H.~Song, ``Applying machine learning to aviation
  big data for flight delay prediction,'' in \emph{2020 IEEE Intl Conf on
  Dependable, Autonomic and Secure Computing, Intl Conf on Pervasive
  Intelligence and Computing, Intl Conf on Cloud and Big Data Computing, Intl
  Conf on Cyber Science and Technology Congress
  (DASC/PiCom/CBDCom/CyberSciTech)}.\hskip 1em plus 0.5em minus 0.4em\relax
  IEEE, 2020, pp. 665--672.

\bibitem{zhang2020tree}
K.~Zhang, Y.~Liu, J.~Wang, H.~Song, and D.~Liu, ``Tree-based airspace capacity
  estimation,'' in \emph{2020 Integrated Communications Navigation and
  Surveillance Conference (ICNS)}.\hskip 1em plus 0.5em minus 0.4em\relax IEEE,
  2020, pp. 5C1--1.

\bibitem{liu2019domain}
Y.~Liu, J.~Li, Z.~Ming, H.~Song, X.~Weng, and J.~Wang, ``Domain-specific data
  mining for residents' transit pattern retrieval from incomplete
  information,'' \emph{Journal of Network and Computer Applications}, vol. 134,
  pp. 62--71, 2019.

\bibitem{SP17}
H.~Song, G.~A. Fink, and S.~Jeschke, \emph{Security and Privacy in
  Cyber-Physical Systems: Foundations, Principles and Applications}.\hskip 1em
  plus 0.5em minus 0.4em\relax Chichester, UK: Wiley-IEEE Press, 2017.

\bibitem{IoT-security}
I.~{Butun}, P.~{Österberg}, and H.~{Song}, ``Security of the internet of
  things: Vulnerabilities, attacks and countermeasures,'' \emph{IEEE
  Communications Surveys Tutorials}, pp. 1--1, 2019.

\bibitem{wurm2016security}
J.~Wurm, K.~Hoang, O.~Arias, A.-R. Sadeghi, and Y.~Jin, ``Security analysis on
  consumer and industrial iot devices,'' in \emph{2016 21st Asia and South
  Pacific Design Automation Conference (ASP-DAC)}.\hskip 1em plus 0.5em minus
  0.4em\relax IEEE, 2016, pp. 519--524.

\bibitem{shwartz2018reverse}
O.~Shwartz, Y.~Mathov, M.~Bohadana, Y.~Elovici, and Y.~Oren, ``Reverse
  engineering iot devices: Effective techniques and methods,'' \emph{IEEE
  Internet of Things Journal}, vol.~5, no.~6, pp. 4965--4976, 2018.

\bibitem{gupta2019iot}
A.~Gupta, \emph{The IoT Hacker's Handbook}.\hskip 1em plus 0.5em minus
  0.4em\relax Springer, 2019.

\bibitem{liu2020manually}
K.~Liu, M.~Yang, Z.~Ling, H.~Yan, Y.~Zhang, X.~Fu, and W.~Zhao, ``On manually
  reverse engineering communication protocols of linux based iot systems,''
  \emph{arXiv preprint arXiv:2007.11981}, 2020.

\bibitem{costa2019vulnerabilities}
L.~Costa, J.~P. Barros, and M.~Tavares, ``Vulnerabilities in iot devices for
  smart home environment,'' in \emph{Proceedings of the 5th International
  Conference on Information Systems Security e Privacy, ICISSP 2019.},
  vol.~1.\hskip 1em plus 0.5em minus 0.4em\relax SCITEPRESS, 2019, pp.
  615--622.

\bibitem{pollack2018aviation}
J.~Pollack and P.~Ranganathan, ``Aviation navigation systems security: Ads-b,
  gps, iff,'' in \emph{Proceedings of the International Conference on Security
  and Management (SAM)}.\hskip 1em plus 0.5em minus 0.4em\relax The Steering
  Committee of The World Congress in Computer Science, Computer~…, 2018, pp.
  129--135.

\bibitem{manesh2017analysis}
M.~R. Manesh and N.~Kaabouch, ``Analysis of vulnerabilities, attacks,
  countermeasures and overall risk of the automatic dependent
  surveillance-broadcast (ads-b) system,'' \emph{International Journal of
  Critical Infrastructure Protection}, vol.~19, pp. 16--31, 2017.

\bibitem{dalaklis2018vulnerabilities}
D.~Dimitrios, M.~Baldauf, and M.~Kitada, ``Vulnerabilities of the automatic
  identification system in the era of maritime autonomous surface ships,'' 06
  2018.

\bibitem{ray2015deais}
C.~Ray, R.~Gallen, C.~Iphar, A.~Napoli, and A.~Bouju, ``Deais project:
  detection of ais spoofing and resulting risks,'' in \emph{OCEANS
  2015-Genova}.\hskip 1em plus 0.5em minus 0.4em\relax IEEE, 2015, pp. 1--6.

\bibitem{yihunie2020assessing}
F.~L. Yihunie, A.~K. Singh, and S.~Bhatia, ``Assessing and exploiting security
  vulnerabilities of unmanned aerial vehicles,'' in \emph{Smart Systems and
  IoT: Innovations in Computing}.\hskip 1em plus 0.5em minus 0.4em\relax
  Springer, 2020, pp. 701--710.

\bibitem{koubaa2019micro}
A.~Koubaa, A.~Allouch, M.~Alajlan, Y.~Javed, A.~Belghith, and M.~Khalgui,
  ``Micro air vehicle link (mavlink) in a nutshell: A survey,'' \emph{IEEE
  Access}, vol.~7, pp. 87\,658--87\,680, 2019.

\bibitem{paul2008physical}
L.~Y. Paul, J.~S. Baras, and B.~M. Sadler, ``Physical-layer authentication,''
  \emph{IEEE Transactions on Information Forensics and Security}, vol.~3,
  no.~1, pp. 38--51, 2008.

\bibitem{wang2016wireless}
W.~Wang, Z.~Sun, S.~Piao, B.~Zhu, and K.~Ren, ``Wireless physical-layer
  identification: Modeling and validation,'' \emph{IEEE Transactions on
  Information Forensics and Security}, vol.~11, no.~9, pp. 2091--2106, 2016.

\bibitem{khattak2019perception}
H.~A. Khattak, M.~A. Shah, S.~Khan, I.~Ali, and M.~Imran, ``Perception layer
  security in internet of things,'' \emph{Future Generation Computer Systems},
  vol. 100, pp. 144--164, 2019.

\bibitem{brik2008wireless}
V.~Brik, S.~Banerjee, M.~Gruteser, and S.~Oh, ``Wireless device identification
  with radiometric signatures,'' in \emph{Proceedings of the 14th ACM
  international conference on Mobile computing and networking}.\hskip 1em plus
  0.5em minus 0.4em\relax ACM, 2008, pp. 116--127.

\bibitem{riddle1090}
J.~Sun, ``An open-access book about decoding mode-s and ads-b data,''
  \url{https://mode-s.org/}, May 2017.

\bibitem{tetreault2005use}
B.~J. Tetreault, ``Use of the automatic identification system (ais) for
  maritime domain awareness (mda),'' in \emph{Proceedings of Oceans 2005
  Mts/Ieee}.\hskip 1em plus 0.5em minus 0.4em\relax IEEE, 2005, pp. 1590--1594.

\bibitem{amanullah2020deep}
M.~A. Amanullah, R.~A.~A. Habeeb, F.~H. Nasaruddin, A.~Gani, E.~Ahmed, A.~S.~M.
  Nainar, N.~M. Akim, and M.~Imran, ``Deep learning and big data technologies
  for iot security,'' \emph{Computer Communications}, vol. 151, pp. 495--517,
  2020.

\bibitem{al2020survey}
M.~A. Al-Garadi, A.~Mohamed, A.~Al-Ali, X.~Du, I.~Ali, and M.~Guizani, ``A
  survey of machine and deep learning methods for internet of things (iot)
  security,'' \emph{IEEE Communications Surveys \& Tutorials}, 2020.

\bibitem{wang2019physical}
N.~Wang, P.~Wang, A.~Alipour-Fanid, L.~Jiao, and K.~Zeng, ``Physical-layer
  security of 5g wireless networks for iot: Challenges and opportunities,''
  \emph{IEEE Internet of Things Journal}, vol.~6, no.~5, pp. 8169--8181, 2019.

\bibitem{baldini2017survey}
G.~Baldini and G.~Steri, ``A survey of techniques for the identification of
  mobile phones using the physical fingerprints of the built-in components,''
  \emph{IEEE Communications Surveys \& Tutorials}, vol.~19, no.~3, pp.
  1761--1789, 2017.

\bibitem{danev2012physical}
B.~Danev and S.~Capkun, ``Physical-layer identification of wireless sensor
  nodes,'' \emph{Technical report/ETH Z{\"u}rich, Department of Computer
  Science}, vol. 604, 2012.

\bibitem{zeng2010non}
K.~Zeng, K.~Govindan, and P.~Mohapatra, ``Non-cryptographic authentication and
  identification in wireless networks,'' \emph{network security}, vol.~1, p.~3,
  2010.

\bibitem{wang2018security}
K.-H. Wang, C.-M. Chen, W.~Fang, and T.-Y. Wu, ``On the security of a new
  ultra-lightweight authentication protocol in iot environment for rfid tags,''
  \emph{The Journal of Supercomputing}, vol.~74, no.~1, pp. 65--70, 2018.

\bibitem{loi2017systematically}
F.~Loi, A.~Sivanathan, H.~H. Gharakheili, A.~Radford, and V.~Sivaraman,
  ``Systematically evaluating security and privacy for consumer iot devices,''
  in \emph{Proceedings of the 2017 Workshop on Internet of Things Security and
  Privacy}, 2017, pp. 1--6.

\bibitem{guaman2020systematic}
D.~S. Guam{\'a}n, J.~M. Del~Alamo, and J.~C. Caiza, ``A systematic mapping
  study on software quality control techniques for assessing privacy in
  information systems,'' \emph{IEEE Access}, vol.~8, pp. 74\,808--74\,833,
  2020.

\bibitem{ren2019information}
J.~Ren, D.~J. Dubois, D.~Choffnes, A.~M. Mandalari, R.~Kolcun, and H.~Haddadi,
  ``Information exposure from consumer iot devices: A multidimensional,
  network-informed measurement approach,'' in \emph{Proceedings of the Internet
  Measurement Conference}, 2019, pp. 267--279.

\bibitem{al2017internet}
S.~Al-Sarawi, M.~Anbar, K.~Alieyan, and M.~Alzubaidi, ``Internet of things
  (iot) communication protocols,'' in \emph{2017 8th International conference
  on information technology (ICIT)}.\hskip 1em plus 0.5em minus 0.4em\relax
  IEEE, 2017, pp. 685--690.

\bibitem{jawhar2018networking}
I.~Jawhar, N.~Mohamed, and J.~Al-Jaroodi, ``Networking architectures and
  protocols for smart city systems,'' \emph{Journal of Internet Services and
  Applications}, vol.~9, no.~1, p.~26, 2018.

\bibitem{metzger2019modeling}
F.~Metzger, T.~Ho{\ss}feld, A.~Bauer, S.~Kounev, and P.~E. Heegaard, ``Modeling
  of aggregated iot traffic and its application to an iot cloud,''
  \emph{Proceedings of the IEEE}, vol. 107, no.~4, pp. 679--694, 2019.

\bibitem{han2010packetshader}
S.~Han, K.~Jang, K.~Park, and S.~Moon, ``Packetshader: a gpu-accelerated
  software router,'' \emph{ACM SIGCOMM Computer Communication Review}, vol.~40,
  no.~4, pp. 195--206, 2010.

\bibitem{hu2014survey}
F.~Hu, Q.~Hao, and K.~Bao, ``A survey on software-defined network and openflow:
  From concept to implementation,'' \emph{IEEE Communications Surveys \&
  Tutorials}, vol.~16, no.~4, pp. 2181--2206, 2014.

\bibitem{rafique2020complementing}
W.~Rafique, L.~Qi, I.~Yaqoob, M.~Imran, R.~ur~Rasool, and W.~Dou,
  ``Complementing iot services through software defined networking and edge
  computing: A comprehensive survey,'' \emph{IEEE Communications Surveys \&
  Tutorials}, 2020.

\bibitem{gao2011restful}
L.~Gao, C.~Zhang, and L.~Sun, ``Restful web of things api in sharing sensor
  data,'' in \emph{2011 International Conference on Internet Technology and
  Applications}.\hskip 1em plus 0.5em minus 0.4em\relax IEEE, 2011, pp. 1--4.

\bibitem{sivanathan2020iot}
A.~Sivanathan, ``Iot behavioral monitoring via network traffic analysis,''
  \emph{arXiv preprint arXiv:2001.10632}, 2020.

\bibitem{sivanathan2017characterizing}
A.~Sivanathan, D.~Sherratt, H.~H. Gharakheili, A.~Radford, C.~Wijenayake,
  A.~Vishwanath, and V.~Sivaraman, ``Characterizing and classifying iot traffic
  in smart cities and campuses,'' in \emph{2017 IEEE Conference on Computer
  Communications Workshops (INFOCOM WKSHPS)}.\hskip 1em plus 0.5em minus
  0.4em\relax IEEE, 2017, pp. 559--564.

\bibitem{marchal2019audi}
S.~Marchal, M.~Miettinen, T.~D. Nguyen, A.-R. Sadeghi, and N.~Asokan, ``Audi:
  Toward autonomous iot device-type identification using periodic
  communication,'' \emph{IEEE Journal on Selected Areas in Communications},
  vol.~37, no.~6, pp. 1402--1412, 2019.

\bibitem{miettinen2017iot}
M.~Miettinen, S.~Marchal, I.~Hafeez, N.~Asokan, A.-R. Sadeghi, and S.~Tarkoma,
  ``Iot sentinel: Automated device-type identification for security enforcement
  in iot,'' in \emph{2017 IEEE 37th International Conference on Distributed
  Computing Systems (ICDCS)}.\hskip 1em plus 0.5em minus 0.4em\relax IEEE,
  2017, pp. 2177--2184.

\bibitem{meidan2017profiliot}
Y.~Meidan, M.~Bohadana, A.~Shabtai, J.~D. Guarnizo, M.~Ochoa, N.~O.
  Tippenhauer, and Y.~Elovici, ``Profiliot: a machine learning approach for iot
  device identification based on network traffic analysis,'' in
  \emph{Proceedings of the symposium on applied computing}, 2017, pp. 506--509.

\bibitem{meidan2017detection}
Y.~Meidan, M.~Bohadana, A.~Shabtai, M.~Ochoa, N.~O. Tippenhauer, J.~D.
  Guarnizo, and Y.~Elovici, ``Detection of unauthorized iot devices using
  machine learning techniques,'' \emph{arXiv preprint arXiv:1709.04647}, 2017.

\bibitem{aksoy2019automated}
A.~Aksoy and M.~H. Gunes, ``Automated iot device identification using network
  traffic,'' in \emph{ICC 2019-2019 IEEE International Conference on
  Communications (ICC)}.\hskip 1em plus 0.5em minus 0.4em\relax IEEE, 2019, pp.
  1--7.

\bibitem{sivanathan2020managing}
A.~Sivanathan, H.~H. Gharakheili, and V.~Sivaraman, ``Managing iot
  cyber-security using programmable telemetry and machine learning,''
  \emph{IEEE Transactions on Network and Service Management}, vol.~17, no.~1,
  pp. 60--74, 2020.

\bibitem{kotak2020iot}
J.~Kotak and Y.~Elovici, ``Iot device identification using deep learning,''
  \emph{arXiv preprint arXiv:2002.11686}, 2020.

\bibitem{lear2019rfc}
E.~Lear, R.~Droms, and D.~Romascanu, ``Rfc 8520: Manufacturer usage description
  specification,'' \emph{Internet Engineering Task Force}, 2019.

\bibitem{iotanalytics}
H.~Hassan, S.~Vijay, H.~Ayyoob, S.~Arunan, and P.~Arman, ``\text{IoT} device
  profiles, attack traces and traffic traces,''
  \url{https://iotanalytics.unsw.edu.au/index}, Accessed on Oct. 2020.

\bibitem{hamza2020verifying}
A.~Hamza, D.~Ranathunga, H.~H. Gharakheili, T.~A. Benson, M.~Roughan, and
  V.~Sivaraman, ``Verifying and monitoring iots network behavior using mud
  profiles,'' \emph{IEEE Transactions on Dependable and Secure Computing},
  2020.

\bibitem{ashibani2018user}
Y.~Ashibani and Q.~H. Mahmoud, ``A user authentication model for iot networks
  based on app traffic patterns,'' in \emph{2018 IEEE 9th Annual Information
  Technology, Electronics and Mobile Communication Conference (IEMCON)}.\hskip
  1em plus 0.5em minus 0.4em\relax IEEE, 2018, pp. 632--638.

\bibitem{ashibani2020design}
------, ``Design and evaluation of a user authentication model for iot networks
  based on app event patterns,'' \emph{Cluster Computing}, pp. 1--14, 2020.

\bibitem{zvonar2001software}
Z.~Zvonar and J.~Mitola, \emph{Software radio technologies: selected
  readings}.\hskip 1em plus 0.5em minus 0.4em\relax Wiley-IEEE Press, 2001.

\bibitem{chatterjee2018rf}
B.~Chatterjee, D.~Das, S.~Maity, and S.~Sen, ``Rf-puf: Enhancing iot security
  through authentication of wireless nodes using in-situ machine learning,''
  \emph{IEEE Internet of Things Journal}, vol.~6, no.~1, pp. 388--398, 2018.

\bibitem{herder2014physical}
C.~Herder, M.-D. Yu, F.~Koushanfar, and S.~Devadas, ``Physical unclonable
  functions and applications: A tutorial,'' \emph{Proceedings of the IEEE},
  vol. 102, no.~8, pp. 1126--1141, 2014.

\bibitem{polak2015wireless}
A.~C. Polak and D.~L. Goeckel, ``Wireless device identification based on rf
  oscillator imperfections,'' \emph{IEEE Transactions on Information Forensics
  and Security}, vol.~10, no.~12, pp. 2492--2501, 2015.

\bibitem{azarmehr2017wireless}
M.~Azarmehr, A.~Mehta, and R.~Rashidzadeh, ``Wireless device identification
  using oscillator control voltage as rf fingerprint,'' in \emph{2017 IEEE 30th
  Canadian Conference on Electrical and Computer Engineering (CCECE)}.\hskip
  1em plus 0.5em minus 0.4em\relax IEEE, 2017, pp. 1--4.

\bibitem{zhuang2018fbsleuth}
Z.~Zhuang, X.~Ji, T.~Zhang, J.~Zhang, W.~Xu, Z.~Li, and Y.~Liu, ``Fbsleuth:
  Fake base station forensics via radio frequency fingerprinting,'' in
  \emph{Proceedings of the 2018 on Asia Conference on Computer and
  Communications Security}.\hskip 1em plus 0.5em minus 0.4em\relax ACM, 2018,
  pp. 261--272.

\bibitem{peng2016differential}
L.~Peng, A.~Hu, Y.~Jiang, Y.~Yan, and C.~Zhu, ``A differential constellation
  trace figure based device identification method for zigbee nodes,'' in
  \emph{2016 8th International Conference on Wireless Communications \& Signal
  Processing (WCSP)}.\hskip 1em plus 0.5em minus 0.4em\relax IEEE, 2016, pp.
  1--6.

\bibitem{peng2018design}
L.~Peng, A.~Hu, J.~Zhang, Y.~Jiang, J.~Yu, and Y.~Yan, ``Design of a hybrid rf
  fingerprint extraction and device classification scheme,'' \emph{IEEE
  Internet of Things Journal}, vol.~6, no.~1, pp. 349--360, 2018.

\bibitem{peng2019deep}
L.~Peng, J.~Zhang, M.~Liu, and A.~Hu, ``Deep learning based rf fingerprint
  identification using differential constellation trace figure,'' \emph{IEEE
  Transactions on Vehicular Technology}, 2019.

\bibitem{zhang2016identification}
G.~Zhang, L.~Xia, S.~Jia, and Y.~Ji, ``Identification of cloned hf rfid
  proximity cards based on rf fingerprinting,'' in \emph{2016 IEEE
  Trustcom/BigDataSE/ISPA}.\hskip 1em plus 0.5em minus 0.4em\relax IEEE, 2016,
  pp. 292--300.

\bibitem{cobb2010physical}
W.~E. Cobb, E.~W. Garcia, M.~A. Temple, R.~O. Baldwin, and Y.~C. Kim,
  ``Physical layer identification of embedded devices using rf-dna
  fingerprinting,'' in \emph{2010-Milcom 2010 Military Communications
  Conference}.\hskip 1em plus 0.5em minus 0.4em\relax IEEE, 2010, pp.
  2168--2173.

\bibitem{wang2017research}
C.~Wang, Y.~Lin, and Z.~Zhang, ``Research on physical layer security of
  cognitive radio network based on rf-dna,'' in \emph{2017 IEEE International
  Conference on Software Quality, Reliability and Security Companion
  (QRS-C)}.\hskip 1em plus 0.5em minus 0.4em\relax IEEE, 2017, pp. 252--255.

\bibitem{bihl2016feature}
T.~J. Bihl, K.~W. Bauer, and M.~A. Temple, ``Feature selection for rf
  fingerprinting with multiple discriminant analysis and using zigbee device
  emissions,'' \emph{IEEE Transactions on Information Forensics and Security},
  vol.~11, no.~8, pp. 1862--1874, 2016.

\bibitem{ramsey2015wireless}
B.~W. Ramsey, T.~D. Stubbs, B.~E. Mullins, M.~A. Temple, and M.~A. Buckner,
  ``Wireless infrastructure protection using low-cost radio frequency
  fingerprinting receivers,'' \emph{international journal of critical
  infrastructure protection}, vol.~8, pp. 27--39, 2015.

\bibitem{dubendorfer2012rf}
C.~K. Dubendorfer, B.~W. Ramsey, and M.~A. Temple, ``An rf-dna verification
  process for zigbee networks,'' in \emph{MILCOM 2012-2012 IEEE Military
  Communications Conference}.\hskip 1em plus 0.5em minus 0.4em\relax IEEE,
  2012, pp. 1--6.

\bibitem{harmer2011using}
P.~K. Harmer, M.~A. Temple, M.~A. Buckner, and E.~Farquahar, ``Using
  differential evolution to optimize'learning from signals' and enhance network
  security,'' in \emph{Proceedings of the 13th annual conference on Genetic and
  evolutionary computation}.\hskip 1em plus 0.5em minus 0.4em\relax ACM, 2011,
  pp. 1811--1818.

\bibitem{zhou2019design}
X.~Zhou, A.~Hu, G.~Li, L.~Peng, Y.~Xing, and J.~Yu, ``Design of a robust rf
  fingerprint generation and classification scheme for practical device
  identification,'' in \emph{2019 IEEE Conference on Communications and Network
  Security (CNS)}.\hskip 1em plus 0.5em minus 0.4em\relax IEEE, 2019, pp.
  196--204.

\bibitem{danev2010attacks}
B.~Danev, H.~Luecken, S.~Capkun, and K.~El~Defrawy, ``Attacks on physical-layer
  identification,'' in \emph{Proceedings of the third ACM conference on
  Wireless network security}.\hskip 1em plus 0.5em minus 0.4em\relax ACM, 2010,
  pp. 89--98.

\bibitem{polak2015identification}
A.~C. Polak and D.~L. Goeckel, ``Identification of wireless devices of users
  who actively fake their rf fingerprints with artificial data distortion,''
  \emph{IEEE Transactions on Wireless Communications}, vol.~14, no.~11, pp.
  5889--5899, 2015.

\bibitem{kose2019rf}
M.~K{\"o}se, S.~Ta{\c{s}}cio{\u{g}}lu, and Z.~Telatar, ``Rf fingerprinting of
  iot devices based on transient energy spectrum,'' \emph{IEEE Access}, vol.~7,
  pp. 18\,715--18\,726, 2019.

\bibitem{zhang2018ensemble}
Z.~Zhang, Y.~Li, C.~Wang, M.~Wang, Y.~Tu, and J.~Wang, ``An ensemble learning
  method for wireless multimedia device identification,'' \emph{Security and
  Communication Networks}, vol. 2018, 2018.

\bibitem{tu2019research}
Y.~Tu, Z.~Zhang, Y.~Li, C.~Wang, and Y.~Xiao, ``Research on the internet of
  things device recognition based on rf-fingerprinting,'' \emph{IEEE Access},
  vol.~7, pp. 37\,426--37\,431, 2019.

\bibitem{ali2019assessment}
A.~M. Ali, E.~Uzundurukan, and A.~Kara, ``Assessment of features and
  classifiers for bluetooth rf fingerprinting,'' \emph{IEEE Access}, vol.~7,
  pp. 50\,524--50\,535, 2019.

\bibitem{yu2019radio}
J.~Yu, A.~Hu, F.~Zhou, Y.~Xing, Y.~Yu, G.~Li, and L.~Peng, ``Radio frequency
  fingerprint identification based on denoising autoencoders,'' \emph{arXiv
  preprint arXiv:1907.08809}, 2019.

\bibitem{romano2018unsupervised}
J.~M.~T. Romano, R.~Attux, C.~C. Cavalcante, and R.~Suyama, \emph{Unsupervised
  signal processing: channel equalization and source separation}.\hskip 1em
  plus 0.5em minus 0.4em\relax CRC Press, 2018.

\bibitem{restuccia2019deepradioid}
F.~Restuccia, S.~D'Oro, A.~Al-Shawabka, M.~Belgiovine, L.~Angioloni,
  S.~Ioannidis, K.~Chowdhury, and T.~Melodia, ``Deepradioid: Real-time
  channel-resilient optimization of deep learning-based radio fingerprinting
  algorithms,'' \emph{arXiv preprint arXiv:1904.07623}, 2019.

\bibitem{xu2018research}
C.~Xu, G.~Huang, and X.~Hou, ``Research on communication fm signal blind
  separation under jamming environments,'' in \emph{2nd International Forum on
  Management, Education and Information Technology Application (IFMEITA
  2017)}.\hskip 1em plus 0.5em minus 0.4em\relax Atlantis Press, 2018.

\bibitem{baliarsingh2016adaptive}
S.~N. Baliarsingh, A.~Senapati, A.~Deb, and J.~S. Roy, ``Adaptive beam
  formation for smart antenna for mobile communication network using new hybrid
  algorithms,'' in \emph{2016 International Conference on Communication and
  Signal Processing (ICCSP)}.\hskip 1em plus 0.5em minus 0.4em\relax IEEE,
  2016, pp. 2146--2151.

\bibitem{hanna2019deep}
S.~S. Hanna and D.~Cabric, ``Deep learning based transmitter identification
  using power amplifier nonlinearity,'' in \emph{2019 International Conference
  on Computing, Networking and Communications (ICNC)}.\hskip 1em plus 0.5em
  minus 0.4em\relax IEEE, 2019, pp. 674--680.

\bibitem{zheng2019fid}
T.~Zheng, Z.~Sun, and K.~Ren, ``Fid: Function modeling-based data-independent
  and channel-robust physical-layer identification,'' in \emph{IEEE INFOCOM
  2019-IEEE Conference on Computer Communications}.\hskip 1em plus 0.5em minus
  0.4em\relax IEEE, 2019, pp. 199--207.

\bibitem{halperin2011tool}
D.~Halperin, W.~Hu, A.~Sheth, and D.~Wetherall, ``Tool release: Gathering
  802.11 n traces with channel state information,'' \emph{ACM SIGCOMM Computer
  Communication Review}, vol.~41, no.~1, pp. 53--53, 2011.

\bibitem{ma2019wifi}
Y.~Ma, G.~Zhou, and S.~Wang, ``Wifi sensing with channel state information: A
  survey,'' \emph{ACM Computing Surveys (CSUR)}, vol.~52, no.~3, p.~46, 2019.

\bibitem{liu2014practical}
H.~Liu, Y.~Wang, J.~Liu, J.~Yang, and Y.~Chen, ``Practical user authentication
  leveraging channel state information (csi),'' in \emph{Proceedings of the 9th
  ACM symposium on Information, computer and communications security}.\hskip
  1em plus 0.5em minus 0.4em\relax ACM, 2014, pp. 389--400.

\bibitem{zaman2018deep}
S.~Zaman, C.~Chakraborty, N.~Mehajabin, M.~Mamun-Or-Rashid, and M.~A. Razzaque,
  ``A deep learning based device authentication scheme using channel state
  information,'' in \emph{2018 International Conference on Innovation in
  Engineering and Technology (ICIET)}.\hskip 1em plus 0.5em minus 0.4em\relax
  IEEE, 2018, pp. 1--5.

\bibitem{zou2017tagfree}
Y.~Zou, Y.~Wang, S.~Ye, K.~Wu, and L.~M. Ni, ``Tagfree: Passive object
  differentiation via physical layer radiometric signatures,'' in \emph{2017
  IEEE International Conference on Pervasive Computing and Communications
  (PerCom)}.\hskip 1em plus 0.5em minus 0.4em\relax IEEE, 2017, pp. 237--246.

\bibitem{hua2018accurate}
J.~Hua, H.~Sun, Z.~Shen, Z.~Qian, and S.~Zhong, ``Accurate and efficient
  wireless device fingerprinting using channel state information,'' in
  \emph{IEEE INFOCOM 2018-IEEE Conference on Computer Communications}.\hskip
  1em plus 0.5em minus 0.4em\relax IEEE, 2018, pp. 1700--1708.

\bibitem{liu2019real}
P.~Liu, P.~Yang, W.-Z. Song, Y.~Yan, and X.-Y. Li, ``Real-time identification
  of rogue wifi connections using environment-independent physical features,''
  in \emph{IEEE INFOCOM 2019-IEEE Conference on Computer Communications}.\hskip
  1em plus 0.5em minus 0.4em\relax IEEE, 2019, pp. 190--198.

\bibitem{wang2018wi}
Z.~Wang, B.~Guo, Z.~Yu, and X.~Zhou, ``Wi-fi csi-based behavior recognition:
  From signals and actions to activities,'' \emph{IEEE Communications
  Magazine}, vol.~56, no.~5, pp. 109--115, 2018.

\bibitem{hong2016wfid}
F.~Hong, X.~Wang, Y.~Yang, Y.~Zong, Y.~Zhang, and Z.~Guo, ``Wfid: Passive
  device-free human identification using wifi signal,'' in \emph{Proceedings of
  the 13th International Conference on Mobile and Ubiquitous Systems:
  Computing, Networking and Services}.\hskip 1em plus 0.5em minus 0.4em\relax
  ACM, 2016, pp. 47--56.

\bibitem{zeng2014your}
Y.~Zeng, P.~H. Pathak, C.~Xu, and P.~Mohapatra, ``Your ap knows how you move:
  fine-grained device motion recognition through wifi,'' in \emph{Proceedings
  of the 1st ACM workshop on Hot topics in wireless}.\hskip 1em plus 0.5em
  minus 0.4em\relax ACM, 2014, pp. 49--54.

\bibitem{wang2014eyes}
Y.~Wang, J.~Liu, Y.~Chen, M.~Gruteser, J.~Yang, and H.~Liu, ``E-eyes:
  device-free location-oriented activity identification using fine-grained wifi
  signatures,'' in \emph{Proceedings of the 20th annual international
  conference on Mobile computing and networking}.\hskip 1em plus 0.5em minus
  0.4em\relax ACM, 2014, pp. 617--628.

\bibitem{wang2016wifall}
Y.~Wang, K.~Wu, and L.~M. Ni, ``Wifall: Device-free fall detection by wireless
  networks,'' \emph{IEEE Transactions on Mobile Computing}, vol.~16, no.~2, pp.
  581--594, 2016.

\bibitem{li2019location}
G.~Li, J.~Yu, Y.~Xing, and A.~Hu, ``Location-invariant physical layer
  identification approach for wifi devices,'' \emph{IEEE Access}, vol.~7, pp.
  106\,974--106\,986, 2019.

\bibitem{xu2020rf}
C.~Xu, B.~Chen, Y.~Liu, F.~He, and H.~Song, ``Rf fingerprint measurement for
  detecting multiple amateur drones based on stft and feature reduction,'' in
  \emph{2020 Integrated Communications Navigation and Surveillance Conference
  (ICNS)}.\hskip 1em plus 0.5em minus 0.4em\relax IEEE, 2020, pp. 4G1--1.

\bibitem{chen2017identification}
S.~Chen, F.~Xie, Y.~Chen, H.~Song, and H.~Wen, ``Identification of wireless
  transceiver devices using radio frequency (rf) fingerprinting based on stft
  analysis to enhance authentication security,'' in \emph{2017 IEEE 5th
  International Symposium on Electromagnetic Compatibility
  (EMC-Beijing)}.\hskip 1em plus 0.5em minus 0.4em\relax IEEE, 2017, pp. 1--5.

\bibitem{reising2015authorized}
D.~R. Reising, M.~A. Temple, and J.~A. Jackson, ``Authorized and rogue device
  discrimination using dimensionally reduced rf-dna fingerprints,'' \emph{IEEE
  Transactions on Information Forensics and Security}, vol.~10, no.~6, pp.
  1180--1192, 2015.

\bibitem{li2018low}
Y.~Li, L.~Chen, J.~Chen, F.~Xie, S.~Chen, and H.~Wen, ``A low complexity
  feature extraction for the rf fingerprinting process,'' in \emph{2018 IEEE
  Conference on Communications and Network Security (CNS)}.\hskip 1em plus
  0.5em minus 0.4em\relax IEEE, 2018, pp. 1--2.

\bibitem{wu2019specific}
X.~Wu, Y.~Shi, W.~Meng, X.~Ma, and N.~Fang, ``Specific emitter identification
  for satellite communication using probabilistic neural networks,''
  \emph{International Journal of Satellite Communications and Networking},
  vol.~37, no.~3, pp. 283--291, 2019.

\bibitem{sun2016specific}
D.~Sun, Y.~Li, and Y.~Xu, ``Specific emitter identification based on normalized
  frequency spectrum,'' in \emph{2016 2nd IEEE International Conference on
  Computer and Communications (ICCC)}.\hskip 1em plus 0.5em minus 0.4em\relax
  IEEE, 2016, pp. 1875--1879.

\bibitem{yuan2014specific}
Y.~Yuan, Z.~Huang, H.~Wu, and X.~Wang, ``Specific emitter identification based
  on hilbert--huang transform-based time--frequency--energy distribution
  features,'' \emph{IET Communications}, vol.~8, no.~13, pp. 2404--2412, 2014.

\bibitem{satija2018specific}
U.~Satija, N.~Trivedi, G.~Biswal, and B.~Ramkumar, ``Specific emitter
  identification based on variational mode decomposition and spectral features
  in single hop and relaying scenarios,'' \emph{IEEE Transactions on
  Information Forensics and Security}, vol.~14, no.~3, pp. 581--591, 2018.

\bibitem{wang2017radio}
X.~Wang, J.~Duan, C.~Wang, G.~Cui, and W.~Wang, ``A radio frequency
  fingerprinting identification method based on energy entropy and color
  moments of the bispectrum,'' in \emph{2017 9th International Conference on
  Advanced Infocomm Technology (ICAIT)}.\hskip 1em plus 0.5em minus 0.4em\relax
  IEEE, 2017, pp. 150--154.

\bibitem{lei2016individual}
Y.-K. Lei, ``Individual communication transmitter identification using
  correntropy-based collaborative representation,'' in \emph{2016 9th
  International Congress on Image and Signal Processing, BioMedical Engineering
  and Informatics (CISP-BMEI)}.\hskip 1em plus 0.5em minus 0.4em\relax IEEE,
  2016, pp. 1194--1200.

\bibitem{sun2017rf}
M.~Sun, L.~Zhang, J.~Bao, and Y.~Yan, ``Rf fingerprint extraction for gnss
  anti-spoofing using axial integrated wigner bispectrum,'' \emph{Journal of
  Information Security and Applications}, vol.~35, pp. 51--54, 2017.

\bibitem{zhang2018research}
Z.~Zhang, Y.~Li, and C.~Wang, ``Research on individual identification of
  wireless devices based on signal's energy distribution,'' in \emph{2018 IEEE
  23rd International Conference on Digital Signal Processing (DSP)}.\hskip 1em
  plus 0.5em minus 0.4em\relax IEEE, 2018, pp. 1--5.

\bibitem{huang2014hilbert}
N.~E. Huang, \emph{Hilbert-Huang transform and its applications}.\hskip 1em
  plus 0.5em minus 0.4em\relax World Scientific, 2014, vol.~16.

\bibitem{wang2017specific}
W.~Wang, H.~Liu, J.~Yang, and H.~Yin, ``Specific emitter identification using
  decomposed hierarchical feature extraction methods,'' in \emph{2017 13th
  International Conference on Natural Computation, Fuzzy Systems and Knowledge
  Discovery (ICNC-FSKD)}.\hskip 1em plus 0.5em minus 0.4em\relax IEEE, 2017,
  pp. 1639--1643.

\bibitem{frei2006intrinsic}
M.~G. Frei and I.~Osorio, ``Intrinsic time-scale decomposition:
  time--frequency--energy analysis and real-time filtering of non-stationary
  signals,'' \emph{Proceedings of the Royal Society A: Mathematical, Physical
  and Engineering Sciences}, vol. 463, no. 2078, pp. 321--342, 2006.

\bibitem{shi2011improved}
Y.~Shi and M.~A. Jensen, ``Improved radiometric identification of wireless
  devices using mimo transmission,'' \emph{IEEE Transactions on Information
  Forensics and Security}, vol.~6, no.~4, pp. 1346--1354, 2011.

\bibitem{vo2016fingerprinting}
T.~D. Vo-Huu, T.~D. Vo-Huu, and G.~Noubir, ``Fingerprinting wi-fi devices using
  software defined radios,'' in \emph{Proceedings of the 9th ACM Conference on
  Security \& Privacy in Wireless and Mobile Networks}.\hskip 1em plus 0.5em
  minus 0.4em\relax ACM, 2016, pp. 3--14.

\bibitem{agadakos2019deep}
I.~Agadakos, N.~Agadakos, J.~Polakis, and M.~R. Amer, ``Deep complex networks
  for protocol-agnostic radio frequency device fingerprinting in the wild,''
  \emph{arXiv preprint arXiv:1909.08703}, 2019.

\bibitem{yu2019robust}
J.~Yu, A.~Hu, G.~Li, and L.~Peng, ``A robust rf fingerprinting approach using
  multi-sampling convolutional neural network,'' \emph{IEEE Internet of Things
  Journal}, 2019.

\bibitem{merchant2018deep}
K.~Merchant, S.~Revay, G.~Stantchev, and B.~Nousain, ``Deep learning for rf
  device fingerprinting in cognitive communication networks,'' \emph{IEEE
  Journal of Selected Topics in Signal Processing}, vol.~12, no.~1, pp.
  160--167, 2018.

\bibitem{riyaz2018deep}
S.~Riyaz, K.~Sankhe, S.~Ioannidis, and K.~Chowdhury, ``Deep learning
  convolutional neural networks for radio identification,'' \emph{IEEE
  Communications Magazine}, vol.~56, no.~9, pp. 146--152, 2018.

\bibitem{huang2017specific}
G.~Huang, Y.~Yuan, X.~Wang, and Z.~Huang, ``Specific emitter identification for
  communications transmitter using multi-measurements,'' \emph{Wireless
  Personal Communications}, vol.~94, no.~3, pp. 1523--1542, 2017.

\bibitem{bandt2002permutation}
C.~Bandt and B.~Pompe, ``Permutation entropy: a natural complexity measure for
  time series,'' \emph{Physical review letters}, vol.~88, no.~17, p. 174102,
  2002.

\bibitem{nikias1987bispectrum}
C.~L. Nikias and M.~R. Raghuveer, ``Bispectrum estimation: A digital signal
  processing framework,'' \emph{Proceedings of the IEEE}, vol.~75, no.~7, pp.
  869--891, 1987.

\bibitem{morin2019transmitter}
C.~Morin, L.~Cardoso, J.~Hoydis, J.-M. Gorce, and T.~Vial, ``Transmitter
  classification with supervised deep learning,'' \emph{arXiv preprint
  arXiv:1905.07923}, 2019.

\bibitem{allahham2019dronerf}
M.~S. Allahham, M.~F. Al-Sa'd, A.~Al-Ali, A.~Mohamed, T.~Khattab, and A.~Erbad,
  ``Dronerf dataset: A dataset of drones for rf-based detection, classification
  and identification,'' \emph{Data in brief}, vol.~26, p. 104313, 2019.

\bibitem{GENESYS}
C.~Kaushik, ``Genesys lab mldatasets,''
  \url{http://genesys-lab.org/mldatasets.au/index}, Accessed on Oct. 2020.

\bibitem{gt9v-kz32-20}
\BIBentryALTinterwordspacing
Y.~Liu, J.~Wang, H.~Song, S.~Niu, and Y.~Thomas, ``A 24-hour signal recording
  dataset with labels for cybersecurity and \text{IoT},'' 2020. [Online].
  Available: \url{http://dx.doi.org/10.21227/gt9v-kz32}
\BIBentrySTDinterwordspacing

\bibitem{baldini2018comparison}
G.~Baldini, C.~Gentile, R.~Giuliani, and G.~Steri, ``Comparison of techniques
  for radiometric identification based on deep convolutional neural networks,''
  \emph{Electronics Letters}, vol.~55, no.~2, pp. 90--92, 2018.

\bibitem{roy2019rfal}
D.~Roy, T.~Mukherjee, M.~Chatterjee, E.~Blasch, and E.~Pasiliao, ``Rfal:
  Adversarial learning for rf transmitter identification and classification,''
  \emph{IEEE Transactions on Cognitive Communications and Networking}, 2019.

\bibitem{gopalakrishnan2019robust}
S.~Gopalakrishnan, M.~Cekic, and U.~Madhow, ``Robust wireless fingerprinting
  via complex-valued neural networks,'' \emph{arXiv preprint arXiv:1905.09388},
  2019.

\bibitem{huang2017communication}
J.~Huang, Y.~Lei, and X.~Liao, ``Communication transmitter individual feature
  extraction method based on stacked denoising autoencoders under small sample
  prerequisite,'' in \emph{2017 7th IEEE International Conference on
  Electronics Information and Emergency Communication (ICEIEC)}.\hskip 1em plus
  0.5em minus 0.4em\relax IEEE, 2017, pp. 132--135.

\bibitem{oyedareestimating}
T.~Oyedare and J.-M.~J. Park, ``Estimating the required training dataset size
  for transmitter classification using deep learning.''

\bibitem{DBLP:journals/corr/abs-1905-12321}
\BIBentryALTinterwordspacing
J.~S. Dramsch, M.~L{\"{u}}thje, and A.~N. Christensen, ``Complex-valued neural
  networks for machine learning on non-stationary physical data,'' \emph{CoRR},
  vol. abs/1905.12321, 2019. [Online]. Available:
  \url{http://arxiv.org/abs/1905.12321}
\BIBentrySTDinterwordspacing

\bibitem{liu2020zero}
Y.~Liu, J.~Wang, J.~Li, H.~Song, T.~Yang, S.~Niu, and Z.~Ming, ``Zero-bias deep
  learning for accurate identification of internet of things (iot) devices,''
  \emph{IEEE Internet of Things Journal}, 2020.

\bibitem{liu2020deep}
Y.~Liu, J.~Wang, S.~Niu, and H.~Song, ``Deep learning enabled reliable identity
  verification and spoofing detection,'' in \emph{International Conference on
  Wireless Algorithms, Systems, and Applications}.\hskip 1em plus 0.5em minus
  0.4em\relax Springer, 2020, pp. 333--345.

\bibitem{li2018generative}
M.~Li, O.~Li, G.~Liu, and C.~Zhang, ``Generative adversarial networks-based
  semi-supervised automatic modulation recognition for cognitive radio
  networks,'' \emph{Sensors}, vol.~18, no.~11, p. 3913, 2018.

\bibitem{marchi2015novel}
E.~Marchi, F.~Vesperini, F.~Eyben, S.~Squartini, and B.~Schuller, ``A novel
  approach for automatic acoustic novelty detection using a denoising
  autoencoder with bidirectional lstm neural networks,'' in \emph{2015 IEEE
  International Conference on Acoustics, Speech and Signal Processing
  (ICASSP)}.\hskip 1em plus 0.5em minus 0.4em\relax IEEE, 2015, pp. 1996--2000.

\bibitem{khan2017detecting}
S.~S. Khan and B.~Taati, ``Detecting unseen falls from wearable devices using
  channel-wise ensemble of autoencoders,'' \emph{Expert Systems with
  Applications}, vol.~87, pp. 280--290, 2017.

\bibitem{shi2019deep}
Y.~Shi, K.~Davaslioglu, Y.~E. Sagduyu, W.~C. Headley, M.~Fowler, and G.~Green,
  ``Deep learning for rf signal classification in unknown and dynamic spectrum
  environments,'' in \emph{2019 IEEE International Symposium on Dynamic
  Spectrum Access Networks (DySPAN)}.\hskip 1em plus 0.5em minus 0.4em\relax
  IEEE, 2019, pp. 1--10.

\bibitem{gritsenko2019finding}
A.~Gritsenko, Z.~Wang, T.~Jian, J.~Dy, K.~Chowdhury, and S.~Ioannidis,
  ``Finding a ‘new’needle in the haystack: Unseen radio detection in large
  populations using deep learning,'' in \emph{2019 IEEE International Symposium
  on Dynamic Spectrum Access Networks (DySPAN)}.\hskip 1em plus 0.5em minus
  0.4em\relax IEEE, 2019, pp. 1--10.

\bibitem{kim2018identifying}
Y.~Kim, S.~An, and J.~So, ``Identifying signal source using channel state
  information in wireless lans,'' in \emph{2018 International Conference on
  Information Networking (ICOIN)}.\hskip 1em plus 0.5em minus 0.4em\relax IEEE,
  2018, pp. 616--621.

\bibitem{wong2018clustering}
L.~J. Wong, W.~C. Headley, S.~Andrews, R.~M. Gerdes, and A.~J. Michaels,
  ``Clustering learned cnn features from raw i/q data for emitter
  identification,'' in \emph{MILCOM 2018-2018 IEEE Military Communications
  Conference (MILCOM)}.\hskip 1em plus 0.5em minus 0.4em\relax IEEE, 2018, pp.
  26--33.

\bibitem{jafari2018iot}
H.~Jafari, O.~Omotere, D.~Adesina, H.-H. Wu, and L.~Qian, ``Iot devices
  fingerprinting using deep learning,'' in \emph{MILCOM 2018-2018 IEEE Military
  Communications Conference (MILCOM)}.\hskip 1em plus 0.5em minus 0.4em\relax
  IEEE, 2018, pp. 1--9.

\bibitem{bergstra2011algorithms}
J.~S. Bergstra, R.~Bardenet, Y.~Bengio, and B.~K{\'e}gl, ``Algorithms for
  hyper-parameter optimization,'' in \emph{Advances in neural information
  processing systems}, 2011, pp. 2546--2554.

\bibitem{pelikan1999boa}
M.~Pelikan, D.~E. Goldberg, E.~Cant{\'u}-Paz \emph{et~al.}, ``Boa: The bayesian
  optimization algorithm,'' in \emph{Proceedings of the genetic and
  evolutionary computation conference GECCO-99}, vol.~1, 1999, pp. 525--532.

\bibitem{young2015optimizing}
S.~R. Young, D.~C. Rose, T.~P. Karnowski, S.-H. Lim, and R.~M. Patton,
  ``Optimizing deep learning hyper-parameters through an evolutionary
  algorithm,'' in \emph{Proceedings of the Workshop on Machine Learning in
  High-Performance Computing Environments}, 2015, pp. 1--5.

\bibitem{elsken2018neural}
T.~Elsken, J.~H. Metzen, and F.~Hutter, ``Neural architecture search: A
  survey,'' \emph{arXiv preprint arXiv:1808.05377}, 2018.

\bibitem{liu2017survey}
W.~Liu, Z.~Wang, X.~Liu, N.~Zeng, Y.~Liu, and F.~E. Alsaadi, ``A survey of deep
  neural network architectures and their applications,'' \emph{Neurocomputing},
  vol. 234, pp. 11--26, 2017.

\bibitem{truong2019towards}
A.~Truong, A.~Walters, J.~Goodsitt, K.~Hines, B.~Bruss, and R.~Farivar,
  ``Towards automated machine learning: Evaluation and comparison of automl
  approaches and tools,'' \emph{arXiv preprint arXiv:1908.05557}, 2019.

\bibitem{AutoMLNeuralArchSearch}
M.~Lindauer, ``Literature on neural architecture search,''
  \url{https://www.automl.org/automl/literature-on-neural-architecture-search/},
  Feb. 2020.

\bibitem{AwesomeNASDxy}
X.~Dong, ``Awesome-nas: A curated list of neural architecture search (nas)
  resources.'' \url{https://github.com/D-X-Y/Awesome-NAS}, Jan. 2020.

\bibitem{zanchettin2006methodology}
C.~Zanchettin and T.~B. Ludermir, ``A methodology to train and improve
  artificial neural networks' weights and connections,'' in \emph{The 2006 IEEE
  International Joint Conference on Neural Network Proceedings}.\hskip 1em plus
  0.5em minus 0.4em\relax IEEE, 2006, pp. 5267--5274.

\bibitem{islam2009new}
M.~M. Islam, M.~A. Sattar, M.~F. Amin, X.~Yao, and K.~Murase, ``A new adaptive
  merging and growing algorithm for designing artificial neural networks,''
  \emph{IEEE Transactions on Systems, Man, and Cybernetics, Part B
  (Cybernetics)}, vol.~39, no.~3, pp. 705--722, 2009.

\bibitem{lam2001tuning}
H.~Lam, S.~Ling, F.~H. Leung, and P.~K.-S. Tam, ``Tuning of the structure and
  parameters of neural network using an improved genetic algorithm,'' in
  \emph{IECON'01. 27th Annual Conference of the IEEE Industrial Electronics
  Society (Cat. No. 37243)}, vol.~1.\hskip 1em plus 0.5em minus 0.4em\relax
  IEEE, 2001, pp. 25--30.

\bibitem{alvarez2016learning}
J.~M. Alvarez and M.~Salzmann, ``Learning the number of neurons in deep
  networks,'' in \emph{Advances in Neural Information Processing Systems},
  2016, pp. 2270--2278.

\bibitem{guo2016dynamic}
Y.~Guo, A.~Yao, and Y.~Chen, ``Dynamic network surgery for efficient dnns,'' in
  \emph{Advances in neural information processing systems}, 2016, pp.
  1379--1387.

\bibitem{han2015learning}
S.~Han, J.~Pool, J.~Tran, and W.~Dally, ``Learning both weights and connections
  for efficient neural network,'' in \emph{Advances in neural information
  processing systems}, 2015, pp. 1135--1143.

\bibitem{islam2003constructive}
M.~M. Islam, X.~Yao, and K.~Murase, ``A constructive algorithm for training
  cooperative neural network ensembles,'' \emph{IEEE Transactions on neural
  networks}, vol.~14, no.~4, pp. 820--834, 2003.

\bibitem{narasimha2008integrated}
P.~L. Narasimha, W.~H. Delashmit, M.~T. Manry, J.~Li, and F.~Maldonado, ``An
  integrated growing-pruning method for feedforward network training,''
  \emph{Neurocomputing}, vol.~71, no. 13-15, pp. 2831--2847, 2008.

\bibitem{ma2003new}
L.~Ma and K.~Khorasani, ``A new strategy for adaptively constructing multilayer
  feedforward neural networks,'' \emph{Neurocomputing}, vol.~51, pp. 361--385,
  2003.

\bibitem{cortes2017adanet}
C.~Cortes, X.~Gonzalvo, V.~Kuznetsov, M.~Mohri, and S.~Yang, ``Adanet: Adaptive
  structural learning of artificial neural networks,'' in \emph{Proceedings of
  the 34th International Conference on Machine Learning-Volume 70}.\hskip 1em
  plus 0.5em minus 0.4em\relax JMLR. org, 2017, pp. 874--883.

\bibitem{liu2018progressive}
C.~Liu, B.~Zoph, M.~Neumann, J.~Shlens, W.~Hua, L.-J. Li, L.~Fei-Fei,
  A.~Yuille, J.~Huang, and K.~Murphy, ``Progressive neural architecture
  search,'' in \emph{Proceedings of the European Conference on Computer Vision
  (ECCV)}, 2018, pp. 19--34.

\bibitem{stanley2002efficient}
K.~O. Stanley and R.~Miikkulainen, ``Efficient evolution of neural network
  topologies,'' in \emph{Proceedings of the 2002 Congress on Evolutionary
  Computation. CEC'02 (Cat. No. 02TH8600)}, vol.~2.\hskip 1em plus 0.5em minus
  0.4em\relax IEEE, 2002, pp. 1757--1762.

\bibitem{stanley2002evolving}
------, ``Evolving neural networks through augmenting topologies,''
  \emph{Evolutionary computation}, vol.~10, no.~2, pp. 99--127, 2002.

\bibitem{liang2019evolutionary}
J.~Liang, E.~Meyerson, B.~Hodjat, D.~Fink, K.~Mutch, and R.~Miikkulainen,
  ``Evolutionary neural automl for deep learning,'' in \emph{Proceedings of the
  Genetic and Evolutionary Computation Conference}, 2019, pp. 401--409.

\bibitem{baker2016designing}
B.~Baker, O.~Gupta, N.~Naik, and R.~Raskar, ``Designing neural network
  architectures using reinforcement learning,'' \emph{arXiv preprint
  arXiv:1611.02167}, 2016.

\bibitem{tan2019mnasnet}
M.~Tan, B.~Chen, R.~Pang, V.~Vasudevan, M.~Sandler, A.~Howard, and Q.~V. Le,
  ``Mnasnet: Platform-aware neural architecture search for mobile,'' in
  \emph{Proceedings of the IEEE Conference on Computer Vision and Pattern
  Recognition}, 2019, pp. 2820--2828.

\bibitem{hsu2018monas}
C.-H. Hsu, S.-H. Chang, J.-H. Liang, H.-P. Chou, C.-H. Liu, S.-C. Chang, J.-Y.
  Pan, Y.-T. Chen, W.~Wei, and D.-C. Juan, ``Monas: Multi-objective neural
  architecture search using reinforcement learning,'' \emph{arXiv preprint
  arXiv:1806.10332}, 2018.

\bibitem{liu2018darts}
H.~Liu, K.~Simonyan, and Y.~Yang, ``Darts: Differentiable architecture
  search,'' \emph{arXiv preprint arXiv:1806.09055}, 2018.

\bibitem{chen2019progressive}
X.~Chen, L.~Xie, J.~Wu, and Q.~Tian, ``Progressive differentiable architecture
  search: Bridging the depth gap between search and evaluation,'' in
  \emph{Proceedings of the IEEE International Conference on Computer Vision},
  2019, pp. 1294--1303.

\bibitem{dong2019searching}
X.~Dong and Y.~Yang, ``Searching for a robust neural architecture in four gpu
  hours,'' in \emph{Proceedings of the IEEE Conference on Computer Vision and
  Pattern Recognition}, 2019, pp. 1761--1770.

\bibitem{ying2019bench}
C.~Ying, A.~Klein, E.~Real, E.~Christiansen, K.~Murphy, and F.~Hutter,
  ``Nas-bench-101: Towards reproducible neural architecture search,''
  \emph{arXiv preprint arXiv:1902.09635}, 2019.

\bibitem{dong2020bench}
X.~Dong and Y.~Yang, ``Nas-bench-102: Extending the scope of reproducible
  neural architecture search,'' \emph{arXiv preprint arXiv:2001.00326}, 2020.

\bibitem{youssef2018machine}
K.~Youssef, L.~Bouchard, K.~Haigh, J.~Silovsky, B.~Thapa, and C.~Vander~Valk,
  ``Machine learning approach to rf transmitter identification,'' \emph{IEEE
  Journal of Radio Frequency Identification}, vol.~2, no.~4, pp. 197--205,
  2018.

\bibitem{scheirer2012toward}
W.~J. Scheirer, A.~de~Rezende~Rocha, A.~Sapkota, and T.~E. Boult, ``Toward open
  set recognition,'' \emph{IEEE transactions on pattern analysis and machine
  intelligence}, vol.~35, no.~7, pp. 1757--1772, 2012.

\bibitem{bendale2016towards}
A.~Bendale and T.~E. Boult, ``Towards open set deep networks,'' in
  \emph{Proceedings of the IEEE conference on computer vision and pattern
  recognition}, 2016, pp. 1563--1572.

\bibitem{gunning2017explainable}
D.~Gunning, ``Explainable artificial intelligence (xai),'' \emph{Defense
  Advanced Research Projects Agency (DARPA), nd Web}, vol.~2, 2017.

\bibitem{tan2018survey}
C.~Tan, F.~Sun, T.~Kong, W.~Zhang, C.~Yang, and C.~Liu, ``A survey on deep
  transfer learning,'' in \emph{International conference on artificial neural
  networks}.\hskip 1em plus 0.5em minus 0.4em\relax Springer, 2018, pp.
  270--279.

\bibitem{parisi2019continual}
G.~I. Parisi, R.~Kemker, J.~L. Part, C.~Kanan, and S.~Wermter, ``Continual
  lifelong learning with neural networks: A review,'' \emph{Neural Networks},
  2019.

\bibitem{shin2017continual}
H.~Shin, J.~K. Lee, J.~Kim, and J.~Kim, ``Continual learning with deep
  generative replay,'' in \emph{Advances in Neural Information Processing
  Systems}, 2017, pp. 2990--2999.

\bibitem{girosi1995regularization}
F.~Girosi, M.~Jones, and T.~Poggio, ``Regularization theory and neural networks
  architectures,'' \emph{Neural computation}, vol.~7, no.~2, pp. 219--269,
  1995.

\bibitem{kirkpatrick2017overcoming}
J.~Kirkpatrick, R.~Pascanu, N.~Rabinowitz, J.~Veness, G.~Desjardins, A.~A.
  Rusu, K.~Milan, J.~Quan, T.~Ramalho, A.~Grabska-Barwinska \emph{et~al.},
  ``Overcoming catastrophic forgetting in neural networks,'' \emph{Proceedings
  of the national academy of sciences}, vol. 114, no.~13, pp. 3521--3526, 2017.

\bibitem{yoon2017lifelong}
J.~Yoon, E.~Yang, J.~Lee, and S.~J. Hwang, ``Lifelong learning with dynamically
  expandable networks,'' \emph{arXiv preprint arXiv:1708.01547}, 2017.

\bibitem{axell2012spectrum}
E.~Axell, G.~Leus, E.~G. Larsson, and H.~V. Poor, ``Spectrum sensing for
  cognitive radio: State-of-the-art and recent advances,'' \emph{IEEE signal
  processing magazine}, vol.~29, no.~3, pp. 101--116, 2012.

\bibitem{sivanathan2020detecting}
A.~Sivanathan, H.~H. Gharakheili, and V.~Sivaraman, ``Detecting behavioral
  change of iot devices using clustering-based network traffic modeling,''
  \emph{IEEE Internet of Things Journal}, 2020.

\bibitem{sivanathan2019inferring}
------, ``Inferring iot device types from network behavior using unsupervised
  clustering,'' in \emph{2019 IEEE 44th Conference on Local Computer Networks
  (LCN)}.\hskip 1em plus 0.5em minus 0.4em\relax IEEE, 2019, pp. 230--233.

\bibitem{ortiz2019devicemien}
J.~Ortiz, C.~Crawford, and F.~Le, ``Devicemien: network device behavior
  modeling for identifying unknown iot devices,'' in \emph{Proceedings of the
  International Conference on Internet of Things Design and Implementation},
  2019, pp. 106--117.

\bibitem{liu2019synchronization}
G.~Liu, R.~Zhang, C.~Wang, and L.~Liu, ``Synchronization-free gps spoofing
  detection with crowdsourced air traffic control data,'' in \emph{2019 20th
  IEEE International Conference on Mobile Data Management (MDM)}.\hskip 1em
  plus 0.5em minus 0.4em\relax IEEE, 2019, pp. 260--268.

\bibitem{monteiro2015detectinglb}
M.~Monteiro, A.~Barreto, T.~Kacem, D.~Wijesekera, and P.~Costa, ``Detecting
  malicious ads-b transmitters using a low-bandwidth sensor network,'' in
  \emph{2015 18th International Conference on Information Fusion
  (Fusion)}.\hskip 1em plus 0.5em minus 0.4em\relax IEEE, 2015, pp. 1696--1701.

\bibitem{xu2012high}
B.~Xu, G.~Sun, R.~Yu, and Z.~Yang, ``High-accuracy tdoa-based localization
  without time synchronization,'' \emph{IEEE Transactions on Parallel and
  Distributed Systems}, vol.~24, no.~8, pp. 1567--1576, 2012.

\bibitem{schafer2014bringing}
M.~Sch{\"a}fer, M.~Strohmeier, V.~Lenders, I.~Martinovic, and M.~Wilhelm,
  ``Bringing up opensky: A large-scale ads-b sensor network for research,'' in
  \emph{Proceedings of the 13th international symposium on Information
  processing in sensor networks}.\hskip 1em plus 0.5em minus 0.4em\relax IEEE
  Press, 2014, pp. 83--94.

\bibitem{liu2007survey}
H.~Liu, H.~Darabi, P.~Banerjee, and J.~Liu, ``Survey of wireless indoor
  positioning techniques and systems,'' \emph{IEEE Transactions on Systems,
  Man, and Cybernetics, Part C (Applications and Reviews)}, vol.~37, no.~6, pp.
  1067--1080, 2007.

\bibitem{guvenc2009survey}
I.~Guvenc and C.-C. Chong, ``A survey on toa based wireless localization and
  nlos mitigation techniques,'' \emph{IEEE Communications Surveys \&
  Tutorials}, vol.~11, no.~3, pp. 107--124, 2009.

\bibitem{han2013localization}
G.~Han, H.~Xu, T.~Q. Duong, J.~Jiang, and T.~Hara, ``Localization algorithms of
  wireless sensor networks: a survey,'' \emph{Telecommunication Systems},
  vol.~52, no.~4, pp. 2419--2436, 2013.

\bibitem{li2017fingerprints}
Q.~Li, H.~Fan, W.~Sun, J.~Li, L.~Chen, and Z.~Liu, ``Fingerprints in the air:
  unique identification of wireless devices using rf rss fingerprints,''
  \emph{IEEE Sensors Journal}, vol.~17, no.~11, pp. 3568--3579, 2017.

\bibitem{zheleva2015txminer}
M.~Zheleva, R.~Chandra, A.~Chowdhery, A.~Kapoor, and P.~Garnett, ``Txminer:
  Identifying transmitters in real-world spectrum measurements,'' in \emph{2015
  IEEE International Symposium on Dynamic Spectrum Access Networks
  (DySPAN)}.\hskip 1em plus 0.5em minus 0.4em\relax IEEE, 2015, pp. 94--105.

\bibitem{nijsure2015adaptive}
Y.~A. Nijsure, G.~Kaddoum, G.~Gagnon, F.~Gagnon, C.~Yuen, and R.~Mahapatra,
  ``Adaptive air-to-ground secure communication system based on ads-b and
  wide-area multilateration,'' \emph{IEEE Transactions on Vehicular
  Technology}, vol.~65, no.~5, pp. 3150--3165, 2015.

\bibitem{monteiro2015detecting}
M.~Monteiro, A.~Barreto, T.~Kacem, J.~Carvalho, D.~Wijesekera, and P.~Costa,
  ``Detecting malicious ads-b broadcasts using wide area multilateration,'' in
  \emph{2015 IEEE/AIAA 34th Digital Avionics Systems Conference (DASC)}.\hskip
  1em plus 0.5em minus 0.4em\relax IEEE, 2015, pp. 4A3--1.

\bibitem{schubert2017dbscan}
E.~Schubert, J.~Sander, M.~Ester, H.~P. Kriegel, and X.~Xu, ``Dbscan revisited,
  revisited: why and how you should (still) use dbscan,'' \emph{ACM
  Transactions on Database Systems (TODS)}, vol.~42, no.~3, pp. 1--21, 2017.

\bibitem{ankerst1999optics}
M.~Ankerst, M.~M. Breunig, H.-P. Kriegel, and J.~Sander, ``Optics: ordering
  points to identify the clustering structure,'' \emph{ACM Sigmod record},
  vol.~28, no.~2, pp. 49--60, 1999.

\bibitem{pelleg2000x}
D.~Pelleg, A.~W. Moore \emph{et~al.}, ``X-means: Extending k-means with
  efficient estimation of the number of clusters.'' in \emph{Icml}, vol.~1,
  2000, pp. 727--734.

\bibitem{hwang2020unsupervised}
R.-H. Hwang, M.-C. Peng, C.-W. Huang, P.-C. Lin, and V.-L. Nguyen, ``An
  unsupervised deep learning model for early network traffic anomaly
  detection,'' \emph{IEEE Access}, vol.~8, pp. 30\,387--30\,399, 2020.

\bibitem{hwang2019detecting}
R.-H. Hwang, M.-C. Peng, and C.-W. Huang, ``Detecting iot malicious traffic
  based on autoencoder and convolutional neural network,'' in \emph{2019 IEEE
  Globecom Workshops (GC Wkshps)}.\hskip 1em plus 0.5em minus 0.4em\relax IEEE,
  2019, pp. 1--6.

\bibitem{alipour2015wireless}
H.~Alipour, Y.~B. Al-Nashif, P.~Satam, and S.~Hariri, ``Wireless anomaly
  detection based on ieee 802.11 behavior analysis,'' \emph{IEEE transactions
  on information forensics and security}, vol.~10, no.~10, pp. 2158--2170,
  2015.

\bibitem{sehatbakhsh2018syndrome}
N.~Sehatbakhsh, M.~Alam, A.~Nazari, A.~Zajic, and M.~Prvulovic, ``Syndrome:
  Spectral analysis for anomaly detection on medical iot and embedded
  devices,'' in \emph{2018 IEEE international symposium on hardware oriented
  security and trust (HOST)}.\hskip 1em plus 0.5em minus 0.4em\relax IEEE,
  2018, pp. 1--8.

\bibitem{hamza2019detecting}
A.~Hamza, H.~H. Gharakheili, T.~A. Benson, and V.~Sivaraman, ``Detecting
  volumetric attacks on lot devices via sdn-based monitoring of mud activity,''
  in \emph{Proceedings of the 2019 ACM Symposium on SDN Research}, 2019, pp.
  36--48.

\bibitem{thing2017ieee}
V.~L. Thing, ``Ieee 802.11 network anomaly detection and attack classification:
  A deep learning approach,'' in \emph{2017 IEEE Wireless Communications and
  Networking Conference (WCNC)}.\hskip 1em plus 0.5em minus 0.4em\relax IEEE,
  2017, pp. 1--6.

\bibitem{sridharan2018wadac}
R.~Sridharan, R.~R. Maiti, and N.~O. Tippenhauer, ``Wadac: Privacy-preserving
  anomaly detection and attack classification on wireless traffic,'' in
  \emph{Proceedings of the 11th ACM Conference on Security \& Privacy in
  Wireless and Mobile Networks}, 2018, pp. 51--62.

\bibitem{luo2018distributed}
T.~Luo and S.~G. Nagarajan, ``Distributed anomaly detection using autoencoder
  neural networks in wsn for iot,'' in \emph{2018 IEEE International Conference
  on Communications (ICC)}.\hskip 1em plus 0.5em minus 0.4em\relax IEEE, 2018,
  pp. 1--6.

\bibitem{meidan2018n}
Y.~Meidan, M.~Bohadana, Y.~Mathov, Y.~Mirsky, A.~Shabtai, D.~Breitenbacher, and
  Y.~Elovici, ``N-baiot—network-based detection of iot botnet attacks using
  deep autoencoders,'' \emph{IEEE Pervasive Computing}, vol.~17, no.~3, pp.
  12--22, 2018.

\bibitem{chen2001one}
Y.~Chen, X.~S. Zhou, and T.~S. Huang, ``One-class svm for learning in image
  retrieval,'' in \emph{Proceedings 2001 International Conference on Image
  Processing (Cat. No. 01CH37205)}, vol.~1.\hskip 1em plus 0.5em minus
  0.4em\relax IEEE, 2001, pp. 34--37.

\bibitem{ding2013anomaly}
Z.~Ding and M.~Fei, ``An anomaly detection approach based on isolation forest
  algorithm for streaming data using sliding window,'' \emph{IFAC Proceedings
  Volumes}, vol.~46, no.~20, pp. 12--17, 2013.

\bibitem{kriegel2009loop}
H.-P. Kriegel, P.~Kr{\"o}ger, E.~Schubert, and A.~Zimek, ``Loop: local outlier
  probabilities,'' in \emph{Proceedings of the 18th ACM conference on
  Information and knowledge management}, 2009, pp. 1649--1652.

\bibitem{parwez2017big}
M.~S. Parwez, D.~B. Rawat, and M.~Garuba, ``Big data analytics for
  user-activity analysis and user-anomaly detection in mobile wireless
  network,'' \emph{IEEE Transactions on Industrial Informatics}, vol.~13,
  no.~4, pp. 2058--2065, 2017.

\bibitem{wei2012intrusion}
M.~Wei and K.~Kim, ``Intrusion detection scheme using traffic prediction for
  wireless industrial networks,'' \emph{journal of communications and
  networks}, vol.~14, no.~3, pp. 310--318, 2012.

\bibitem{tandiya2018deep}
N.~Tandiya, A.~Jauhar, V.~Marojevic, and J.~H. Reed, ``Deep predictive coding
  neural network for rf anomaly detection in wireless networks,'' in \emph{2018
  IEEE International Conference on Communications Workshops (ICC
  Workshops)}.\hskip 1em plus 0.5em minus 0.4em\relax IEEE, 2018, pp. 1--6.

\bibitem{rajendran2018saife}
S.~Rajendran, W.~Meert, V.~Lenders, and S.~Pollin, ``Saife: Unsupervised
  wireless spectrum anomaly detection with interpretable features,'' in
  \emph{2018 IEEE International Symposium on Dynamic Spectrum Access Networks
  (DySPAN)}.\hskip 1em plus 0.5em minus 0.4em\relax IEEE, 2018, pp. 1--9.

\bibitem{lotter2016deep}
W.~Lotter, G.~Kreiman, and D.~Cox, ``Deep predictive coding networks for video
  prediction and unsupervised learning,'' \emph{arXiv preprint
  arXiv:1605.08104}, 2016.

\bibitem{sankhe2019oracle}
K.~Sankhe, M.~Belgiovine, F.~Zhou, S.~Riyaz, S.~Ioannidis, and K.~Chowdhury,
  ``Oracle: Optimized radio classification through convolutional neural
  networks,'' in \emph{IEEE INFOCOM 2019-IEEE Conference on Computer
  Communications}.\hskip 1em plus 0.5em minus 0.4em\relax IEEE, 2019, pp.
  370--378.

\bibitem{sankhe2019no}
K.~Sankhe, M.~Belgiovine, F.~Zhou, L.~Angioloni, F.~Restuccia, S.~D’Oro,
  T.~Melodia, S.~Ioannidis, and K.~Chowdhury, ``No radio left behind: Radio
  fingerprinting through deep learning of physical-layer hardware
  impairments,'' \emph{IEEE Transactions on Cognitive Communications and
  Networking}, vol.~6, no.~1, pp. 165--178, 2019.

\bibitem{soltanimore}
N.~Soltani, K.~Sankhe, J.~Dy, S.~Ioannidis, and K.~Chowdhury, ``More is better:
  Data augmentation for channel-resilient rf fingerprinting.''

\bibitem{pan2009survey}
S.~J. Pan and Q.~Yang, ``A survey on transfer learning,'' \emph{IEEE
  Transactions on knowledge and data engineering}, vol.~22, no.~10, pp.
  1345--1359, 2009.

\bibitem{niu2020transfer}
S.~Niu, J.~Wang, Y.~Liu, and H.~Song, ``Transfer learning based data-efficient
  machine learning enabled classification,'' in \emph{2020 IEEE Intl Conf on
  Dependable, Autonomic and Secure Computing, Intl Conf on Pervasive
  Intelligence and Computing, Intl Conf on Cloud and Big Data Computing, Intl
  Conf on Cyber Science and Technology Congress
  (DASC/PiCom/CBDCom/CyberSciTech)}.\hskip 1em plus 0.5em minus 0.4em\relax
  IEEE, 2020, pp. 620--626.

\bibitem{nguyen2017matthan}
P.~Nguyen, H.~Truong, M.~Ravindranathan, A.~Nguyen, R.~Han, and T.~Vu,
  ``Matthan: Drone presence detection by identifying physical signatures in the
  drone's rf communication,'' in \emph{Proceedings of the 15th Annual
  International Conference on Mobile Systems, Applications, and
  Services}.\hskip 1em plus 0.5em minus 0.4em\relax ACM, 2017, pp. 211--224.

\bibitem{wang2020counter}
J.~Wang, Y.~Liu, and H.~Song, ``Counter-unmanned aircraft system (s)(c-uas):
  State of the art, challenges and future trends,'' \emph{arXiv preprint
  arXiv:2008.12461}, 2020.

\bibitem{shi2019generative}
Y.~Shi, K.~Davaslioglu, and Y.~E. Sagduyu, ``Generative adversarial network for
  wireless signal spoofing,'' in \emph{Proceedings of the ACM Workshop on
  Wireless Security and Machine Learning}.\hskip 1em plus 0.5em minus
  0.4em\relax ACM, 2019, pp. 55--60.

\bibitem{mohanti2020airid}
S.~M. Mohanti, N.~S. Soltani, K.~Sankhe, D.~Jaisinghani, M.~Di~Felice, and
  K.~Chowdhury, ``Airid: Injecting a custom rf fingerprint for enhanced uav
  identification using deep learning,'' in \emph{IEEE Global Communications
  Conference}, 2020.

\end{thebibliography}
%

%

\begin{IEEEbiography}[{\includegraphics[width=1in,height=1.25in,clip,keepaspectratio]{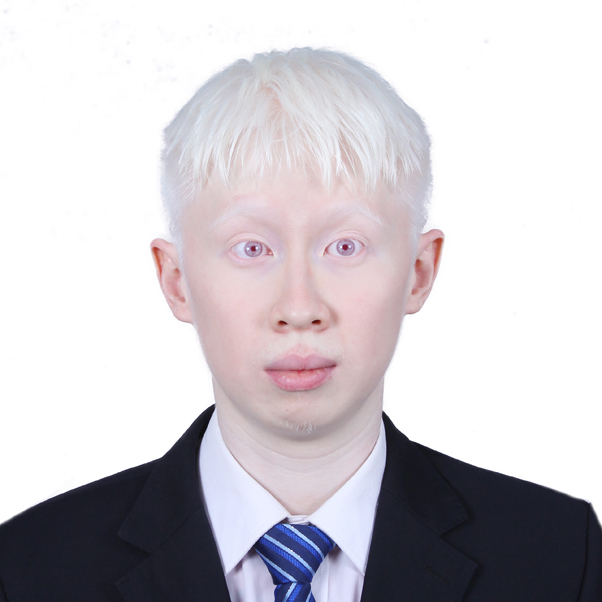}}]{YongXin Liu}
(LIU11@my.erau.edu) received his first Ph.D. from South China University of Technology (SCUT) and currently working towards his second Ph.D. in the Department of Electrical Engineering and Computer Science, Embry-Riddle Aeronautical University, Daytona Beach, FL. His major research interests include data mining, wireless networks, the Internet of Things, and unmanned aerial vehicles. 

\end{IEEEbiography}

\begin{IEEEbiography}[{\includegraphics[width=1in,height=1.25in,clip,keepaspectratio]{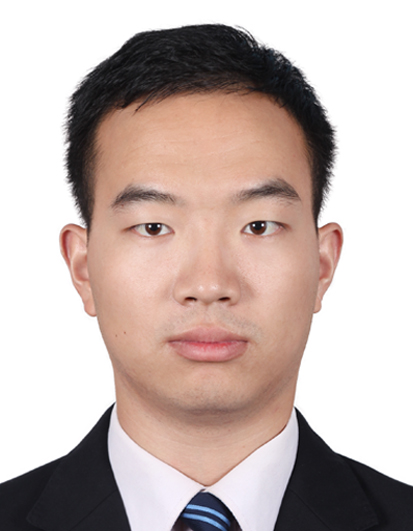}}]{Jian Wang}
(wangj14@my.erau.edu) is a Ph.D. student in the Department of Electrical, Computer, Software, and Systems Engineering (ECSSE), Embry-Riddle Aeronautical University (ERAU), Daytona Beach, Florida, and a graduate research assistant in the Security and Optimization for Networked Globe Laboratory (SONG Lab, www.SONGLab.us). He received his M.S. from South China Agricultural University (SCAU) in 2017 and B.S. from Nanyang Normal University in 2014. His major research interests include wireless networks, unmanned aerial systems, and machine learning.
\end{IEEEbiography}

\begin{IEEEbiography}[{\includegraphics[width=1in,height=1.25in,clip,keepaspectratio]{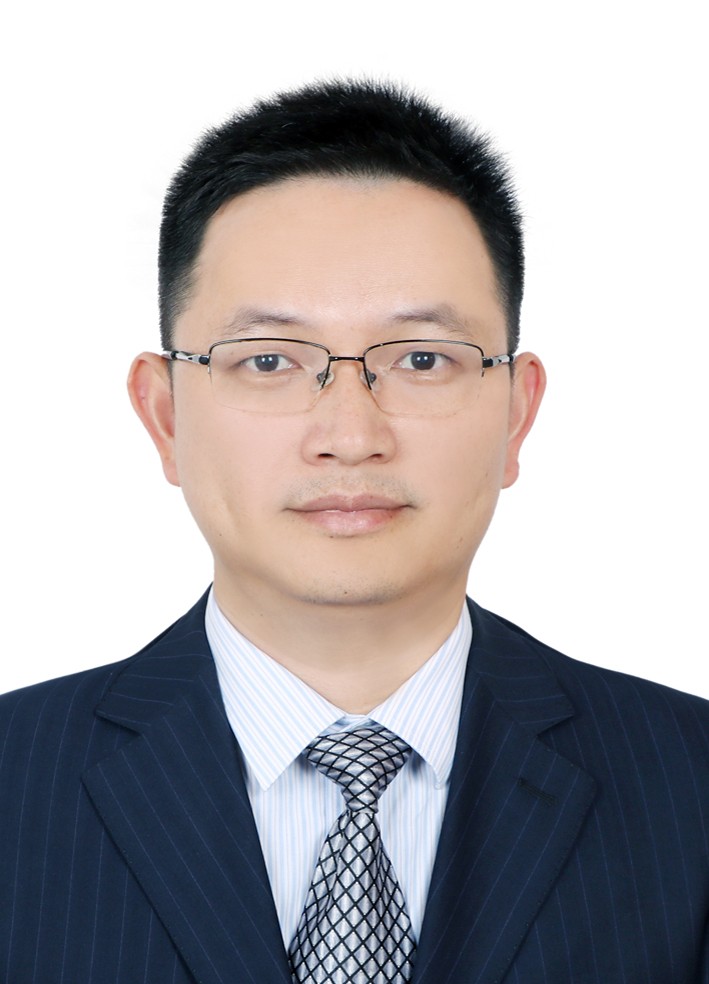}}]{Jianqiang Li}
(lijq@szu.edu.cn) received his B.S. and Ph.D.
degrees from the South China University of
Technology in 2003 and 2008, respectively. He is a Professor with the College of Computer
and Software Engineering, Shenzhen University,
Shenzhen, China. He is leading two projects funded
by the National Natural Science Foundation of
China and two projects funded by the Natural
Science Foundation of Guangdong, China. His major
research interests include Internet of Things, robotic,
hybrid systems, and embedded systems.
\end{IEEEbiography}

\begin{IEEEbiography}[{\includegraphics[width=1in,height=1.25in,clip,keepaspectratio]{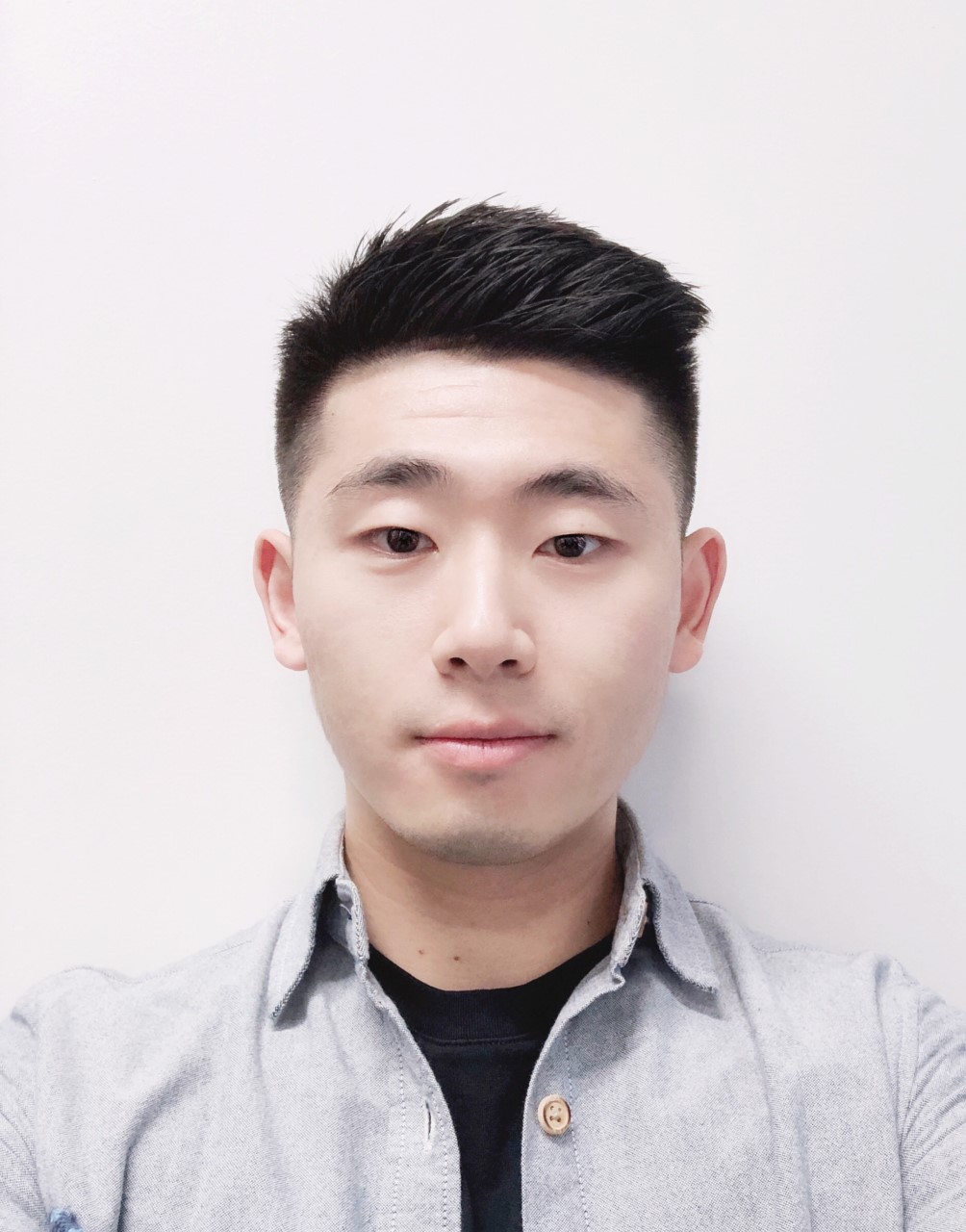}}]{Shuteng Niu}
(shutengn@my.erau.edu) is a Ph.D. student in the Department of Electrical, Computer, Software, and Systems Engineering (ECSSE), Embry-Riddle Aeronautical University (ERAU), Daytona Beach, Florida, and a graduate research assistant in the Security and Optimization for Networked Globe Laboratory (SONG Lab, www.SONGLab.us). He received his M.S. from Embry-Riddle Aeronautical University (ERAU) in 2018 and B.S. from Civil Aviation University of China (CAUC) in 2015. His major research interests include machine learning, data mining, and signal processing.
\end{IEEEbiography}

\begin{IEEEbiography}[{\includegraphics[width=1in,height=1.25in,clip,keepaspectratio]{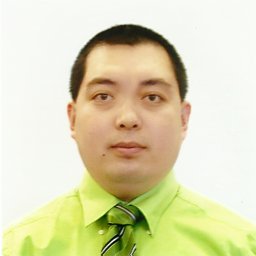}}]{Houbing Song} (M'12-SM'14) received the Ph.D. degree in electrical engineering from the University of Virginia, Charlottesville, VA, in August 2012.

In August 2017, he joined the Department of Electrical Engineering and Computer Science, Embry-Riddle Aeronautical University, Daytona Beach, FL, where he is currently an Assistant Professor and the Director of the Security and Optimization for Networked Globe Laboratory (SONG Lab, www.SONGLab.us). He has served as an Associate Technical Editor for IEEE Communications Magazine (2017-present), an Associate Editor for IEEE Internet of Things Journal (2020-present) and IEEE Journal on Miniaturization for Air and Space Systems (J-MASS) (2020-present), and a Guest Editor for IEEE Journal on Selected Areas in Communications (J-SAC), IEEE Internet of Things Journal, IEEE Transactions on Industrial Informatics, IEEE Sensors Journal, IEEE Transactions on Intelligent Transportation Systems, and IEEE Network. He is the editor of six books, including Big Data Analytics for Cyber-Physical Systems: Machine Learning for the Internet of Things, Elsevier, 2019,  Smart Cities: Foundations, Principles and Applications, Hoboken, NJ: Wiley, 2017, Security and Privacy in Cyber-Physical Systems: Foundations, Principles and Applications, Chichester, UK: Wiley-IEEE Press, 2017, Cyber-Physical Systems: Foundations, Principles and Applications, Boston, MA: Academic Press, 2016, and Industrial Internet of Things: Cybermanufacturing Systems, Cham, Switzerland: Springer, 2016.  He is the author of more than 100 articles. His research interests include cyber-physical systems, cybersecurity and privacy, internet of things, edge computing, AI/machine learning, big data analytics, unmanned aircraft systems, connected vehicle, smart and connected health, and wireless communications and networking. His research has been featured by popular news media outlets, including IEEE GlobalSpec's Engineering360, USA Today, U.S. News \& World Report, Fox News, Association for Unmanned Vehicle Systems International (AUVSI), Forbes, WFTV, and New Atlas.

Dr. Song is a senior member of ACM and an ACM Distinguished Speaker. Dr. Song was a recipient of the Best Paper Award from the 12th IEEE International Conference on Cyber, Physical and Social Computing (CPSCom-2019), the Best Paper Award from the 2nd IEEE International Conference on Industrial Internet (ICII 2019), the Best Paper Award from the 19th Integrated Communication, Navigation and Surveillance technologies (ICNS 2019) Conference, the Best Paper Award from the 6th IEEE International Conference on Cloud and Big Data Computing (CBDCom 2020), and the Best Paper Award from the 15th International Conference on Wireless Algorithms, Systems, and Applications (WASA 2020).

\end{IEEEbiography}




\end{document}